\documentclass[twocolumn,longbibliography,floatfix,prb]{revtex4-1}
\usepackage{graphicx}
\usepackage{bm}
\usepackage{color}
\usepackage[normalem]{ulem}
\usepackage{mathtools}

\newcommand{\bk}{{\bf k}}
\newcommand{\bq}{{\bf q}}
\newcommand{\bdq}{\Delta {\bf q}}
\newcommand{\bqt}{\frac{\bf q}2}

\newcommand{\br}{{\bf r}}

\begin{document}
\title{Structure of the Charge-Density Wave in Cuprate Superconductors: Lessons from NMR}
\author{W.  A. Atkinson$^1$} \email{billatkinson@trentu.ca}
\author{S. Ufkes$^{1,2}$} 
\author{A. P. Kampf$^3$} 
\affiliation{$^1$Department of Physics and Astronomy, Trent University, Peterborough Ontario, Canada, K9L0G2}
\affiliation{$^2$Department of Physics, University of Toronto, Toronto Ontario, Canada, M5S1A7}
\affiliation{$^3$Theoretical Physics III, Center for Electronic Correlations and Magnetism, Institute of Physics, University of Augsburg, 86135 Augsburg, Germany}
\date{\today}
\begin{abstract}
Using a mix of numerical and analytic methods, we show that recent NMR $^{17}$O measurements provide detailed information about the structure of the charge-density wave (CDW) phase  in underdoped YBa$_2$Cu$_3$O$_{6+x}$.  We perform Bogoliubov-de Gennes (BdG) calculations of both the local density of states and the orbitally resolved charge density,  which are closely related to the magnetic and electric quadrupole contributions to the NMR spectrum, using a microscopic model that was shown previously to agree closely with x-ray experiments.    The BdG results reproduce qualitative features of the experimental spectrum extremely well.  These results are interpreted in terms of a generic ``hotspot'' model that allows one to trace the origins of the NMR lineshapes.  We find that four quantities---the orbital character of the Fermi surface at the hotspots, the Fermi surface curvature at the hotspots, the CDW correlation length, and the magnitude of the subdominant CDW component---are key in determining the  lineshapes.
\end{abstract}
\maketitle

\section{Introduction}
Charge-density waves occupy a significant portion of the  phase diagram of cuprate high-temperature superconductors and, along with $d$-wave superconductivity, antiferromagnetism, and the pseudogap, appear to be a generic feature of the underdoped cuprates.  The discovery of widespread charge ordering has led to interesting questions about the role of strong correlations in CDW formation,\cite{Bejas:2012wa,Faye:2017tx}  the connection to the pseudogap,\cite{Hamidian:2015us,Caprara:2017he,Verret:2017th,Chatterjee:2016ij} and the possible entanglement of  multiple order parameters.\cite{Efetov:2013,Hayward:2014eo,Tsvelik:2014ce,Wang:2014fr,Wang:2015iq,Atkinson:2016ba}  Despite the ubiquity of the CDWs,  their detailed  structure has only been established in certain special cases.  

CDWs were originally observed by  scanning tunneling microscopy experiments in the Bi-based cuprates, 
Bi$_2$Sr$_2$CaCu$_2$O$_{8+\delta}$ (Bi2212) and Bi$_{2-y}$Pb$_y$Sr$_{2-z}$La$_z$CuO$_{6+x}$ (Bi2201),  where both the periodicity and the intra-unit cell structure of the local density of states (LDOS) have been mapped out.\cite{Hoffman:2002bk,Kohsaka:2007hx,Wise:2008cd,Fujita:2014kg,Mesaros:2016uz}    Experimentally,  the CDW wavevectors $\bq^\ast$ lie along the Cu-O bond directions, so that Cu-O bonds perpendicular and parallel to $\bq^\ast$ are inequivalent.  Tunneling and x-ray experiments have further established that the LDOS modulations on the two inequivalent O sites are out of phase, so that when one is enhanced, the other is reduced.\cite{Fujita:2014kg,Mesaros:2016uz,Comin:2014vq,Comin:2016}  This unusual intra-unit cell structure is consistent with a   charge transfer between neighboring O sites, whose amplitude is modulated with period $2\pi/|\bq^\ast|$.  The notation ``$d$-wave CDW''  is commonly used to describe this state, though one does not generally expect a pure $d_{x^2-y^2}$ form factor when $|\bq^\ast|$ is not zero.\cite{Atkinson:2015hd} 

The universality of this  structure has  been harder to establish in  other cuprate families.  While x-ray experiments have obtained a comprehensive picture of $\bq^\ast$ as a function of doping  in YBa$_2$Cu$3$O$_{6+x}$ (YBCO$_{6+x}$),\cite{Ghiringhelli:2012bw,Chang:2012vf,Blackburn:2013bs,BlancoCanosa:2014ul,Chang:2016gz}  HgBa$_2$CuO$_{4+\delta}$ (Hg1201),\cite{Tabis:2014kb,Tabis:2017vp} and Bi2201,\cite{Comin:2013ck,Peng:2016jr} the form factor is more difficult to determine.  In YBCO$_{6+x}$, both elastic\cite{Forgan:2015ux} and resonant x-ray scattering\cite{Comin:2014vq}  point to an admixture of $d_{x^2-y^2}$ and $s$-symmetry charge densities.   We note that stripe order in the La-based cuprates appears to be distinct from the CDW order described above.  Recently, it was shown that La$_{1.875}$Ba$_{0.125}$CuO$_4$  exhibits charge correlations with wavevectors similar to those found in other materials at temperatures $T\sim 90$~K;\cite{Miao:2017bv} however, as $T$ is lowered additional quasistatic spin-stripe correlations develop that  modify the charge order.    Of particular relevance to this work, the doping dependence of the stripe and CDW wavevectors are completely different: while the CDW wavevectors evolve with doping in a way that is quantitatvely consistent with Fermi surface nesting scenarios,  the stripe wavevector evolves oppositely to what one would expect in a nesting scenario.\cite{Vojta:2009}  Because of these complications, it is unlikely that the model presented here applies to La-based cuprates.

Important  information, complementary to x-ray scattering, is available from NMR experiments.\cite{Wu:2011ke,Wu:2013,Wu:2015bt,Zhou:2016,Kawasaki:2017ue}  Because it is a local probe, NMR is sensitive to  inhomogeneities in  both the charge density and the LDOS, and historically NMR was the first technique to identify the existence of charge order in YBCO$_{6+x}$.\cite{Wu:2011ke}   In some well-known CDW materials, the structure of the CDW has been inferred from an analysis of NMR lineshapes.\cite{Butaud:1985eh,Ross:1990in,Skripov:1995ts,Berthier:2001ga,Ghoshray:2009dq}  Indeed, Kharkov and Sushkov were able to extract the size of the  $d$- and $s$-symmetry CDW components in YBCO$_{6+x}$,
without reference to their microscopic origins,  from an analysis of the  electric quadrupole data.\cite{Kharkov:2016tf} YBCO$_{6+x}$ presents a {particular} complication, {found also in other cuprates,\cite{Haase:2000vb}} in that the  quadrupole  and magnetic broadenings are comparable, and must be disentangled for a full analysis. 

To this end, we explore the NMR lineshapes in the context of a microscopic Hamiltonian.  In particular, we isolate the quantities that are principally responsible for determining the lineshapes. These are:  the degree of orthorhombicity (which selects the dimensionality of the CDW), the orbital character of the Fermi surface, the Fermi surface curvature at the hotspots, and the CDW correlation length.  As found elsewhere, disorder plays a key role both in nucleating the CDW at high $T$ and in limiting the correlation length at low $T$.\cite{Wu:2013,Campi:2015cva,Caplan:2015jm} 
 

A large body of theoretical work has shown that $d$-wave CDW instabilities emerge naturally from weak-coupling theories.\cite{Metlitski:2010vf,Holder:2012ks,Bejas:2012wa,Efetov:2013,Sachdev:2013bo,Sau:2013vw,Bulut:2013bz,Wang:2014fr,Pepin:2014tb} While these theories obtain an intra-unit cell structure similar to experiments, they generically predict that $\bq^\ast$ lies along the diagonal, rather than axial, directions.    In notable exceptions, it was shown via functional-renormalization group\cite{Yamakawa:2015hb,Tsuchiizu:2015vz} and Monte Carlo\cite{Wang:2017wk} calculations that for spin-fluctuation-mediated CDWs, axial order becomes dominant close to the spin-density wave quantum critical point.   While relevant at low doping, it is not clear that this mechanism is applicable throughout the CDW phase since spin correlation lengths are only a few lattice constants at higher doping levels.  It is also possible that strong correlations may influence the CDW:    axial order was shown to emerge for particular choices of model parameters within a Gutzwiller variational ansatz.\cite{Allais:2014kg}  Alternatively, it was shown that axial order appears when when the antinodal Fermi surface is removed, either by $d$-wave superconductivity\cite{Chowdhury:2014} or by
a Fermi surface reconstruction (FSR) mimicking the pseudogap.\cite{Atkinson:2014} In the latter case,  $\bq^\ast$ was found to agree quantitatively with experiments on YBCO$_{6+x}$ across a wide doping range.  The implication that the CDW emerges from the pseudogap, rather than causing it,  is supported by the observation that $\bq^\ast$ connects the tips of the Fermi arcs in the pseudogap phase, rather than nesting the sections of Fermi surface obliterated by the pseudogap.\cite{Comin:2013ck}  We  adopt the FSR model here.

\begin{figure}
\includegraphics[width=\columnwidth]{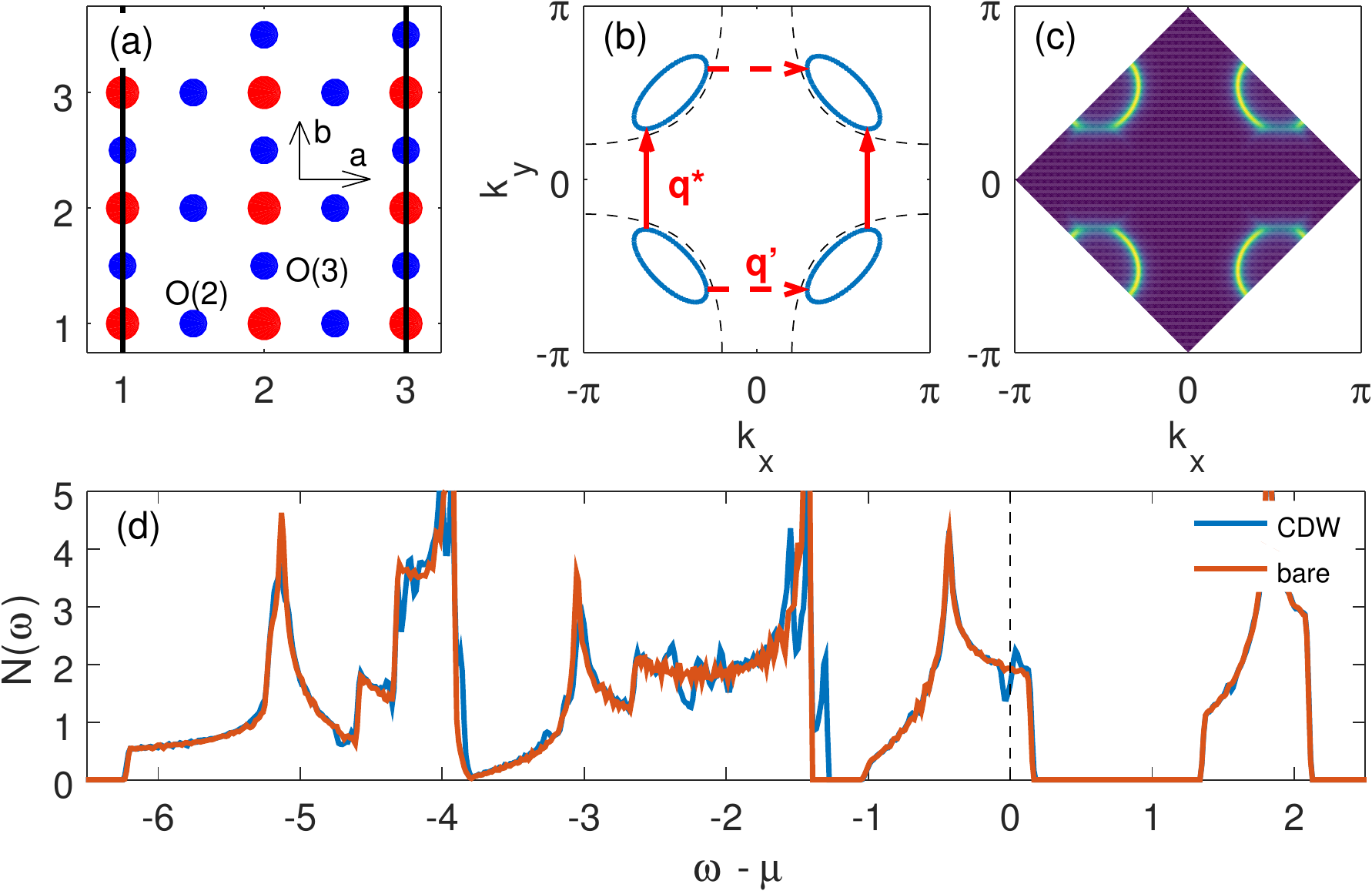}
\caption{Weak-coupling CDW model  in the clean limit.  (a) Schematic of the CuO$_2$ plane  for ideal (ortho-II) YBCO$_{6.5}$.   Cu sites (red circles), and O(2) and O(3) sites (blue) are indicated, as are the crystalline $a$ and $b$ axes.  The ortho-II structure is orthorhombic because  of 1D CuO chains running parallel to the $b$ axis above alternating rows of Cu sites (straight black lines). (b)  Unreconstructed (dashed black lines)  and reconstructed (solid blue lines) CuO$_2$ Fermi surfaces.  The leading charge instability for the reconstructed Fermi surface has a wavevector $\bq^\ast$ (solid arrows) that connects parallel Fermi surface sections. A second set of nesting wavevectors ${\bf q'}$ (dashed arrows)  connects additional hotspots; these are subdominant when the crystal is orthorhombic.  (c)  Spectral function at the chemical potential in the antiferromagnetic Brillouin zone for a {uniaxial} CDW with wavevector $\bq^\ast$.  The CDW washes out spectral intensity near the hotspots.  (d)   Density of states  for the bare band structure (red) and in the CDW phase (blue).  The model has six bands in total, with the conduction band  extending over $-1 < \omega -\mu < 0.2$.  The depletion of the hotspots by the CDW opens a partial gap at the chemical potential.  The energy scale is in units of the $p$-$d$ orbital hopping, $t_{pd} \sim 1$~eV. }
\label{fig:cleanlimit}
\end{figure}

The essential elements of the FSR model are shown in Fig.~\ref{fig:cleanlimit}.  
Figure~\ref{fig:cleanlimit}(a) shows a cartoon of the CuO$_2$ unit cell for an idealized YBCO$_{6.5}$ crystal with the ``ortho-II'' structure.   In YBCO$_{6+x}$, the tetragonal symmetry of the CuO$_2$ planes is broken by one-dimensional (1D) CuO chains that run parallel to the $b$-axis and sit in between CuO$_2$ bilayers.  In the ideal ortho-II structure, there is a chain above every second planar CuO$_2$ unit cell, as indicated by solid black lines in columns 1 and 3 of   Fig.~\ref{fig:cleanlimit}(a).  

The Fermi surface obtained from the CuO$_2$ planes is shown in Fig.~\ref{fig:cleanlimit}(b) (dashed curves);  in the FSR model, a staggered magnetic moment is imposed on the Cu sites, which reconstructs the Fermi surface to form four hole pockets (solid blue ellipses).  As discussed elsewhere,\cite{Atkinson:2014} this provides a useful phenomenology that captures the ``Fermi arc'' structure of the pseudogap phase.   For a generic short-range interaction, the leading CDW instability of the reconstructed Fermi surface couples  ``hotspots'' at the tips and tails of the modulation wavevectors $\bq^\ast$ shown in the figure.  For a tetragonal CuO$_2$ unit cell, a second instability with wavevector ${\bf q'}$ is degenerate with the first  for a total of eight hotspots.  This instability is subdominant for the orthorhombic case. 

The key point of weak-coupling models is that the physics of the CDW state is determined entirely by the structure of the bands in the neighborhood of these eight hotspots.  The key role of the hotspots is illustrated in Fig.~\ref{fig:cleanlimit}(c), which shows the spectral function for a self-consistently calculated {uniaxial} CDW (see appendix for details).    The spectral intensity in the neighborhood of the hotspots coupled by $\bq^\ast$ is washed out  by the CDW, while other regions of the Fermi surface are unaffected.  The depletion of spectral weight at the hotspots opens a partial gap in the density of states at the chemical potential [Fig.~\ref{fig:cleanlimit}(d)].\cite{caveat2} Experimentally, this gap is distinct from the pseudogap,\cite{Wu:2013} which affects a much larger  fraction of the Fermi surface.  [We note a limitation of the FSR model that is evident in Fig.~\ref{fig:cleanlimit}(d):  the density of states does not have a pseudogap at the Fermi level; rather, the  staggered magnetic moment on the Cu sites opens a Mott-like gap above the Fermi level.  This is not a significant problem here since we are focused on the physics of the hotspots, which lie away from the regions of Fermi surface associated with the pseudogap.]

\begin{figure}
\includegraphics[width=\columnwidth]{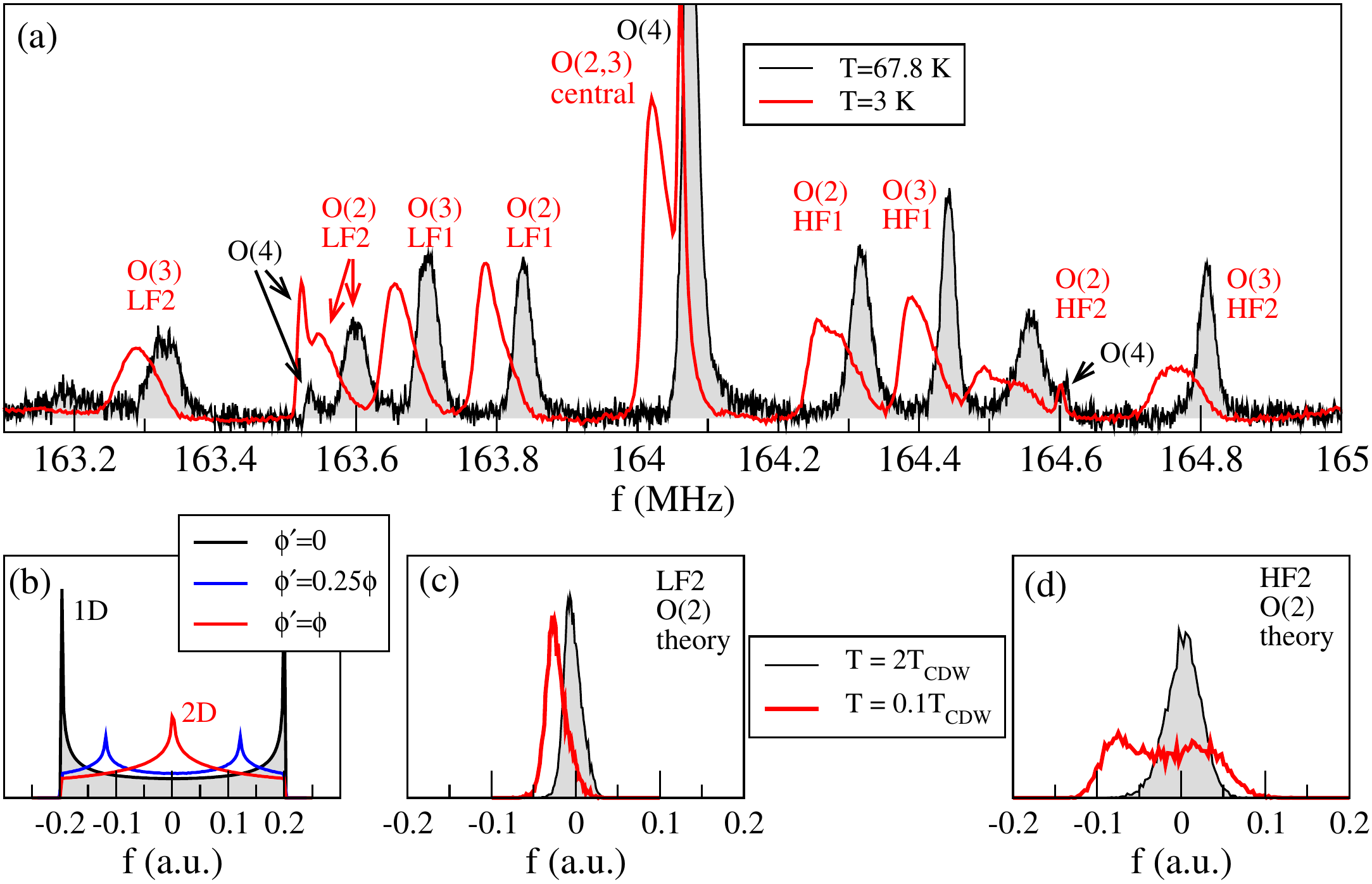}
\caption{Comparison of experimental and theoretical lineshapes.  
(a) Oxygen quadrupole satellites for YBCO$_{6.56}$ (Ref.~\onlinecite{Wu:2017}). Results were measured in a 28.5~T magnetic field tilted from the c-axis by  $16^\circ$  along the $a$ direction.  This field strength is believed sufficient to suppress superconductivity.  Results are shown at $T=67.8$K, where the CDW has short-range correlations, and $T=3$K, where the CDW is long-range ordered.  Following Ref.~\onlinecite{Zhou:2016}, the satellites are labeled LF (low frequency), central, and HF  (high frequency).
(b) Expected lineshapes for charge density waves with different $\phi'/\phi$.  $\phi' = 0$ ($\phi'=\phi$) corresponds to an ideal {uniaxial} ({biaxial}) CDW.   The peak splitting on the O(2) lines in (a), most visible for the HF2 O(2) spectrum, is an indication of quasi-{uniaxial} CDW correlations with $\phi \gg \phi'$.  (c) and (d) Simulated lineshapes obtained from BdG calculations averaged over 50 disorder configurations.  Plots are taken from histograms of (c) 
 $\delta N_0(\br) + 0.5 \delta n(\br)$ and (d) $\delta N_0(\br) - 0.5 \delta n(\br)$ [see Eq.~(\ref{eq:hist}) and surrounding discussion].  Results are shown at temperatures below and above the clean-limit transition temperature $T_\mathrm{CDW}$.
 }
\label{fig:main_results}
\end{figure}

One of the main points made in this work is that the hotspots continue to play a central role when it comes to understanding the NMR lineshapes.  For a nucleus with total angular momentum quantum number $I$, there are $2I$ transitions between pairs of nuclear states $|Im\rangle$ and $|I\, m+1\rangle$.    Crystal fields break the degeneracy of these transitions so that
${}^{17}$O ($I=5/2$) has five quadrupole satellites.  This can be seen in the oxygen NMR spectra {for YBCO$_{6.56}$,  which are shown in Fig.~\ref{fig:main_results}(a).\cite{Wu:2017}}  Lines labelled O(2) and O(3) refer to inequivalent O sites in the CuO$_2$ planes [Fig.~\ref{fig:cleanlimit}(a)], while O(4) refers to the apical oxygen immediately above planar Cu sites.   When the external magnetic field is oriented along a nuclear principal axis, the transition energies take a simple form,
\begin{eqnarray}
\Delta E_{m\rightarrow m+1} &=&  \gamma \hbar H (1+K)  - \frac{3eQ (2m+1) }{4I(2I-1)} V_{ZZ}. 
\label{eq:NMR}
\end{eqnarray}
The expression is slightly more complicated in the general case (see, for example, Ref.~\onlinecite{Haase:2004jk}); however, Eq.~(\ref{eq:NMR}) is sufficient for our purposes.
In Eq.~(\ref{eq:NMR}), $\gamma$ is the nuclear gyromagnetic ratio, $K$ is the Knight shift, $H$ the applied magnetic field,
 Q is the quadrupole moment, and $-V_{ZZ} \equiv -\partial^2 V/\partial Z^2$ is the the electric field gradient along the principal axis.  The first term in Eq.~(\ref{eq:NMR}) is  the so-called magnetic contribution, and the second is the electric quadrupole contribution.

The experimental line shapes  are
histograms of $\Delta E_{m\rightarrow m+1}$ values, and therefore of $K$ and $V_{ZZ}$.  
In a simple metal, $K(\br)$ is proportional to $N_0(\br)$, the LDOS at the chemical potential at position $\br$, while $V_{ZZ}(\br)$ is a function of the electron density in the neighborhood of the atomic nucleus.  Both disorder and the CDW  induce shifts in the Knight shift and  electric field gradient that vary from atom to atom.  
For simplicity, we assume that the change $\delta V_{ZZ}(\br)$ at a particular atomic nucleus is proportional to the change $\delta n(\br)$ in the orbital charge density for that atom. Our analysis of the NMR spectrum, therefore, focuses on $N_0(\br)$ and $\delta n(\br)$ as proxies for the magnetic and electric quadrupole terms in Eq.~(\ref{eq:NMR}).

With this in mind, we now summarize what has been previously inferred from NMR spectra using Fig.~\ref{fig:main_results}(a) as a representative example, and further describe some of the puzzles that have emerged from the experiments.   In Fig.~\ref{fig:main_results}(a), the data at $T=67.8$K are at a temperature above the onset of long range CDW correlations.\cite{Wu:2013}  The lines are broadened by disorder and by short-range CDW correlations that develop below a high onset temperature $T_\mathrm{onset} \sim 150$K.\cite{Wu:2015bt}  As $T$ is lowered,  there is a pronounced leftward shift of the O(2) and O(3) satellites.  The magnitude of the shift is the same for all satellites, which from Eq.~(\ref{eq:NMR}) indicates that it is a Knight shift and can be tied to a depletion of states at the chemical potential.  This depletion is mainly due to the pseudogap, rather than the onset of the CDW.\cite{Zhou:2016}  (Note that the data are measured in a magnetic field of 28.5~T, which is believed sufficient to suppress superconductivity.{\cite{Grissonnanche:us,Zhou:2017}})

The onset of long-range order at $T_\mathrm{CDW} \approx 50$-60K is signalled by a splitting of the quadrupole satellites as $T$ is lowered.  This can be seen most clearly in the HF2 O(2) satellite at $T=3$K.  This splitting (observed first for Cu nuclei\cite{Wu:2011ke}) was originally interpreted as a signature of commensurate order, but  is also consistent with incommensurate quasi-{uniaxial} CDWs.  To understand this latter point, we consider a potential $\phi(\br) = \phi \cos(q^\ast y)$ generated by a CDW along the $b$-axis.  If the shifts $\delta N_0(\br)$ and $\delta n(\br)$ due to the CDW are both proportional to $\phi(\br)$, then the lineshape obtained from a histogram of $\Delta E_{m\rightarrow m+1}$ will look like the histogram of $\cos(q^\ast y)$.  This corresponds to the 1D case shown in Fig.~\ref{fig:main_results}(b).  Such lineshapes have been observed in  the well-known quasi-1D CDW materials Rb$_{0.3}$MoO$_3$\cite{Butaud:1985eh} and NbSe$_3$.\cite{Ross:1990in}  For comparison, we also include a secondary component with $\phi(\br) = \phi' \cos(q' x)+ \phi \cos(q^\ast y)$ [Fig.~\ref{fig:main_results}(b)].   When $0 < \phi' < \phi$, the two peaks move inwards with increasing $\phi'$, and finally merge to form a single peak in the two-dimensional ({biaxial}) limit, $\phi' = \phi$.  Note that the  histograms are independent of $q'$ and $q^\ast$ provided both wavevectors are incommensurate with the lattice.

As mentioned above, the splitting of the oxygen peaks in Fig.~\ref{fig:main_results}(a) is not equally apparent in all satellites.  This has been attributed to the quadrupole and magnetic terms in Eq.~(\ref{eq:NMR}) having comparable magnitude.\cite{Zhou:2016}  We illustrate this point qualitatively by plotting in Fig.~\ref{fig:main_results}(c) and (d) histograms of linear combinations of $\delta N_0(\br)$ and $\delta n(\br)$ for the O(2) sites (representing magnetic and quadrupole terms, respectively).  The LF2 and HF2 satellites in Fig.~\ref{fig:main_results}(a), corresponding to $m = -\frac{5}{2}$ and $m = \frac 32$ in Eq.~(\ref{eq:NMR}), have equal magnetic contributions and equal-but-opposite quadrupole contributions.   To model this, we calculate the distributions  
\begin{equation}
P\left [\delta N_0(\br) \pm A \delta n(\br)\right ].
\label{eq:hist}
\end{equation}
  In this expression, the LDOS and charge densities are taken from self-consistent Bogoliubov-de Gennes (BdG) calculations for the CDW (described in Sec.~\ref{sec:BdG}).  Results are shown at two temperatures:
  $T = 2T_\mathrm{CDW}$, which lies above the the transition to long-range order at $T_\mathrm{CDW}$, and $T = 0.1T_\mathrm{CDW}$,   which lies deep in the long-range ordered phase. To obtain a near cancellation for the upper sign in Eq.~(\ref{eq:hist}), we take $A = 0.5$; this generates the narrow single peak shown in Fig.~\ref{fig:main_results}(c),  similar to the LF2 line in Fig.~\ref{fig:main_results}(a).  Then, for the same set of data, the lower sign in Eq.~(\ref{eq:hist}) yields the pair of low-$T$ peaks in Fig.~\ref{fig:main_results}(d), similar to the HF2 line in Fig.~\ref{fig:main_results}(a).  The key point is that because $\delta N_0(\br)$ and $\delta n(\br)$ are  correlated,  one may obtain different lineshapes for the upper and lower signs in Eq.~(\ref{eq:hist}).  

The fact that the magnetic and quadrupole terms have comparable magnitude complicates the interpretation of the NMR spectrum, but also presents a unique opportunity to obtain simultaneous information about the charge density and LDOS in the CDW phase.   Empirically:
\begin{itemize}
\item The short range CDW correlations that develop below $T_\mathrm{onset}$ { can, in some instances,} affect the quadrupole and magnetic terms differently.\cite{Wu:2015bt} Whereas the O(2) line broadening comes equally from both magnetic and quadrupole terms, the O(3) line broadening {in YBCO$_{6.56}$ comes almost exclusively from the quadrupole term.  This discrepancy is puzzling  because the O(2) and O(3) lineshapes are determined by the same Fermi surface hotspots.   Differences between O(2) and O(3) sites are harder to identify in YBCO$_{6+x}$ samples with higher oxygen content. }

\item {Similarly,} the long-range correlations that develop below $T_\mathrm{CDW}$ affect the O(2) magnetic and quadrupole terms equally.\cite{Wu:2011ke,Wu:2013}   That is, the splitting that is clearly resolved in the O(2) HF lines comes from both quadrupole and magnetic contributions.  Although details of the O(3) lines are difficult to resolve experimentally, ortho-II YBCO appears to exhibit a dichotomy between the O(2) and O(3) sites similar to that  above $T_\mathrm{CDW}$.\cite{Wu:2017}
 
\item In addition to line-splitting, the long-range correlations below $T_\mathrm{CDW}$ also produce a  lineshape asymmetry that grows with decreasing temperature. This asymmetry is manifested in both the tails and the heights of the two peaks (for those satellites that are split),\cite{Zhou:2016}  and is clearly visible for both the O(2) and O(3) lines in the 3K spectrum of Fig.~\ref{fig:main_results}(a).   The asymmetry is clearly tied to long-range order, and indeed has an order parameter-like $T$-dependence; nonetheless, it is distinct from the splitting because it comes entirely from the magnetic contribution (namely, all satellites have identical skewness).  Asymmetric lineshapes are rare in NMR.   They have  been observed in hexagonal 2H-NbSe$_2$,\cite{Skripov:1995ts,Berthier:2001ga} where the CDW is two-dimensional (2D) with three distinct wavevectors with comparable weight; however, there is no  evidence for this mechanism in ortho-II YBCO.  In the La-cuprates, lineshape asymmetry was attributed to glassy spin stripes.\cite{Haase:2000vb,Hunt:2001jp}
Asymmetry has also been seen in Zn-doped YBCO$_{6+x}$,\cite{Ouazi:2006ep} where it is attributed to a combination of near-unitary Zn resonances and locally induced antiferromagnetism.\cite{Harter:2007da} Ref.~\onlinecite{Zhou:2016} did indeed find that the left and right linewidths scale with the amount of disorder in the crystal; however, it is also clear that long-range CDW order is  prerequisite for this effect, suggesting a different mechanism. 
\end{itemize}

In this work, we address these observations via a mix of numerical BdG calculations (Sec.~\ref{sec:BdG}) and analytic calculations (Sec.~\ref{sec:analytic}).  Our main results are: 
\begin{itemize}
\item Differences between O(2) and O(3) lineshapes can be traced back to the orbital character of the hotspots.  As there is only a single Fermi surface per CuO$_2$ plane, and the structure of the CDW is entirely determined by that Fermi surface in the neighborhood of the nesting hotspots, the charge modulations on the Cu, O(2), and O(3) orbitals are not independent.  Rather, it is the admixture of the different orbitals making up the Bloch states at the hotspots that determines both the amplitude and phase of the CDW on each orbital.  In many cases, the differences between O(2) and O(3) lineshapes are mild; however, in some cases the lines may have qualitatively different shapes.
\item The CDW potential introduces a homogeneous shift in the density of states (i.e.\ a partial gap at the Fermi energy) and an inhomogeneous modulation of the LDOS.  The Knight shift distribution is equally sensitive to both of these; however, the quadrupole term is mostly determined by the inhomogeneous modulation.  For this reason, the two terms probe different aspects of the CDW.
\item Weak disorder plays a key role because it induces spatial variations of the CDW wavevector.   The Knight shift distribution is especially sensitive to these variations, which sample the band dispersion near the hotspots. In particular, the asymmetry of the lineshapes can be traced back to a combination of the distribution of CDW wavevectors and the curvature of the Fermi surface near the hotspots.
\item The presence of a secondary CDW component, with amplitude $\phi'$, qualitatively changes the shape of the NMR line.   Lineshapes in the clean limit generically have two peaks whenever $\phi'\neq \phi$, as in Fig.~\ref{fig:main_results}(b).  However, in the presence of weak disorder, there is a wide range of $\phi'$ values for which the line has a single peak.  We find that orthorhombicity reduces $\phi'$, while disorder enhances it. 
\end{itemize}

\section{Bogoliubov-de Gennes Calculations}
\label{sec:BdG}
In this section, we describe self-consistent solutions of the BdG equations for a CDW  on a finite lattice.  These calculations allow us to explore numerically the various factors---orthorhombicity, disorder, band structure, etc.---that influence the structure of the CDW, and therefore of the  NMR spectrum. 

\subsection{Model}
\begin{figure}[tb]
\includegraphics[width=\columnwidth]{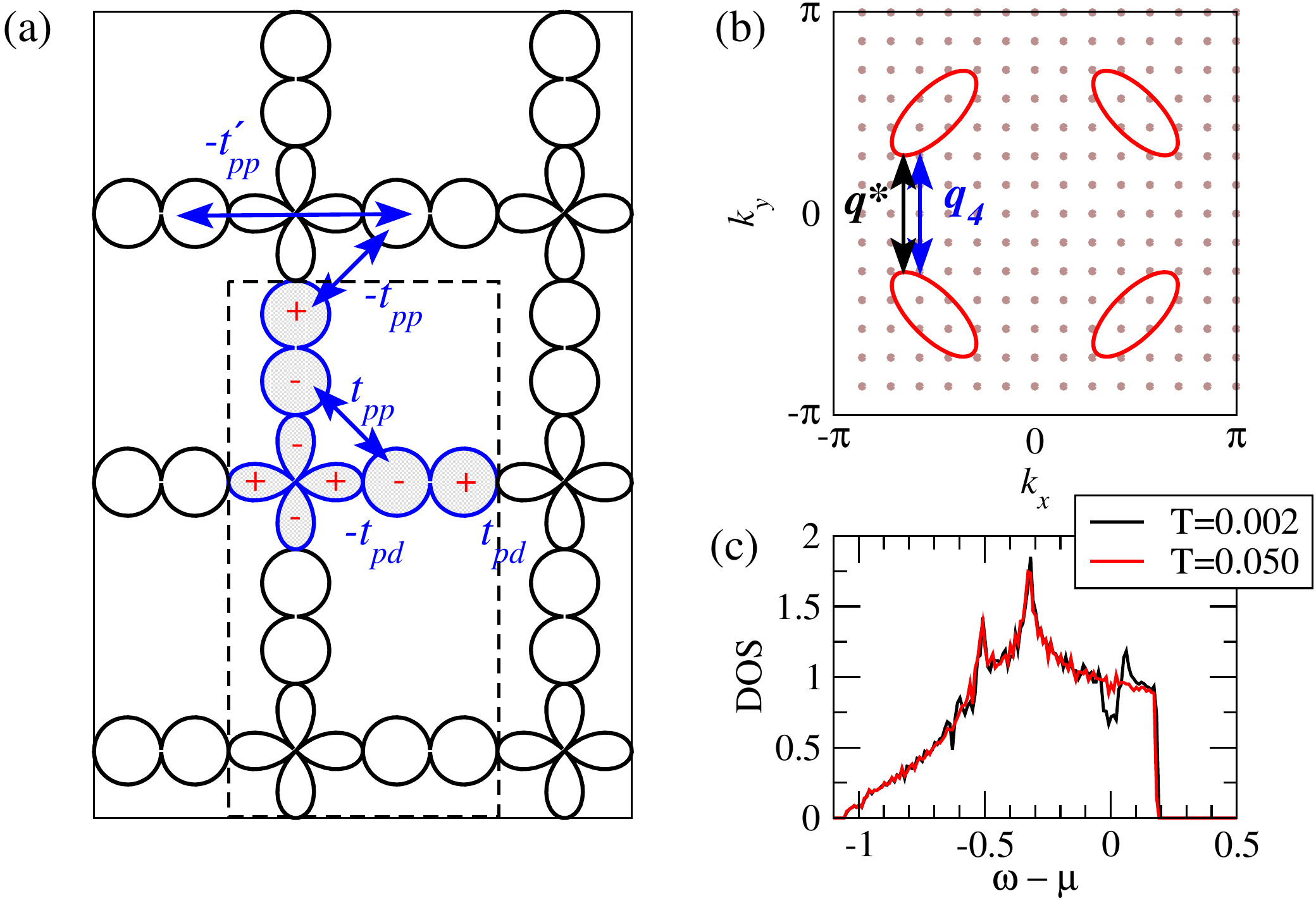}
\caption{Structure of the  ALJP model.  (a) Hopping matrix
  elements are $\pm t_{pd}$ between nearest-neighbor Cu$d$ and O$p$
  orbitals, $\pm t_{pp}$ between nearest-neighbor O$p$ orbitals, and
  $-t'_{pp}$ between next-nearest-neighbor O$p$ orbitals.  The signs
  of the  matrix elements are given by the phase difference
  between the overlapping portions of the orbitals involved in the
  hopping process.   The AF potential on the Cu sites
  doubles the unit cell, which is indicated by a dashed box. (b) For a finite $L\times L$ lattice, $k$-space
  comprises a discrete set of points separated by $\Delta k = 2\pi/L$
  (gray dots).  Because the periodic charge modulation must be
  commensurate with the lattice, allowed CDW wavevectors
  $\bq_m$, corresponding to period $L/m$ modulations, must connect
  distinct $\bk$-points.  The wavevector $\bq_4$ shown in (b) is close
  to the optimal wavevector $\bq^\ast$ for which the CDW amplitude is
  maximized. The difference between $\bq_m$ and
  $\bq^\ast$ is minimized by   adjusting the filling.  (c) When the nesting wavevector is close to
  $\bq^\ast$,  the density
  of states  at the Fermi energy is partially gapped, similar to Fig.~\ref{fig:cleanlimit}.  
  The results in (c) are for a uniaxial CDW with disorder potential $w=0.2$ at temperatures
  above ($T=0.050$) and below ($T=0.002$) the onset of long-range order at $T_\mathrm{CDW} \approx 0.020$. 
 Throughout this work, the charge density is $n=4.82$ electrons per unit cell and
  $L=14$.}
\label{fig:unitcell}
\end{figure}

Our approach is to solve a simple mean-field three-orbital model for a
single CuO$_2$ plane, similar to that of
Ref.~\onlinecite{Atkinson:2014}, in real space on an $L\times L$
lattice with periodic boundary conditions.  
The model includes the Cu $d_{x^2-y^2}$ orbitals and those O$p$ orbitals that form $\sigma$-bonds with the Cu sites, as shown in Fig.~\ref{fig:unitcell}(a).  Additionally, there is 
an implicit Cu$4s$ orbital that has been downfolded into the Hamiltonian
matrix elements.\cite{Atkinson:2014}  We refer to this as the ALJP model, after Andersen {\em et al.}
who first pointed out the importance of the 4$s$ orbital.\cite{Andersen:1995}   As discussed in the previous section, we further include an antiferromagnetic (AF) moment on the Cu sites as a means to generate a Fermi surface reconstruction.  As in the real materials, the local moments reduce double occupancy on the Cu sites, and change the character of the Fermi surface from primarily Cu-like to primarily O-like.  The AF moments double the unit cell, as shown in Fig.~\ref{fig:unitcell}(a).

In Ref.~\onlinecite{Atkinson:2014}, we proposed that the CDW may 
be driven by short-range Coulomb repulsion between
neighboring O sites.   However, because the density of states contributing to CDW formation is quite small (involving only states near the hotspots),  large values for the Coulomb interactions are required to induce CDW order.  Other interactions, notably antiferromagnetic
superexchange, are also attractive in the charge ordering channel (at
least in one-band models\cite{Sau:2013vw,Atkinson:2015hd}), and a quantitative description of the CDW
may indeed require multiple interactions.  We do not address this
point here, and simply treat the interactions in our model as a phenomenological means
to generate CDW order.

The Hamiltonian has the form
\begin{eqnarray}
\hat H &=& \sum_{i\alpha \sigma}  \tilde \epsilon_{i\alpha \sigma} \hat n_{i\alpha\sigma} 
+ \sum_{ i\alpha,j\beta }\sum_\sigma \tilde t_{i\alpha,j\beta} c^\dagger_{i\alpha\sigma} c_{j\beta\sigma}
\label{eq:H}
\end{eqnarray}
where $\hat n_{i\alpha \sigma} = c^\dagger_{i\alpha\sigma} c_{i\alpha\sigma}$ is the number operator for spin-$\sigma$  electrons in orbital $\alpha$ of unit cell $i$, and  $c_{i\alpha\sigma}$ ($c^\dagger_{i\alpha\sigma}$) is the corresponding electron annihilation (creation) operator.  The site-energies $\tilde \epsilon_{i\alpha\sigma}$ and hopping
matrix elements $\tilde t_{i\alpha,j\beta}$ are renormalized by mean-field
interactions, and are calculated self-consistently.     
The CDW appears as a periodic modulation of both $\tilde \epsilon_{i\alpha\sigma}$ and $\tilde t_{i\alpha,j\beta}$. 

The renormalized site energies are
\begin{equation}
\tilde \epsilon_{i\alpha \sigma} = \epsilon_\alpha - \sigma (-1)^i \delta_{\alpha,d} M + U_{\alpha} \langle \Delta \hat n_{i\alpha} \rangle + \sum_{j\beta} V_{i\alpha,j\beta} \langle \Delta \hat n_{j\beta}\rangle,
\label{eq:epsilon}
\end{equation}
where $\epsilon_\alpha$ is the bare site energy and $M$ is the AF potential on the Cu sites. 
The remaining terms in Eq.~(\ref{eq:epsilon}) describe the intraorbital
($U_\alpha$) and nearest-neighbor ($V_{i\alpha,j\beta}$) interactions in
the charge channel.  The main effect of the  Hubbard interactions $U_d$ and $U_p$ is to reduce charge modulations on the Cu and O sites, respectively.  The nearest-neighbor interactions are 
\begin{equation}
V_{i\alpha,j\beta} = \left \{ \begin{array}{ll}
   V_{pd} & \text{Cu-O pairs},\\
   V_{pp} & \text{O-O pairs}.\\
\end{array} \right .
\end{equation}
$V_{pp}$ drives the CDW transition, and we treat it as an adjustable parameter.   Values for model parameters are given in Table~\ref{tab:1}.   To avoid
double-counting of interactions, we have included in Eq.~(\ref{eq:epsilon}) only contributions
due to the deviation
\begin{eqnarray}
 \Delta \hat n_{id} &=& \hat n_{id} - n_\text{Cu}, \\
 \Delta \hat n_{ix/y} &=& \hat n_{ix/y} - n_\text{O},
\end{eqnarray}
from the average charge densities  $n_\text{Cu}$  and $n_\text{O}$.  

\begin{table}
\begin{tabular}{l|l}
Parameter & Value\\
\hline
$\epsilon_d - \epsilon_p$  & 0.5 \\
$t_{pd}$ & 1.0 \\
$t_{pp}$ & -0.6 \\
$t_{pp}'$ & 0.6 \\
$U_{d}$ & 6.0 \\
$U_{p}$ & 2.0 \\
$V_{pd}$ & 1.0 \\
$V_{pp}$ & 2.6 \\
$M$ & 1.5 
\end{tabular}
\caption{Model parameters used in the BdG calculations.  Energies are in units of $t_{pd}$, which is of order 1~eV.}
\label{tab:1}
\end{table}

Similarly, the renormalized hopping matrix elements are
\begin{equation}
 \tilde t_{i\alpha,j\beta} = t_{i\alpha,j\beta} - \frac 12 V_{i\alpha,j\beta} \sum_\sigma \langle \Delta c^\dagger_{j\beta \sigma} c_{i\alpha \sigma} \rangle,
\end{equation}
where $t_{i\alpha,j\beta}$  are the bare hopping matrix elements,
\begin{equation}
t_{i\alpha,j\beta} = \left \{ \begin{array}{ll} 
\pm t_{pd} & \mbox{ nearest-neighbor $p$ and $d$ orbitals,} \\
\pm t_{pp} & \mbox{ nearest-neighbor $p$ orbitals,} \\
-t_{pp}^\prime & \mbox{ next-nearest-neighbor $p$ orbitals.} 
\end{array} \right .
\label{eq:tij}
\end{equation}
The $\pm$ prefix in Eq.~(\ref{eq:tij}) is bond-dependent, and is given
by the phase difference between adjacent orbitals, as pictured in
Fig.~\ref{fig:unitcell}(a).  The matrix element $t'_{pp}$ describes
indirect hopping between O$p$ orbitals via the Cu4$s$ orbital.  
Density-functional theory calculations by ALJP show that this is large
and cannot be neglected.\cite{Andersen:1995}   

The expectation value $ \langle \Delta c^\dagger_{j\beta \sigma}
c_{i\alpha \sigma} \rangle$ measures the deviation of the exchange
energy along the bond $(j\beta,i\alpha)$ from the system average of
all bonds of that type:
\begin{equation}
 \Delta c^\dagger_{j\beta \sigma} c_{i\alpha \sigma} = c^\dagger_{j\beta \sigma} c_{i\alpha \sigma}
 - \mbox{sgn}(t_{j\beta,i\alpha}) X_{\beta,\alpha},
\end{equation}
where 
\begin{equation}
X_{\beta,\alpha} = \overline{| \langle c^\dagger_{j\beta \sigma} c_{i\alpha \sigma}\rangle |}
\end{equation}
is averaged over sites and spins.  

 The BdG calculation proceeds as follows:  the Hamiltonian in Eq.~(\ref{eq:H}) is expressed as a matrix in the space of orbitals and unit cells, and is diagonalized to find eigenenergies and eigenstates; these are used to evaluate $\langle \Delta \hat n_{i\alpha}\rangle$ and $\langle \Delta c^\dagger_{j\beta\sigma} c_{i\alpha \sigma}\rangle$, which are then used to update $\tilde \epsilon_{i\alpha \sigma}$ [Eq.~\ref{eq:epsilon}] and $\tilde t_{i\alpha,j\beta}$ [Eq.~(\ref{eq:tij})]; the renormalized orbital energies and hopping matrix elements are then fed back into Eq.~(\ref{eq:H}) to obtain an updated Hamiltonian, and the cycle is repeated until self-consistency of  $\tilde \epsilon_{i\alpha \sigma}$ and $\tilde t_{i\alpha,j\beta}$ is achieved.

We find that, for a tetragonal band structure, self-consistent calculations obtain a biaxial CDW as the 
leading instability.  Ortho-II YBCO is orthorhombic, however, and to model this we introduce an asymmetry
 in the bare hopping parameters by increasing $t_{pd}$ by 5\% along the $b$-direction, parallel to the chain direction.  As we 
 show below, this preferentially selects CDW order along $\bq^\ast$ (i.e.\ along the $b$ direction); however, a weak subdominant CDW, which is enhanced by disorder, appears along $\bq'$.

Disorder plays a central role in our calculations.  At high $T$, disorder nucleates local charge order,
while at low $T$ it weakly distorts the long-range CDW.  We adopt the simplest possible 
disorder model, consisting of a random shift of all bare site energies by amounts 
\begin{equation}
w_{i\alpha} \in
[-w/2,w/2].  
\label{eq:w}
\end{equation}
Unless otherwise stated, results in this work are for $w = 0.1$, which is an order
of magnitude smaller than the conduction bandwidth $W\sim 1.2 t_{pd}$. Within
a Born approximation, the scattering rate for box-distributed disorder is
\begin{equation}
  \gamma = \pi \frac{w^2}{12} N_0,
\end{equation}
where $N_0$ is the single-spin density of states at the chemical potential.  For
$w = 0.1$, this gives an elastic mean-free path $\ell = \hbar
v_F/\gamma \approx W/\pi \gamma \sim 300$ unit cells.  The bare
disorder potential is an extremely weak source of quasiparticle
scattering, but is an important source of pinning for the CDW.

Obtaining self-consistent solutions is severely constrained by
finite-size effects.  Charge order emerges from a nesting of Fermi
surface hotspots.  These hotspots correspond to parallel sections
of different Fermi surface pockets, as shown in Fig.~\ref{fig:unitcell}(b), and
the nesting wavevector $\bq^\ast$ that connects distinct hotspots
determines the periodicity $\lambda = 2\pi/|\bq^\ast|$ of the charge
modulation.  For periodic boundary conditions,
however, the charge modulation must be commensurate with the supercell
(that is, $m\lambda =L$, where $m$ is an integer), such that allowed
values $\bq_m$ of the modulation are generally not close to the
optimal value $\bq^\ast$. On an $L\times L$ lattice with periodic
boundary conditions, the $k$-space resolution is $\Delta k = 2\pi/L$
so that $\bq_m = (0,m\Delta k)$.  We minimize the difference $|\bq^\ast-\bq_m|$ by tuning the filling.  Even
small deviations from optimal filling introduce finite-size effects,
such as spurious first order transitions and reentrant behavior at
low $T$.

Other finite-size effects may occur when the typical energy level spacing
is greater than  other relevant energy scales,  such as the CDW gap.
 This situation can be improved by treating the $L\times L$ lattice as a
supercell in a periodic array of $N_K\times N_K$ supercells.  Then,
the eigenstates of the system are Bloch states of the superlattice,
and are characterized by a superlattice wavevector ${\bf K} =
2\pi(m,n)/(L N_K)$.  The effective size of the system is thus $L N_K$, and the
spectrum is correspondingly denser.  In the clean limit, this process
provides an exact description of the system; however, in the
disordered case, the disorder potential is the same in each supercell
(for a given configuration), which is unphysical.
To address this, we have checked for a few representative cases that
doubling $L$ while keeping $L N_k$ fixed does not change the results
shown below. 

The results shown in this work
are for an $L = 14$ lattice with $N_K = 4$ supercell $k$-points in
each dimension.  The filling is tuned  such that the nesting wavector
is $|\bq^\ast| \approx 8\pi/14$.  For our model parameters, this corresponds to
$n = 4.82$ electrons per unit cell, or a hole doping of $p=0.18$.  This lies outside
the range where CDWs are observed in cuprates; rather, it is chosen here to minimize the finite-size issues 
described above.  We emphasize that the hotspot physics that is central
to this work is independent of the filling.

\subsection{Results}
\label{sec:BdGresults}
The BdG calculations contain a number of simplifications that make it unlikely that all features of the experimental NMR lines can be replicated.  Notably: we have made no attempt to incorporate a realistic model for the CuO chains, but rather make the system orthorhombic by introducing a hopping anisotropy; our disorder model [Eq.~(\ref{eq:w})] is chosen for computational convenience and does not capture the leading source of disorder in YBCO$_{6+x}$, namely oxygen disorder in the CuO chains; and to compensate for finite-size effects we have inflated $V_{pd}$, which leads to an overly-large CDW amplitude and transition temperature.  For these reasons, we use the BdG calculations as a qualitative tool to establish the generic physics of the weak-coupling model.  On the basis of the BdG results, we then develop  a microscopic phenomenology to explain the lineshapes in Sec.~\ref{sec:analytic}.

\begin{figure}[tb]
\includegraphics[width=\columnwidth]{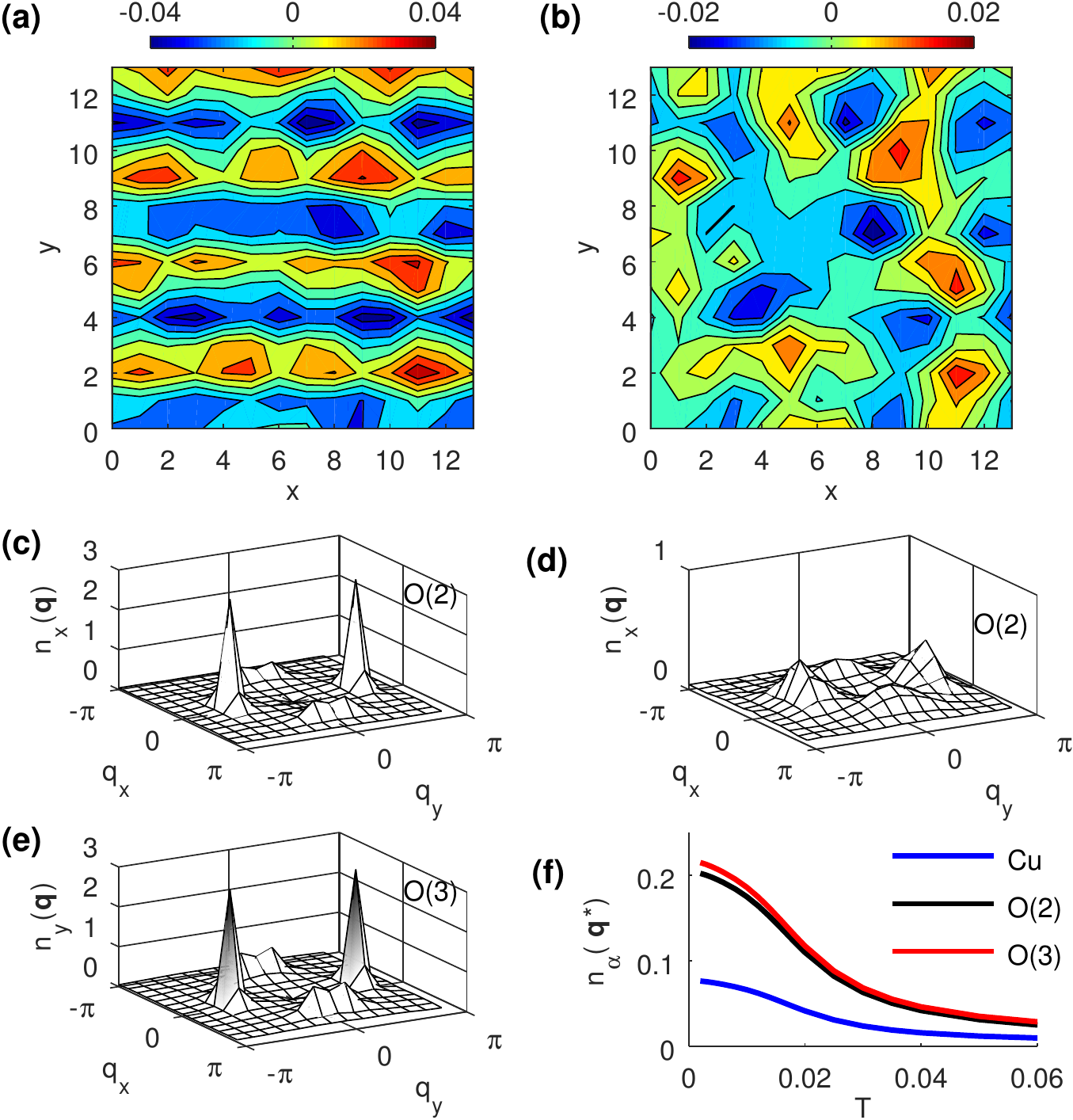}
\caption{Self-consistently calculated charge density.  The $d_{x^2-y^2}$
component of the charge density is 
  shown for a single disorder configuration at (a) low temperature
  ($T=0.002$) and (b) high temperature ($T=0.050$).  
  Fourier transforms of $n_{i\alpha}$ ($\alpha = x,y$), averaged
  over 50 disorder configurations, are shown for (c) O(2) sites at $T=0.002$, (d) 
  O(2) sites at
  $T=0.050$, (e) O(3) sites at $T=0.002$.  Peaks at $\bq=(0,0)$ are removed 
  for clarity.
  Peak heights $n_\alpha(\bq^\ast)$ for the different  orbital  types are plotted versus temperature in (f).    }
\label{fig:orthoI}
\end{figure}

Figure~\ref{fig:orthoI} illustrates typical results for our BdG calculations.  
Self-consistent solutions for the CDW find an admixture of $s$, $p$, and $d$ symmetries with similar amplitudes.  To help visualize the CDW,  
panels  (a) and (b) show the $d_{x^2-y^2}$  component constructed from the O orbital charge densities,
\begin{equation}
D_i = \frac 14  [ n_{ix} + n_{i-\hat x\,x}  - n_{iy}-n_{i-\hat y\,y}],
\label{eq:D}
\end{equation}
for a single disorder configuration below ($T=0.002$) and above ($T = 0.050$) the clean-limit transition temperature $T_\mathrm{CDW} = 0.020$.
  (All energies and temperatures are given in units of the Cu-O(2) hopping matrix element $t_{pd}\sim 1$~eV.  $T_\mathrm{CDW}$ is therefore $\sim 230$K, which is inflated by a factor of four over the experimental value.)   
While (a) shows a long-range ordered {uniaxial} CDW that is weakly distorted by the disorder potential, (b) shows a heavily disordered CDW that is {\em induced} by the disorder potential.   { That static CDW correlations may be pinned by disorder is well-understood in canonical CDW materials,\cite{McMillan:1975,Berthier:2001ga,Ghoshray:2009dq,Arguello:2014} and has been inferred to happen above $T_\mathrm{CDW}$ in YBCO$_{6+x}$.\cite{Wu:2015bt}
}

The evolution from short- to long-range CDW correlations  is further illustrated in Fig.~\ref{fig:orthoI}(c)-(e), which show the root-mean-square disorder-average of the Fourier transformed orbital charge:
\begin{equation}
n_{\alpha}(\bq) = \sqrt{ \overline{ |\langle n_{i\alpha} \rangle_\bq|^2}},
\end{equation}
 where the overline refers to a disorder average and $\langle \ldots\rangle_\bq$ to a Fourier transform.  At high $T$, the charge density $n_x(\bq)$ on the O(2) sites has broad peaks along both $x$ and $y$ axes, showing that CDW correlations are {biaxial}.   This is consistent with a recent x-ray study of YBCO$_{6.54}$ at temperatures slightly above $T_\mathrm{CDW}$,\cite{Forgan:2015ux} which found {biaxial} CDW correlations with similar amplitudes along $\bq^\ast$ and $\bq'$ directions.
As the temperature is lowered below $T_\mathrm{CDW}$, peaks along the $y$ direction narrow and grow in height as long-range {uniaxial} order develops.  These peaks correspond to $\bq^\ast$ in Fig.~\ref{fig:cleanlimit}(b), while the secondary peaks along the $x$ axis correspond to  $\bq'$.  The peak heights at $\bq^\ast$ are plotted for each of the three orbital types  in Fig.~\ref{fig:orthoI}(f) as a function of temperature.


\begin{figure}[tb]
\includegraphics[width=\columnwidth]{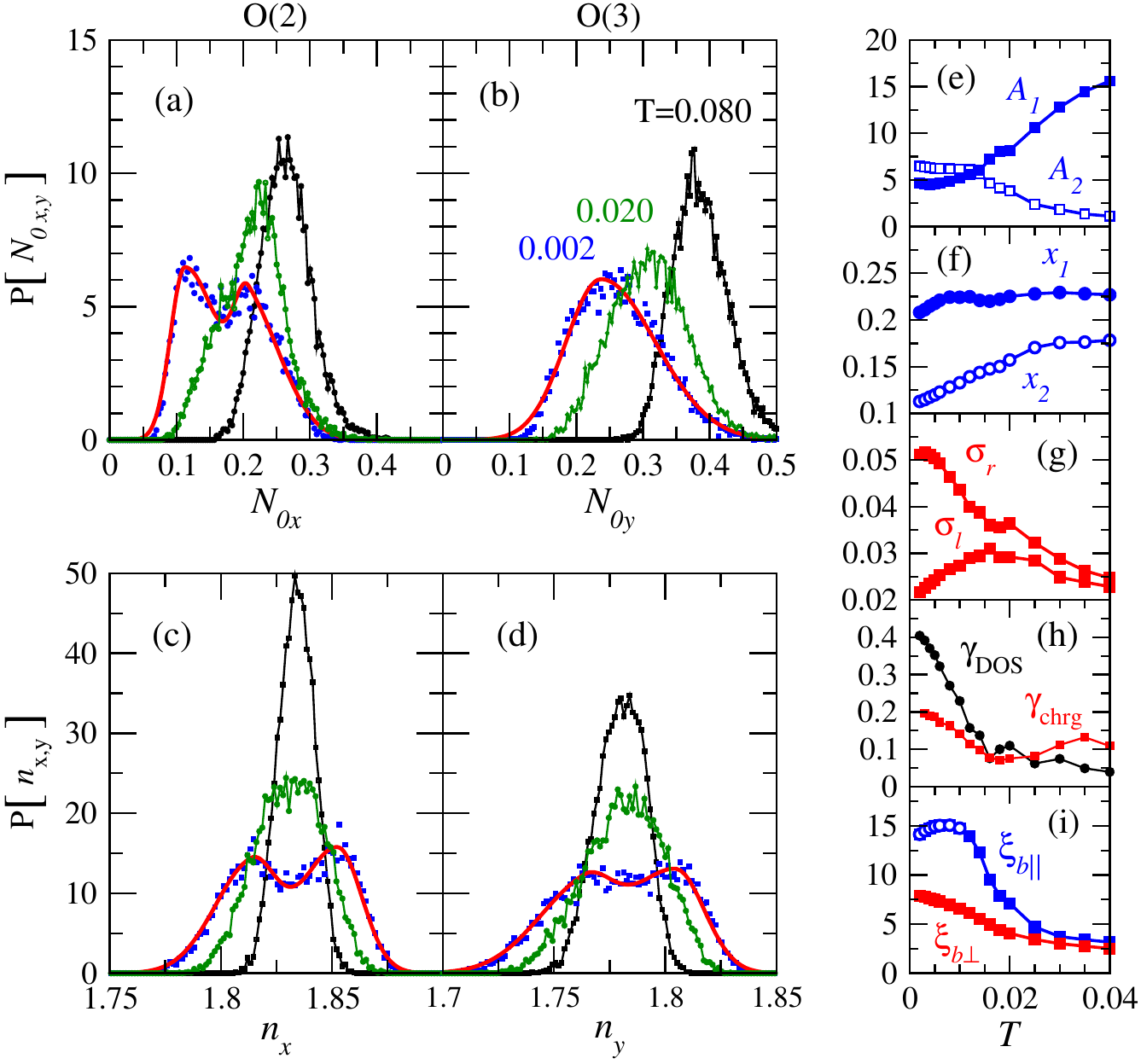}
\caption{Charge and LDOS distributions on oxygen sites.  Distributions of (a), (b) the density of states at the chemical potential and (c), (d) the orbital charge are shown at high ($T=0.080$), intermediate ($T=T_\mathrm{CDW} = 0.020$), and low ($T=0.002$) temperature.     In (a)-(d), the low-$T$ distributions are fitted to
  pairs of bi-gaussians [Eq.~(\ref{eq:doublegaussian})].
Temperature dependencies of (e) the fitted peak heights, (f) peak positions, and (g) fitted left and right peak widths are shown for the O(2) LDOS distributions.  The skewness for the O(2) LDOS, $\gamma_\mathrm{DOS}$, and O(2) orbital charge, $\gamma_\mathrm{chrg}$ are shown in (h). Panel (i) shows the correlation lengths associated with the main peak in $n_{x}(\bq)$.  Because the main peak at $\bq^\ast$ is anisotropic, it is characterized by two correlation lengths, one parallel to $\bq^\ast$ ($\xi_{b\|}$) and one perpendicular to $\bq^\ast$ ($\xi_{b\perp}$).  Empty symbols indicate that $\xi_{b\|}$ exceeds (and is truncated by) the system size $L$.  Results are averaged over 50 disorder configurations.}
\label{fig:orthoII}
\end{figure}

Figure~\ref{fig:orthoII} shows histograms of the LDOS and orbital charge for the O(2) and O(3) sites at low, intermediate, and high temperatures.  Although the NMR lines may be simulated by taking linear combinations of the LDOS and orbital charge (as we did in Fig.~\ref{fig:main_results}), we find it useful to separate the two in order to study their qualitative features.    Several features of these histograms are readily apparent.

  First, there is a progressive leftward shift of the LDOS histograms as $T$ is lowered, corresponding to an overall Knight shift due to the opening of the CDW gap.  As noted in Ref.~\onlinecite{Wu:2013} this is a small shift owing to the small region of Fermi surface affected by the CDW. 
  
  Second, the {O(2)} LDOS histograms in (a) broaden very little between $T=0.080$ and $T_\mathrm{CDW}$ (the histograms are normalized, so the peak height is inversely proportional to the linewidth).  Conversely, the histogram for the O(2) orbital charge, [Fig.~\ref{fig:orthoII}(c)] approximately doubles in width over the same temperature range.  Thus, the quadrupole broadening is more sensitive than the magnetic broadening to the development of short-range CDW correlations.    Interestingly, this is not the case for the O(3) site [Figs.~\ref{fig:orthoII}(b) and (d)], where the LDOS and charge histograms broaden by nearly the same factor between  $T=0.080$ and $T_\mathrm{CDW}$.  This dichotomy between O(2) and O(3) is very much like  experiments, except that the roles of the O(2) and O(3) sites are reversed:  experimentally, the magnetic broadening is small on the O(3) site and large on the O(2) site. Nonetheless, the important message here is that quantitatively different behavior is possible for the two sites, even though the physics of the NMR lineshapes is determined by the same set of hotspots.  As we discuss in Sec.~\ref{sec:analytic}, this shows that orbital matrix elements play a key role in the cuprates.

Third, below $T_\mathrm{CDW}$, both the O(2) LDOS and O(2) orbital charge distributions split into a pair of peaks, reflecting the onset of {uniaxial} CDW correlations. Conversely, we see that the O(3) LDOS histogram remains as a single peak down to the lowest temperature.  Again, the dichotomy between O(2) and O(3) sites demonstrates the key role of orbital matrix elements.

Fourth, the histograms become skewed below $T_\mathrm{CDW}$, with the onset of long-range charge order.  For the  model parameters used in Fig.~\ref{fig:orthoII}, the LDOS and orbital charge distributions are  skewed by comparable amounts.  We show in Sec.~\ref{sec:analytic} that in general, the skewness of the charge distribution may be much less than that of the LDOS, depending on the band structure and level of disorder.


Thus, we find that all of the main qualitative features of the NMR experiments can be found from a weak-coupling BdG calculation, although there are some discrepancies in the details.  To quantifiy our results, we have made  fits of the histograms to pairs of bi-gaussian functions.  Bi-gaussians have different left and right widths, which allows us to fit the line asymmetry, and we take a sum of two bi-gaussians to allow for peak splitting in the CDW phase.  We write 
\begin{eqnarray}
  P(x) &=& \sum_{i=1,2} A_i \Big [ e^{-(x- x_i)^2/2\sigma_\ell^2} \Theta(x_i-x) \nonumber \\
  && + 
  e^{-(x- x_i)^2/2\sigma_r^2} \Theta(x-x_i) \Big ], 
  \label{eq:doublegaussian}
\end{eqnarray}
where $A_i$ and $x_i$ are the height and location of the $i$th peak, $\Theta(x)$ is a step function,  $\sigma_\ell$ and $\sigma_r$ are the left and right widths of the peaks, and $x$ is either the LDOS or orbital charge as appropriate.  To constrain the fitting procedure, we require that the left and right widths be the same for each of the bi-gaussians.  Once the left and right widths are known, we can define the skewness of each peak in the distribution,\cite{skewness}
\begin{equation}
\gamma = \frac{|\sigma_\ell -\sigma_r|}{\sigma_\ell + \sigma_r}.
\end{equation}
Examples of the fits are shown by the solid red curves in Figs.~\ref{fig:orthoII}(a)-(d) for $T=0.002$, and the temperature dependencies of the fitting parameters  are shown in (e)-(h) for the O(2) LDOS.  

From Fig.~\ref{fig:orthoII}(e)-(h), it is clear that there is a qualitative distinction between  $T< T_\mathrm{CDW}$
and $T> T_\mathrm{CDW}$.  This point is emphasized in Fig.~\ref{fig:orthoII}(i), which shows the  correlation lengths obtained from the widths of the main  peak of  $n_x(\bq)$ [recall Fig.~\ref{fig:orthoI}(c)].  The peak widths are obtained from the second moment, and because the main peak at $\bq^\ast$ is anisotropic, we obtain two distinct correlation lengths: one in the direction parallel to $\bq^\ast$ ($\xi_{b\|}$) and one transverse to $\bq^\ast$  ($\xi_{b\perp}$).
This figure confirms that $T_\mathrm{CDW}$ marks the onset of a rapid rise in the correlation length as temperature is lowered.  $\xi_{b\perp}$ rises to a maximum of around 8 lattice constants at low $T$, while $\xi_{b\|}$ exceeds the system size $L=14$ at $T = 0.012$;  values of $\xi_{b\|}$ are truncated by $L$ below this temperature.  Experimentally, $\xi_{b\|} \sim 15$ lattice constants, and $\xi_{b\perp} \sim 7$ lattice constants in ortho-II YBCO at $T_\mathrm{CDW}$.\cite{Comin:2015fo}

Below $T_\mathrm{CDW}$, we see that the two peaks have approximately the same height ($A_1\sim A_2$), and that their separation $x_1-x_2$ grows as $T$ decreases [Fig.~\ref{fig:orthoII} (e) and (f)]. This is exactly what one expects for a {uniaxial} CDW with long-range order.  However, unlike the clean limit, the two peaks do not merge at $T_\mathrm{CDW}$; instead, their separation saturates and the height of the left peak drops towards zero as $T$ increases.  This rather unusual behavior occurs  at temperatures where the CDW crosses over from {uniaxial} to {biaxial} [recall Figs.~\ref{fig:orthoI}(c) and (d)], and indeed  the single peak at high $T$ is consistent with a {biaxial} CDW.  (It is no longer possible to resolve a second peak when $T>0.040$.)

 If the crossover occurred homogeneously, that is if the two CDW components $\phi$ and  $\phi'$ were spatially homogeneous, then the lineshape evolution would be similar to that shown in Fig.~\ref{fig:main_results}(b), with two equal-weight peaks merging to form a single peak in the {biaxial} limit.  Instead,  Fig.~\ref{fig:orthoII} is consistent with an inhomogeneous crossover in which coexisting domains of {uniaxial} and {biaxial} order span the range $0.020 \lesssim T \lesssim 0.040$, and in which the fraction of the sample occupied by {uniaxial} domains shrinks with increasing $T$. Experimentally, the degree to which the transition at $T_\mathrm{CDW}$ is homogeneous or inhomogeneous depends on the level of disorder.

Figure~\ref{fig:orthoII}(g) shows the left and right bi-gaussian widths.  Above $T_\mathrm{CDW}$ the lines are symmetric ($\sigma_\ell \approx \sigma_r$) and the linewidth grows with decreasing $T$.  We emphasize that this broadening is not simply an unresolved splitting, but rather that {\it each} of the two peaks making up the LDOS histograms broadens with decreasing $T$.   Below $T_\mathrm{CDW}$, $\sigma_\ell$ and $\sigma_r$ are distinctly different, and the individual peaks become skewed.  Figure~\ref{fig:orthoII}(h) shows that the  skewness for the O(2) LDOS, $\gamma_\mathrm{DOS}$, and orbital charge, $\gamma_\mathrm{chrg}$, are small and equal above $T_\mathrm{CDW}$, and that $\gamma_\mathrm{DOS}$  is approximately twice $\gamma_\mathrm{chrg}$ below $T_\mathrm{CDW}$.   Like experiments, then, we find that the lineshape asymmetry comes predominantly from the Knight shift distribution, although the difference between magnetic and quadrupole contributions is larger in experiments. We revisit this point in Sec.~\ref{sec:analytic}, where we unpack the factors controlling the two quantities.


\begin{figure}
\includegraphics[width=\columnwidth]{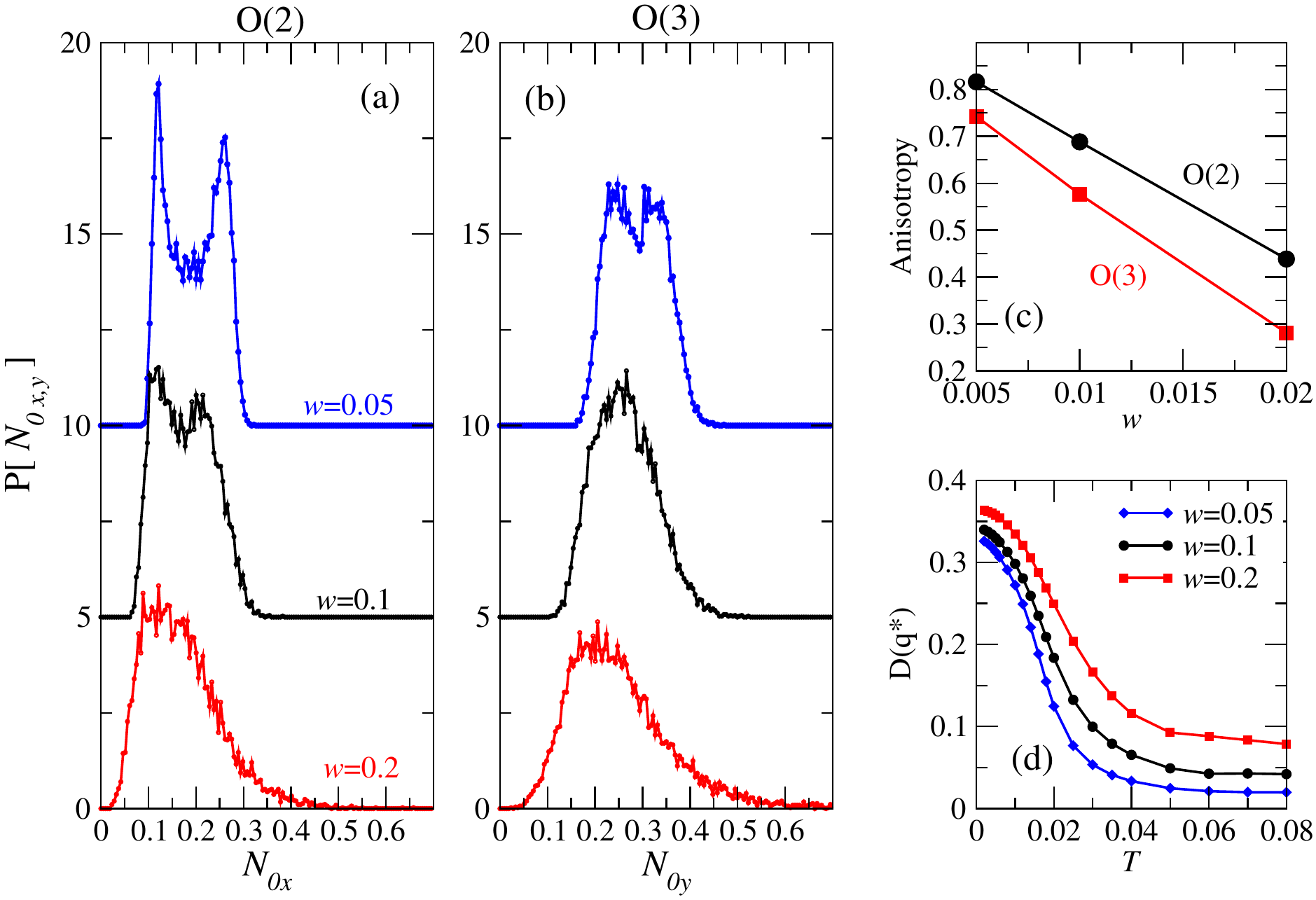}
\caption{Effect of disorder on the CDW.  LDOS distributions on the (a) O(2) and (b) O(3) sites for three different disorder strengths at $T=0.002$.    (c)  CDW anisotropy [Eq.~(\ref{eq:anisotropy})]   for a fixed crystalline orthorhombicity, showing a {uniaxial}-{biaxial} crossover as $w$ increases.  
(d)  Peak height of the Fourier transformed $d_{x^2-y^2}$ charge density,
$D(\bq) =\sqrt{ \overline{|\langle D_i \rangle_\bq|^2}}$, with $D_i$ given by Eq.~(\ref{eq:D}).
All results are averaged over 50 disorder configurations.}
\label{fig:disorder}
\end{figure}

To clarify the role of disorder, we show LDOS histograms for three different disorder strengths 
in Fig.~\ref{fig:disorder} (a) and (b).  The results are for the lowest temperature, $T=0.002$, in the long-range ordered phase.  These plots show two important trends as the disorder increases:  first, there is a crossover from two peaks to a single peak; second, the lineshape asymmetry increases.  

The first of these is tied to the dimensionality of the CDW.  We define the  CDW  anisotropy by
\begin{equation}
\left | \frac{ n_\alpha(\bq^\ast)-n_\alpha(\bq')}{ n_\alpha(\bq^\ast)+n_\alpha(\bq') } \right |.
\label{eq:anisotropy}
\end{equation}
This anisotropy is one for a purely {uniaxial} CDW and is zero for a purely {biaxial} CDW.  The anisotropy is plotted in Fig.~\ref{fig:disorder}(c), and shows a smooth crossover between the two limits with varying $w$. Thus, dimensionality is affected both by the orthorhombicity of the unit cell and by the strength of disorder, with the latter making the CDW more {biaxial}.

One intriguing feature of these results is that the crossover in lineshape happens at different points for the O(2) and O(3) sites:  the anisotropy is always higher for the O(2) sites than for the O(3) sites, which means that the CDW appears more 2D in the latter case.  We thus have a situation for $w=0.1$ in which the CDW appears quasi-{uniaxial} to the O(2) sites and quasi-{biaxial} to the O(3) sites. 

As noted already, Fig.~\ref{fig:disorder} (a) and (b) demonstrate that disorder is key to the lineshape asymmetry.  We emphasize that the mechanism for this asymmetry is different from that proposed by Zhou~{\em et al.},\cite{Zhou:2016}  who suggested that near-unitary impurity resonances generate an asymmetric LDOS distribution.  The scattering potential used in this work is far too weak to generate such resonances, and instead we propose below that its main role is to disorder the CDW, which in turn generates skewed lineshapes.

For completeness, we show in Fig.~\ref{fig:disorder}(d) the effect of disorder on the charge order.  For simplicity, we plot the $d_{x^2-y^2}$ component of the charge density as a function of temperature at three different disorder potentials.  Previously, we defined a $d_{x^2-y^2}$ real-space component $D_i$ [Eq.~(\ref{eq:D})], and here we show the root-mean-square disorder average $D(\bq)= \sqrt{ \overline{|\langle D_i \rangle_\bq|^2}}$, evaluated at the main peak $\bq=\bq^\ast$.  This figure shows clearly that disorder induces charge order at high $T$, but has little effect on the CDW amplitude at low $T$.

\section{Analytic Results for a Hotspot Model}
\label{sec:analytic}
Having established that the main features of the NMR spectrum can be obtained from a BdG calculation, we now analyze these calculations in the context of a hotspot scenario.  We consider a simple model in which electrons are scattered between  hotspot regions by a potential $\phi_{\bq}$ that is generated by a {uniaxial} CDW.  In this model, the hotspots are separated by $\bq^\ast$, and the CDW wavevector is  ${\bq}\equiv \bq^\ast+\bdq$, where $\bdq$ represents a static variation of the CDW away from $\bq^\ast$ due to a weak disorder potential.  Cuprate superconductors have only a single Fermi surface per CuO$_2$ plane, and $\phi_\bq$ therefore represents the CDW potential felt by Bloch electrons in the conduction band.

In principle, three distinct kinds of CDW disorder must be considered: amplitude and phase variations of $\phi_{\bq}$,  and variations of $\bq$.    Because a complex phase corresponds to a translation of the CDW,  phase disorder has no effect on the LDOS and charge distributions, and can be ignored; we therefore take $\phi_{\bq}$ to be real. 

\begin{figure}[tb]
\includegraphics[width=\columnwidth]{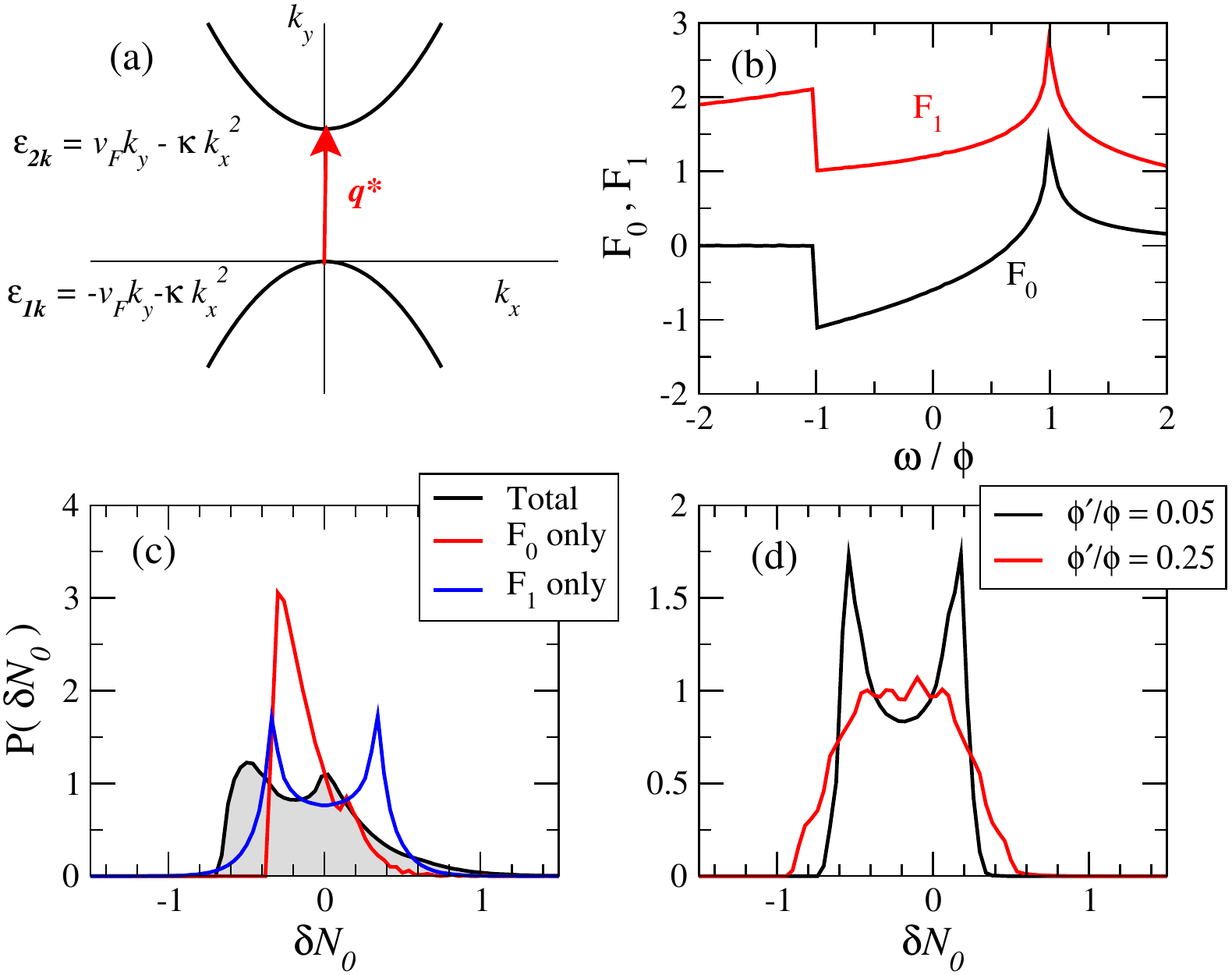}
\caption{Local density of states distributions in the hotspot model. (a) Basic model ingredients for a {uniaxial} CDW:  Fermi surfaces (black lines) are connected at the hotspots by the nesting wavevector $\bq^\ast$ (red arrow).  The dispersions near each hotspot expanded to leading order in $k_x$ and $k_y$, and depend on the Fermi velocity $v_F$ and curvature $\kappa$ at the hotspot.  (b) Universal functions $F_0(\omega/\phi)$ and $F_1(\omega/\phi)$ from Eq.~(\ref{eq:deltaN}).  (c) Histogram of the dimensionless LDOS, $\delta N_0  \equiv 2\pi^2 v_F \delta N(\br,0)$ at $\omega=0$ from Eq.~(\ref{eq:deltaN}).  For convenience, the weight factor $|\psi_\alpha|^2$ is set to one.  For comparison, histograms are also obtained separately for terms proportional to $F_0$ and $F_1$.  We take $\kappa=0.5v_F$,  $\phi = 0.1v_F$, $\xi_{b\|} = 10\pi a_0$ and $\xi_{b\perp} = 5\pi a_0$, where $a_0$ is the lattice constant.  (d) Histograms of $\delta N_0$ for a weakly {biaxial} CDW.  The main peak (amplitude $\phi$) is assumed to have infinite correlation length, while the secondary peak (amplitude $\phi'$) has a finite correlation length with the same anisotropic correlation lengths as in (b).   For simplicity, we have set $|\psi_\alpha|^2 = |\psi_\alpha'|^2 = 1$. Model parameters are otherwise as in (b).  (See text for details.)    }
\label{fig:orthoIII}
\end{figure}

 Figure~\ref{fig:orthoIII}(a) illustrates the structure of the model:  we consider a single pair of hotspots connected by $\bq^\ast$, and expand the dispersion $\epsilon_\bk$ to leading order in $k_x$ and $k_y$ around each of these hotspots to obtain two effective bands.  (We have aligned the coordinate system so that $k_x$ is parallel and $k_y$ is perpendicular to the Fermi surface at the hotspots.)
We let the dispersion near the lower and upper hotspots be $\epsilon_{1\bk}$ and $\epsilon_{2\bk}$ respectively, with
\begin{eqnarray}
\epsilon_{1\bk} &=& \epsilon_\bk = -v_F k_y - \kappa k_x^2 \\
\epsilon_{2\bk} &=& \epsilon_{\bk + \bq^\ast} =v_F k_y - \kappa k_x^2.
\end{eqnarray}
where $v_F$ is the Fermi velocity at the hotspot and $\kappa$ is the Fermi surface curvature.  

Within the subspace of two hotspots coupled by $\phi_\bq$, we can write an effective Hamiltonian for the conduction band:
\begin{equation}
\hat H = \sum_{\bk}  \left( \begin{array}{cc} c^\dagger_{1\bk}, c^\dagger_{2\,\bk+\bdq} \end{array} \right )
\left [ \begin{array}{cc} \epsilon_{1\bk} & \phi_{\bq} \\ \phi_{\bq} & \epsilon_{2\,\bk + \bdq} \end{array} \right ]
\left( \begin{array}{c} c_{1\bk} \\ c_{2\,\bk+\bdq} \end{array} \right ).
\end{equation}
The Green's function in this space is
\begin{eqnarray}
{\bf G}(\bk,\bq,\omega) &= &\frac{1}{(\omega-\epsilon_{2\,\bk+\bdq} )(\omega-\epsilon_{1\bk}) - |\phi_{\bq}|^2} 
\nonumber \\ &&\times 
\left [ \begin{array}{cc}
\omega-\epsilon_{2\,\bk+\bdq} & \phi_{\bq} \\ \phi_{\bq} & \omega -\epsilon_{1\bk } 
\end{array}
\right ]
\label{eq:Gk}
\end{eqnarray}
with the diagonal elements corresponding to $G(\bk,\bk,\omega)$ and $G(\bk+\bq, \bk+\bq, \omega)$,
and the off-diagonal elements corresponding to $G(\bk,\bk+\bq,\omega)$ and $G(\bk+\bq,\bk,\omega)$.

To obtain an orbitally resolved local density of states, we project the Green's function onto individual orbitals, and then Fourier transform to real space:
\begin{eqnarray}
G_{\alpha\alpha} (\br,\br,\omega) &=& \frac{1}{L^2} \sum_{\bk} \Big [
|\psi_{\bk\alpha}|^2 G_{11}(\bk,\bq,\omega) 
\nonumber \\ && 
+ |\psi_{\bk+\bq \alpha}|^2 G_{22}(\bk,\bq,\omega)\nonumber \\
&&+  \psi_{\bk\alpha}^\ast \psi_{\bk +\bq \alpha} G_{12}(\bk,\bq,\omega)e^{-i\bq\cdot \br} \nonumber \\
&& +  \psi_{\bk +\bq \alpha}^\ast \psi_{\bk\alpha} G_{21}(\bk,\bq,\omega)e^{i\bq\cdot \br} \Big ],
\label{eq:Gr}
\end{eqnarray}
where $|\psi_{\bk\alpha}|^2$ is the weight of orbital $\alpha$ in the conduction band at wavevector $\bk$.  Assuming that the orbital character of the Fermi surface does not vary strongly in the neighborhood of the hotspots,
 we make the approximations $|\psi_{\bk\alpha}| = |\psi_{\bk+\bq^\ast \alpha}| \approx |\psi_\alpha|$, and  $\psi_{\bk\alpha}^\ast \psi_{\bk +\bq^\ast \alpha}  = |\psi_\alpha|^2 e^{-i\theta_\alpha}$, where $\psi_\alpha$ is the weight of the orbital $\alpha$ at the hotspots associated with $\bq^\ast$ and $\theta_\alpha$ is the phase difference between the two hotspots. Then
\begin{eqnarray}
G_{\alpha\alpha} (\br,\br,\omega) =  \frac{|\psi_\alpha|^2}{L^2} \sum_{\bk} \Big [
 G_{11}(\bk,\bq,\omega) +  G_{22}(\bk,\bq,\omega)\nonumber \\
+   G_{12}(\bk,\bq,\omega)e^{-i(\bq\cdot \br + \theta_\alpha) } +  G_{21}(\bk,\bq,\omega)e^{i(\bq\cdot \br + \theta_\alpha)} \Big ]. \nonumber \\
\end{eqnarray}

The LDOS is then  obtained from the imaginary part of the Green's function as
\begin{equation}
N_\alpha(\br,\omega) = -\frac{|\psi_\alpha|^2}{\pi} \left[g_0''(\omega) + g_r''(\omega) \cos(\bq\cdot \br + \theta_\alpha)\right ]
\label{eq:Nrw}
\end{equation}
where (taking the cutoff for the $k_y$ integration to infinity)
\begin{eqnarray}
g_0''(\omega) &=& \mbox{Im } \frac{1}{L^2}\sum_{\bk} \left [ G_{11}(\bk,\bq,\omega) +G_{22}(\bk,\bq,\omega)\right ] \nonumber \\
&=& \mbox{Im }  \frac{2}{L^2}\sum_\bk \frac{\omega - (\epsilon_{1\bk} +\epsilon_{2\bk+\bdq})/2}
{(\omega - \epsilon_{1\bk})(\omega-\epsilon_{2\bk+\bdq}) - |\phi_{\bq}|^2} \nonumber \\
&=& -\frac{1}{2\pi v_F} \int dk_x  \frac {|\overline \omega + \kappa k_x^2|}
{\sqrt{ (\overline\omega + \kappa k_x^2)^2 -|\phi_{\bq}|^2}}, 
\label{eq:g0} 
\end{eqnarray}
and
\begin{eqnarray}
g_r''(\omega) &=&  \mbox{Im }\frac{1}{L^2} \sum_{\bk} \left[ G_{12}(\bk,\bq,\omega) +G_{21}(\bk,\bq,\omega)  \right ]\nonumber \\
&=&  -\frac{1}{2\pi v_F}    \int dk_x \frac {|\phi_{\bq}|}
{\sqrt{ (\overline\omega + \kappa k_x^2)^2 -|\phi_{\bq}|^2}}, 
\label{eq:gr}
\end{eqnarray}
with
\begin{equation}
\overline \omega =\omega -\frac{ v_F \Delta q_y}{2} +\kappa  \frac{(\Delta q_x)^2}{4}.
\label{eq:wbar}
\end{equation}
From Eq.~(\ref{eq:Nrw}), $g_0''(\omega)$ determines the spatially uniform shift in the LDOS (i.e.\ the CDW gap), while $g_r''(\omega)$ is the amplitude of the LDOS modulation.  Furthermore, it is clear from Eq.~(\ref{eq:Nrw}) that the phase difference between O(2) and O(3) sites determines the admixture of symmetries in the CDW:  the charge modulations on the O(2) and O(3) sites are out-of-phase when $\theta_x = \theta_y \pm \pi$, and in-phase when $\theta_x = \theta_y$.  The values of $\theta_x$ and $\theta_y$ are set by the band structure.
The two key points about Eq.~(\ref{eq:wbar}) are that (i) the variations of $\bq$ appear as shifts in the energy $\omega$ and (ii) these shifts are not evenly distributed between positive and negative values because $\Delta q_x$ enters quadratically (i.e.\ it is always a positive energy shift).  The quadratic dependence on $\Delta q_x$ reflects the curvature of the Fermi surfaces in Fig.~\ref{fig:orthoIII}(a), and the resultant asymmetry in $\overline \omega$ is the reason for the skewed lineshape.

The integral in Eq.~(\ref{eq:g0}) does not converge quickly, and it is therefore  convenient to define $\delta g_0''(\omega)$, the difference between the normal and CDW phases:
\begin{equation}
\delta g_0''(\omega) = -\frac{1}{2\pi v_F} \int dk_x \left [  \frac {|\overline \omega + \kappa k_x^2|}
{\sqrt{ (\overline\omega + \kappa k_x^2)^2 -|\phi_{\bq}|^2}} -1 \right ].
\end{equation}
This integral is restricted to the region around the hotspot.  $\Delta g_0''(\omega)$ naturally vanishes in the normal state.  Note that $g_r''(\omega)$, being proportional to $\phi_\bq$, also vanishes in the normal state.

 It is now straightforward to show that the LDOS in the CDW phase takes on a universal form.  The difference $\delta N_\alpha(\br,\omega)$ between the CDW and normal phases is 
\begin{eqnarray}
\delta N_\alpha(\br,\omega) &=& \frac{|\psi_\alpha|^2}{2\pi^2 v_F}  \nonumber \\
&&\times \sqrt{\frac{\phi}{\kappa}} \Bigg [ F_0
\left (\frac{\overline \omega}{\phi}\right )  + F_1 \left (\frac{\overline \omega}{\phi}\right )\cos (\bq \cdot \br + \theta_\alpha) \Bigg ] 
\nonumber \\
\label{eq:deltaN}
\end{eqnarray}
where $\phi \equiv |\phi_\bq|$, and 
\begin{eqnarray}
F_0(y) &=&\mbox{Re } \int_{0}^\infty dx \left[
\frac{|x^2+y|}{\sqrt{(x^2+y)^2-1}} - 1 \right ],  \\
F_1 (y ) &=& \mbox{Re }\int_{0}^\infty dx
\frac{1}{\sqrt{(x^2+y)^2-1}}.
\end{eqnarray}
$F_0$ and $F_1$  are plotted in Fig.~\ref{fig:orthoIII}(b).  From Eq.~(\ref{eq:deltaN}), $F_0(\omega/\phi)$ determines the spatially uniform shift in the LDOS due to the CDW.  The feature extending from $[-1,1]$ in Fig.~\ref{fig:orthoIII}(b) is thus the homogeneous component of the  CDW gap that would be seen in a tunneling experiment, and we note that it has the same asymmetric structure as the CDW gap shown in Fig.~\ref{fig:cleanlimit}(c).   Similarly, $F_1(\omega/\phi)$ determines the amplitude of the spatial modulation of the LDOS.  

Equation~(\ref{eq:deltaN}) is the first main result of this section.  It applies in the case of a strictly {uniaxial} CDW.  In this limit, the orbital matrix elements $\psi_\alpha$ appear as a simple prefactor that modifies the amplitude of $\delta N_\alpha(\br,\omega)$.  Because of this, Eq.~(\ref{eq:deltaN})  implies that the LDOS histograms for the O(2) and O(3) sites  have the same shape, albeit with different widths and heights.  Furthermore, both the depth of the CDW gap and the amplitude of the spatial LDOS modulations are proportional to $|\psi_\alpha|^2$. (The width of the gap is set by $\phi$.)  Since the depth of the gap determines the Knight shift, and the amplitude of the modulations determines the line splitting, we have the testable prediction that the ratio of O(2) and O(3) Knight shifts should equal the ratio of O(2) and O(3)  magnetic line splittings.  

For a given value of $\overline\omega$,  the LDOS distribution $P[\delta N_\alpha(\br,\overline \omega)]$  obtained from Eq.~(\ref{eq:deltaN}) has the form of an ideal {uniaxial} CDW.  To explore the effects of disorder,  we consider in turn variations of the wavevector $\bq$ and of the amplitude $\phi_\bq$.  In Eq.~(\ref{eq:deltaN}) $\Delta \bq$ appears in the cosine term, in the amplitude $\phi_\bq$,  and in $\overline \omega$.  Provided $\bq$ is incommensurate with the lattice, the distribution of cosine values is independent of  $\bq$ (and $\theta_\alpha$). $\Delta \bq$ therefore affects $P[\delta N_\alpha(\br,\overline \omega)]$ implicitly through the amplitude $\phi_\bq$ and explicitly through $\overline\omega$.  We discuss the latter influence first.

Keeping $\phi_\bq$ fixed, we average $P[\delta N_\alpha(\br,\overline \omega)]$ over the interval 
\begin{equation}
\Delta q_y  \in \left [\frac{-2\pi}{\xi_{b \| }},\frac{2\pi}{\xi_{b\| }} \right ], \quad \Delta q_x  \in \left [\frac{-2\pi}{\xi_{b\perp}},\frac{2\pi}{\xi_{b\perp}} \right ].
\label{eq:q_interval}
\end{equation}
The results are shown at the Fermi energy ($\omega=0$) in Fig.~\ref{fig:orthoIII}(c).  When both $F_0$ and $F_1$ are included in the calculation, we obtain a skewed structure that has two peaks of unequal height, much like the BdG results from Fig.~\ref{fig:orthoII}(a).  To understand the separate roles of $F_0$ and $F_1$, we can set each to zero in turn and calculate the resultant histogram.  When $F_0 = 0$, $P[\delta N_\alpha]$ has a symmetric two-peaked structure, much like the ideal {uniaxial} case, but with broadened peaks.  When $F_1 = 0$, $P[\delta N_\alpha]$ has a single broad peak that is strongly skewed to the right.

Where does the skewness come from?  Because $\bdq$ appears as an effective energy shift in  $\overline \omega$, averaging over nonzero values of $\bdq$ amounts to sampling the curve $F_0(\omega/\phi)$ over some window around $\omega=0$.   The form of $\overline \omega$ shows that $\Delta q_y$ samples positive and negative values equally, but that $\Delta q_x$ samples positive values only.  This skews the distribution towards higher values. Typical variations are $\Delta q_x \sim \xi_\perp^{-1}$ and $\Delta q_y \sim \xi_\|^{-1}$, so that the asymmetry in the sampling of $\overline \omega$  is a function of the dimensionless ratio $\kappa \xi_\| / v_F \xi_\perp^2$.

One can make a similar analysis of the role of amplitude disorder.  In this case, we fix $\bq$, let $\phi = \phi_0+\Delta \phi$, and allow $\Delta \phi$ to vary.  Provided the variations are not too large ($\Delta\phi \ll \phi_0$),  Eq.~(\ref{eq:deltaN}) is symmetrically distributed around its mean value if $\Delta \phi$ is symmetrically distributed around zero.  Amplitude disorder, therefore, broadens NMR lines but does not cause skewness.  

The second important result of this section, then, is that the skewness of the distribution of Knight shifts comes from the {\em homogeneous} shift in the LDOS due to the opening of the CDW gap.  The gap, represented by $F_0$ in Fig.~\ref{fig:orthoIII}(b), is asymmetric and is preferentially sampled to the right by variations around the nesting wavevector $\bq^\ast$.

The third important result is that, because $F_0$ and $F_1$ appear with equal weight in Eq.~(\ref{eq:deltaN}) and have comparable magnitude near $\omega=0$, they make similar contributions to the LDOS distributions.  Thus, the magnitudes of the splitting and the skewness in the Knight shift distributions are comparable.

So far, the analysis has been for a single Fourier component $\bq$ near $\bq^\ast$.  To study the dimensional crossover that was observed in Fig.~\ref{fig:disorder}, we extend the analysis to include two orthogonal CDW components, at $\bq^\ast$ and at $\bq'$.  For simplicity, we assume that $\phi_{\bq}$ has infinite correlation length and we fix $\bq = \bq^\ast$ for this component; the subdominant component $\phi'_{\bq}$, however, is assumed to be disordered.  For this latter component, we take $\bq = \bq' + \Delta \bq$, and average over the same window of $\Delta \bq$ values as before [Eq.~(\ref{eq:q_interval})]. 
Equation~(\ref{eq:deltaN}) must then be modified to give
\begin{align}
 \delta N_\alpha&(\br,\omega) = \nonumber\\
 &\frac{|\psi_\alpha|^2}{2\pi^2 v_F} \sqrt{\frac{\phi}{\kappa}} \left [ F_0 \left (\frac{\omega}{\phi}\right )
+ F_1 \left (\frac{\omega}{\phi}\right )\cos (q^\ast y+\theta_\alpha) \right ] \nonumber \\
&+ \frac{|\psi'_\alpha|^2}{2\pi^2 v_F}  \sqrt{\frac{\phi'}{\kappa}} \left [ F_0 \left (\frac{\overline \omega}{\phi'}\right )
+ F_1 \left (\frac{\overline \omega}{\phi'}\right )
\cos(\bq\cdot \br + \theta_\alpha') \right ],  \nonumber \\
\label{eq:deltaN2}
\end{align}
where $|\psi'_\alpha|^2$ is the weighting factor for orbital $\alpha$ at the hostpots associated with $\bq'$.  Because these hotspots occupy different regions of the Fermi surface, we can in general expect $|\psi'_\alpha|^2 \neq |\psi_\alpha|^2$.

Results are shown in Fig.~\ref{fig:orthoIII}(d) for two different amplitudes,  $\phi'/\phi = 0.05$ and $\phi'/\phi = 0.25$.  The former has a two-peak structure, while the latter has only a single peak.  This is very different from the ideal case (with $\Delta \bq$ set to zero) shown in Fig.~\ref{fig:main_results}(b), which has two peaks for any $\phi'/\phi<1$.   Thus, it is the confluence of a subdominant CDW component and disorder that gives a single-peaked distribution.

In Fig.~\ref{fig:orthoIII}, we have set $|\psi_\alpha|^2 = |\psi^\prime_\alpha|^2 = 1$ for simplicity; however, the role of orbital matrix elements is straightforward to understand from Eq.~(\ref{eq:deltaN2}).  First, we recall that the $|\psi_\alpha|^2$ and $|\psi^\prime_\alpha|^2$ refer to hotspots associated with charge modulations along the $b$- and $a$-axis, respectively.
Then, the distinction between the O(2) and O(3) sites in Fig.~\ref{fig:orthoII} reflects the different weights $|\psi^\prime_x|^2$ and $|\psi^\prime_y|^2$ at the secondary hotspots.  These weights determine the projection of the $a$-axis CDW component onto the O(2) and O(3) sites, respectively.  An analysis of the bands shows that the Fermi surface near the secondary hotspots has stronger O(3) character than O(2) character.  The secondary CDW component therefore shows up more strongly on the O(3) sites, which in turn look more 2D than the O(2) sites.

The orbital weightings in Eq.~(\ref{eq:deltaN2}) have interesting implications.  
For example, while NMR experiments at low $T$ have been interpreted in terms of a {uniaxial} CDW,\cite{Kharkov:2016tf}   quantum oscillation experiments at similar fields and temperatures are consistent with quasi-{biaxial} charge order.\cite{Sebastian:2012wh}  Equation~(\ref{eq:deltaN2}) shows that a {biaxial} CDW, with $|\phi'| \sim |\phi|$, may look {uniaxial} to NMR if  $|\psi_x|^2$ and $|\psi_y|^2$ are much different than $|\psi_x'|^2$ and $|\psi_y'|^2$, respectively.  Physically, this scenario corresponds to two orbitally selective CDWs, one of which predominantly involves the O(2) sites, and the other of which predominantly involves the O(3) sites.  Depending on the true magnitude of the orbital anisotropy of the hotspots, such a scenario might reconcile the NMR and quantum oscillation experiments. 

To understand the electric quadrupole broadening, we calculate the local charge density from 
\begin{equation}
\delta n_\alpha(\br) = \int_{-\Lambda}^0 d\omega\, \delta N_\alpha(\br,\overline \omega)
\end{equation}
where $-\Lambda$ is a cutoff on the order of the bandwidth.  Because $F_0(\omega/\phi)$ vanishes for $\omega < -\phi$, the integral is dominated by $F_1$,   and for the case of a {uniaxial} CDW,
\begin{equation}
\delta n_\alpha(\br) \approx \frac{\phi^{3/2} |\psi_\alpha|^2}{2\pi^2 v_F \kappa^{1/2}} \int_{-\Lambda/\phi}^0 d\left( \frac{\omega}{\phi} \right)\,
F_1 \left (\frac{\overline \omega}{\phi}\right )\cos (\bq \cdot \br  +\theta_\alpha). 
\label{eq:dnF1}
\end{equation}
The integral gives a function of $\bdq$ that determines the amplitude of the cosine modulation.  For a fixed $\bq$, the histogram $P[\delta n(\br)]$ is that of an ideal {uniaxial} CDW.  Averaging over $\bq$ symmetrically broadens this histogram, so that the resultant distribution resembles the ``$F_1$ only'' curve from Fig.~\ref{fig:orthoIII}(c).    $P[\delta n_\alpha(\br)]$ thus has a two-peaked structure with a splitting proportional to $\phi^{3/2}/ v_F \kappa^{1/2}$.   The contribution to $\delta n_\alpha(\br)$ due to $F_0$, which is neglected in Eq.~(\ref{eq:dnF1}),  skews the histogram; however, this effect is weak:  the analagous integral to Eq.~(\ref{eq:dnF1}) for $F_0$ has a lower cutoff of $\overline \omega/\phi = -1$, beyond which $F_0(\overline \omega/\phi)$ vanishes.  Consequently, the contribution from $F_0$ to $\delta n_\alpha(\br)$ is a factor $\phi/\Lambda$ smaller than that from $F_1$.

The final important result of this section, then, is that the electric quadrupole distribution is most sensitive to the {\em inhomogeneous} component of $\delta n_\alpha(\br)$, and should therefore show a symmetric splitting with little skewness.   This resolves an experimental puzzle first pointed out in Ref.~\onlinecite{Zhou:2016}, that both the line splitting and the skewness are clearly tied to the onset of long-range CDW order, but that only the former appears in the quadrupole contributions to the lineshape.  Furthermore, we can now understand why it is that we {\em did} find asymmetric charge histograms in the BdG calculations shown in Fig.~\ref{fig:orthoII}:  To compensate for finite size effects, we inflated the interaction strengths in our BdG calculations, and consequently the CDW potential is a substantial fraction of the bandwidth.  Indeed, taking the $\phi$ to be the Hartree potential on the oxygen sites, and $\Lambda$ to be the bandwidth, our BdG calculations obtain $\phi \approx \Lambda/3$, which is not small.  The $F_0$ contribution to $\delta n_\alpha(\br)$ is therefore not negligible in this case.   

\section{Conclusions}
We have used a multi-orbital model {with a Fermi surface reconstructed by staggered moments on the Cu sites} to describe the $^{17}$O NMR spectrum for the CDW phase of cuprate superconductors.   { Because the most complete set of experimental results is available for ortho-II YBCO$_{6.56}$, the Hamiltonian was made weakly orthorhombic to account for the influence of CuO chains.}   In the clean limit, the model has a mean-field phase transition at a temperature $T_\mathrm{CDW}$.  Above this temperature, disorder induces CDW modulations (or, equivalently, pins CDW fluctuations); below $T_\mathrm{CDW}$, disorder disrupts the long-range CDW correlations.  { Because of the orthorhombicity, the correlations at low $T$ are predominantly uniaxial, with a weak secondary component.}  With this model, we have identified and explained nearly all features of the experimental NMR lineshapes that are characteristic of these two temperature regimes.

Above $T_\mathrm{CDW}$, our numerical calculations find symmetric, single-peaked lines whose width grows with decreasing temperature.  This structure is traced back to the fact that the CDW correlations at high temperature are {biaxial}, and that the linewidth is a measure of the typical CDW amplitude.  In general, we find that the temperature-dependent broadening appears in both the magnetic and electric quadrupole contributions to the lineshape.

As the temperature is lowered through $T_\mathrm{CDW}$, we observe that the NMR peaks split to form two-peak structures, which are associated with a long-range-ordered quasi-{uniaxial} CDW.  We find that this splitting appears in both the magnetic and quadrupole terms.  In addition to splitting, the measured NMR peaks develop an asymmetry below $T_\mathrm{CDW}$.  We connect this asymmetry to the disordering of the CDW by impurities; variations of the CDW wavevector asymmetrically broaden the NMR lines by an amount proportional to the Fermi surface curvature.  This asymmetry appears primarily in the magnetic contribution to the lineshape.

Our calculations have  identified a particular role for orbital matrix elements.  This leads to the prediction, for example, that { for a purely uniaxial CDW}, the ratio of O(2) and O(3) Knight shifts should equal the ratio of O(2) and O(3) magnetic line splittings.  We have also shown that differences between O(2) and O(3) lineshapes can be traced back to the orbital character of the Fermi surface hotspots.    In some cases, these lineshapes may be qualitatively different, even though they are generated by the same set of hotspots. This offers a potentially valuable perspective, namely that NMR (which is generally considered a real-space probe) provides a unique tool for measuring the character of the Fermi surface near the hotspots.  

 { While our model was developed explicitly for YBCO$_{6.56}$, the NMR signatures identified above, and their connection to the microscopic Hamiltonian, should be the same in all of the cuprates.  In tetragonal materials with a biaxial CDW,  the lines remain unsplit, and the onset of long-range CDW correlations below $T_\mathrm{CDW}$ is marked only by the growth of the lineshape asymmetry. }

In summary, we have found that four quantities---the orbital character of the hotspots, the Fermi surface curvature at the hotspots, the dimensionality of the CDW, and the CDW correlation length---determine the shapes of the quadropole satellites in NMR experiments.   

\section*{Acknowledgments}
We thank Marc-Henri Julien, Rui Zhou,  and Igor Vinograd for extensive comments and discussion, and for sharing their data with us. W.A.A.\ acknowledges support by the Natural Sciences and Engineering Research 
Council (NSERC) of Canada.   A.P.K.\ is supported by the Deutsche Forschungsgemeinschaft through TRR80. 

\appendix
\section*{Appendix:  Charge order in the clean limit}
Here, we outline the procedure for obtaining a self-consistent solution for the CDW in the clean limit ($w=0$). The solution is approximate, but  valid for the limit in which the CDW potential is much less than the bandwidth.   To start, we obtain the bare Hamiltonian, in the absence of both disorder and a CDW.  Equation~(\ref{eq:H}) is then  written conveniently in $\bk$-space.  In this case, the unit cell comprises two CuO$_2$ plaquettes or six orbitals because of the staggered moment on the Cu sites  [Fig.~\ref{fig:unitcell}(a)].

After substituting $c_{i\alpha\sigma} = L^{-1} \sum \exp(i\bk\cdot \br_{i\alpha}) c_{\alpha \bk\sigma}$, where  $\alpha \in \{1d,1x,1y,2d,2x,2y\}$ labels the six  orbitals making up the unit cell, we obtain
\begin{equation}
\hat H_0 = \sum_\bk\sum_\sigma {\bf \hat \Psi^\dagger_{\bk\sigma}} {\bf h_{0\sigma}(\bk)} {\bf \hat \Psi_{\bk\sigma}},
\end{equation}
where the $\bk$-sum is over the magnetic Brillouin zone, and
${\bf \hat \Psi}^\dagger_{\bk\sigma} = \left [ \begin{array}{cccccc} c^\dagger_{1d\bk\sigma},&
c^\dagger_{1y\bk\sigma}, & c^\dagger_{1x\bk\sigma}, & c^\dagger_{2d\bk\sigma}, & c^\dagger_{2y\bk\sigma}, &
c^\dagger_{2x\bk\sigma} \end{array} \right ]$.
The Hamiltonian matrix is
\begin{equation}
{\bf h}_{0\sigma}(\bk) = \left [ \begin{array}{cc} 
{\bf h}_{1\sigma}(\bk) & {\bf h}_2(\bk)\\
{\bf h}_2(\bk)^\dagger & {\bf h}_{1\overline \sigma}(\bk)
\end{array} \right ],
\label{eq:h0}
\end{equation}
where
\begin{eqnarray}
{\bf h}_{1\sigma}(\bk) &=& \left [ \begin{array}{ccc}
 \epsilon_d  -M\sigma  & t_{pd} e^{ik_y/2} & -t_{pd} e^{ik_x/2}  \\
 t_{pd}e^{-ik_y/2} &  \epsilon_p & 2 t_{pp}c_-  \\
-t_{pd}e^{-ik_x/2} & 2t_{pp}c_- &  \epsilon_p 
\end{array} \right  ], \\
{\bf h}_2(\bk) &=& \left [ \begin{array}{ccc}
0 & -t_{pd} e^{-ik_y/2} & t_{pd} e^{-ik_x/2} \\ 
 -t_{pd} e^{ik_y/2} & -2 t_{pp} \cos k_y& -2 t_{pp}c_+  \\
 t_{pd} e^{ik_x/2} & -2 t_{pp} c_+ & -2t_{pp} \cos k_x
\end{array} \right ], \nonumber \\
\end{eqnarray}
and $\overline \sigma = -\sigma$.
The primitive lattice constant is $a_0 = 1$, and 
$c_\pm = \cos(\frac{k_x}{2} \pm \frac{k_y}{2})$.  The signs of
the off-diagonal matrix elements $h_{0,\alpha\beta}(\bk)$ are
determined by the product of signs of the closest lobes of orbitals
$\alpha$ and $\beta$, as shown in Fig.~\ref{fig:unitcell}(a).  Because
the supercell contains two primitive unit cells, the Brillouin zone is
halved and the Fermi surface is folded into the (reduced) antiferromagnetic Brillouin zone.

In the absence of disorder, we proceed with the assumption that the CDW potential has only a single $\bq$-vector.  The CDW potential is then a $6\times 6$ matrix that scatters electrons between $\bk$ and $\bk \pm \bq$.   Following Ref.~\onlinecite{Atkinson:2016ba}, we write the CDW potential energy as
\begin{equation}
\hat V_\mathrm{CDW} = \sum_{\bk\sigma} \sum_{\alpha,\beta} \sum_{\pm} P_{\alpha \beta\sigma}(\bk,\pm \bq) c^\dagger_{\alpha \bk \pm \frac{\bq}{2} \sigma} c_{\beta \bk \mp \frac{\bq}{2}\sigma}
\end{equation}
where $P_{\alpha\beta\sigma}(\bk,\bq)$ is a $6\times 6$ matrix in orbital space.  For short-range interactions, the $\bk$ and $\bq$ dependence is simplified by expanding in terms of a set of 38 basis functions $g^\ell_{\alpha \beta}(\bk)$ [see Table~I of Ref.~\onlinecite{Atkinson:2016ba}],
\begin{equation}
P_{\alpha\beta\sigma}(\bk,\bq) 
= \sum_{\ell} \tilde P^\ell_\sigma(\bq) g^\ell_{\alpha\beta}(\bk).
\label{eq:P2}
\end{equation}
The matrix elements $\tilde P^\ell_\sigma(\bq)$ in this basis can then be obtained from the self-consistent equation
\begin{equation}
\tilde P^\ell_\sigma(\bq) = \frac{1}{L^2}\sum_{\bk} \sum_{\ell',\mu,\nu}
\tilde V^{\ell\ell'}(\bq) g^{\ell'}_{\mu\nu}(\bk+\bqt)^\ast \langle
c^\dagger_{\bk\nu \sigma}c_{\bk+\bq\mu \sigma} \rangle,
\label{eq:Ptilde}
\end{equation}
where $\tilde V^{\ell\ell'}(\bq) $ is the projection of the electron-electron interactions onto the basis functions $g^\ell_{\alpha \beta}(\bk)$.  Explicit expressions for $\tilde V^{\ell\ell'}(\bq)$ are given in Ref.~\onlinecite{Atkinson:2016ba}.

For general $\bq$, the self-consistent equation (\ref{eq:Ptilde}) for $\tilde P^\ell_\sigma(\bq)$ can be solved only approximately;  the simplest approach is to work within a restricted subspace that considers scattering between $\bk$ and $\bk\pm \bq$, but ignores higher order scattering, e.g.\ between $\bk$ and $\bk \pm 2\bq$.  In this subspace, the Hamiltonian is approximately
\begin{widetext}
\begin{eqnarray}
\hat H &\approx& \sum_{\bk\sigma} \left [\begin{array}{ccc} 
	{\bf \hat \Psi}^\dagger_{\bk  \sigma}  & {\bf \hat \Psi}^\dagger_{\bk + \bq \sigma} &{\bf \hat \Psi}^\dagger_{\bk -  \bq \sigma}
	\end{array} \right ]
	\left [ \begin{array}{ccc}
	{\bf h}_{0\sigma}\left (\bk \right ) & {\bf P}^\dagger_\sigma(\bk+\bqt,\bq) &  {\bf P}_\sigma(\bk-\bqt,\bq) \\
	{\bf P}_\sigma(\bk+\bqt,\bq) & {\bf h}_{0\sigma}\left (\bk + \bq \right ) & 0 \\
	{\bf P}^\dagger_\sigma(\bk-\bqt,\bq)&0& {\bf h}_{0\sigma}\left (\bk - \bq \right )
	\end{array} \right ] 
         \left [\begin{array}{c} 
	{\bf \hat \Psi}_{\bk \sigma} \\ {\bf \hat \Psi}_{\bk + \bq \sigma} \\ {\bf \hat \Psi}_{\bk -  \bq \sigma}
	\end{array} \right ].
	\label{eq:Happrox}
\end{eqnarray}
It is then straightforward to obtain the correlations $\langle c^\dagger_{\bk\nu \sigma}c_{\bk+\bq\mu \sigma} \rangle$
that are required for Eq.~(\ref{eq:Ptilde}).
Once the $6\times 6$ matrix ${\bf P}_\sigma(\bk,\bq)$ is obtained, the spectral function and density of states shown in Fig.~\ref{fig:cleanlimit} is calculated directly from  Eq.~(\ref{eq:Happrox}):
\begin{equation}
{\bf A}(\bk,\omega) = -\frac{1}{\pi} \mbox{Im }
\left [ \begin{array}{ccc}
	\omega +i\eta - {\bf h}_{0\sigma}\left (\bk \right ) & - {\bf P}^\dagger_\sigma(\bk+\bqt,\bq) &  -{\bf P}_\sigma(\bk-\bqt,\bq) \\
	-{\bf P}_\sigma(\bk+\bqt,\bq) & \omega+i\eta - {\bf h}_{0\sigma}\left (\bk + \bq \right ) & 0 \\
	-{\bf P}^\dagger_\sigma(\bk-\bqt,\bq)&0& \omega + i\eta - {\bf h}_{0\sigma}\left (\bk - \bq \right )
	\end{array} \right ]^{-1}_{11},
\end{equation}
where $[\ldots]^{-1}_{11}$ indicates the top-left  $6\times 6$ block of the matrix inverse, and $\eta$ is a positive infinitesimal.  The spectral function shown in Fig.~\ref{fig:cleanlimit}(c) results from the trace (i.e.\ sum over orbitals) of the $6\times 6$ maxtrix ${\bf A}(\bk,\omega)$.
\end{widetext}


\begin{thebibliography}{72}%
\makeatletter
\providecommand \@ifxundefined [1]{%
 \@ifx{#1\undefined}
}%
\providecommand \@ifnum [1]{%
 \ifnum #1\expandafter \@firstoftwo
 \else \expandafter \@secondoftwo
 \fi
}%
\providecommand \@ifx [1]{%
 \ifx #1\expandafter \@firstoftwo
 \else \expandafter \@secondoftwo
 \fi
}%
\providecommand \natexlab [1]{#1}%
\providecommand \enquote  [1]{``#1''}%
\providecommand \bibnamefont  [1]{#1}%
\providecommand \bibfnamefont [1]{#1}%
\providecommand \citenamefont [1]{#1}%
\providecommand \href@noop [0]{\@secondoftwo}%
\providecommand \href [0]{\begingroup \@sanitize@url \@href}%
\providecommand \@href[1]{\@@startlink{#1}\@@href}%
\providecommand \@@href[1]{\endgroup#1\@@endlink}%
\providecommand \@sanitize@url [0]{\catcode `\\12\catcode `\$12\catcode
  `\&12\catcode `\#12\catcode `\^12\catcode `\_12\catcode `\%12\relax}%
\providecommand \@@startlink[1]{}%
\providecommand \@@endlink[0]{}%
\providecommand \url  [0]{\begingroup\@sanitize@url \@url }%
\providecommand \@url [1]{\endgroup\@href {#1}{\urlprefix }}%
\providecommand \urlprefix  [0]{URL }%
\providecommand \Eprint [0]{\href }%
\providecommand \doibase [0]{http://dx.doi.org/}%
\providecommand \selectlanguage [0]{\@gobble}%
\providecommand \bibinfo  [0]{\@secondoftwo}%
\providecommand \bibfield  [0]{\@secondoftwo}%
\providecommand \translation [1]{[#1]}%
\providecommand \BibitemOpen [0]{}%
\providecommand \bibitemStop [0]{}%
\providecommand \bibitemNoStop [0]{.\EOS\space}%
\providecommand \EOS [0]{\spacefactor3000\relax}%
\providecommand \BibitemShut  [1]{\csname bibitem#1\endcsname}%
\let\auto@bib@innerbib\@empty
\bibitem [{\citenamefont {Bejas}\ \emph {et~al.}(2012)\citenamefont {Bejas},
  \citenamefont {Greco},\ and\ \citenamefont {Yamase}}]{Bejas:2012wa}%
  \BibitemOpen
  \bibfield  {author} {\bibinfo {author} {\bibfnamefont {Mat\'ias}\
  \bibnamefont {Bejas}}, \bibinfo {author} {\bibfnamefont {Andr\'es}\
  \bibnamefont {Greco}}, \ and\ \bibinfo {author} {\bibfnamefont {Hiroyuki}\
  \bibnamefont {Yamase}},\ }\bibfield  {title} {\enquote {\bibinfo {title}
  {{Possible Charge Instabilities in Two-Dimensional Doped Mott Insulators}},}\
  }\href {\doibase 10.1103/PhysRevB.86.224509} {\bibfield  {journal} {\bibinfo
  {journal} {Phys. Rev. B}\ }\textbf {\bibinfo {volume} {86}},\ \bibinfo
  {pages} {224509} (\bibinfo {year} {2012})}\BibitemShut {NoStop}%
\bibitem [{\citenamefont {Faye}\ and\ \citenamefont
  {S{\'e}n{\'e}chal}(2017)}]{Faye:2017tx}%
  \BibitemOpen
  \bibfield  {author} {\bibinfo {author} {\bibfnamefont {J.~P.~L.}\
  \bibnamefont {Faye}}\ and\ \bibinfo {author} {\bibfnamefont {D.}~\bibnamefont
  {S{\'e}n{\'e}chal}},\ }\bibfield  {title} {\enquote {\bibinfo {title}
  {{Interplay Between $d$-Wave Superconductivity and a Bond Density Wave in the
  One-Band Hubbard Model}},}\ }\href@noop {} {\bibfield  {journal} {\bibinfo
  {journal} {Phys. Rev. B}\ }\textbf {\bibinfo {volume} {95}},\ \bibinfo
  {pages} {115127} (\bibinfo {year} {2017})}\BibitemShut {NoStop}%
\bibitem [{\citenamefont {Hamidian}\ \emph {et~al.}(2015)\citenamefont
  {Hamidian}, \citenamefont {Edkins}, \citenamefont {Kim}, \citenamefont
  {Davis}, \citenamefont {Mackenzie}, \citenamefont {Eisaki}, \citenamefont
  {Uchida}, \citenamefont {Lawler}, \citenamefont {Kim}, \citenamefont
  {Sachdev},\ and\ \citenamefont {Fujita}}]{Hamidian:2015us}%
  \BibitemOpen
  \bibfield  {author} {\bibinfo {author} {\bibfnamefont {M.~H.}\ \bibnamefont
  {Hamidian}}, \bibinfo {author} {\bibfnamefont {S.~D.}\ \bibnamefont
  {Edkins}}, \bibinfo {author} {\bibfnamefont {Chung~Koo}\ \bibnamefont {Kim}},
  \bibinfo {author} {\bibfnamefont {J.~C.~S{\'e}amus}\ \bibnamefont {Davis}},
  \bibinfo {author} {\bibfnamefont {A.~P.}\ \bibnamefont {Mackenzie}}, \bibinfo
  {author} {\bibfnamefont {H.}~\bibnamefont {Eisaki}}, \bibinfo {author}
  {\bibfnamefont {S.}~\bibnamefont {Uchida}}, \bibinfo {author} {\bibfnamefont
  {M.~J.}\ \bibnamefont {Lawler}}, \bibinfo {author} {\bibfnamefont {E.~A.}\
  \bibnamefont {Kim}}, \bibinfo {author} {\bibfnamefont {Subir}\ \bibnamefont
  {Sachdev}}, \ and\ \bibinfo {author} {\bibfnamefont {K.}~\bibnamefont
  {Fujita}},\ }\bibfield  {title} {\enquote {\bibinfo {title} {{Atomic-scale
  Electronic Structure of the Cuprate $d$-Symmetry Form Factor Density Wave
  State}},}\ }\href@noop {} {\bibfield  {journal} {\bibinfo  {journal} {Nature
  Phys.}\ }\textbf {\bibinfo {volume} {12}},\ \bibinfo {pages} {150--156}
  (\bibinfo {year} {2015})}\BibitemShut {NoStop}%
\bibitem [{\citenamefont {Caprara}\ \emph {et~al.}(2017)\citenamefont
  {Caprara}, \citenamefont {Di~Castro}, \citenamefont {Seibold},\ and\
  \citenamefont {Grilli}}]{Caprara:2017he}%
  \BibitemOpen
  \bibfield  {author} {\bibinfo {author} {\bibfnamefont {S.}~\bibnamefont
  {Caprara}}, \bibinfo {author} {\bibfnamefont {C.}~\bibnamefont {Di~Castro}},
  \bibinfo {author} {\bibfnamefont {G.}~\bibnamefont {Seibold}}, \ and\
  \bibinfo {author} {\bibfnamefont {M.}~\bibnamefont {Grilli}},\ }\bibfield
  {title} {\enquote {\bibinfo {title} {{Dynamical Charge Density Waves Rule the
  Phase Diagram of Cuprates}},}\ }\href@noop {} {\bibfield  {journal} {\bibinfo
   {journal} {Phys. Rev. B}\ }\textbf {\bibinfo {volume} {95}},\ \bibinfo
  {pages} {224511} (\bibinfo {year} {2017})}\BibitemShut {NoStop}%
\bibitem [{\citenamefont {Verret}\ \emph {et~al.}(2017)\citenamefont {Verret},
  \citenamefont {Charlebois}, \citenamefont {S\'en\'echal},\ and\ \citenamefont
  {Tremblay}}]{Verret:2017th}%
  \BibitemOpen
  \bibfield  {author} {\bibinfo {author} {\bibfnamefont {S.}~\bibnamefont
  {Verret}}, \bibinfo {author} {\bibfnamefont {M.}~\bibnamefont {Charlebois}},
  \bibinfo {author} {\bibfnamefont {D.}~\bibnamefont {S\'en\'echal}}, \ and\
  \bibinfo {author} {\bibfnamefont {A.-M.~S.}\ \bibnamefont {Tremblay}},\
  }\bibfield  {title} {\enquote {\bibinfo {title} {{Subgap Structures and
  Pseudogap in Cuprate Superconductors: Role of Density Waves}},}\ }\href
  {\doibase 10.1103/PhysRevB.95.054518} {\bibfield  {journal} {\bibinfo
  {journal} {Phys. Rev. B}\ }\textbf {\bibinfo {volume} {95}},\ \bibinfo
  {pages} {054518} (\bibinfo {year} {2017})}\BibitemShut {NoStop}%
\bibitem [{\citenamefont {Chatterjee}\ and\ \citenamefont
  {Sachdev}(2016)}]{Chatterjee:2016ij}%
  \BibitemOpen
  \bibfield  {author} {\bibinfo {author} {\bibfnamefont {Shubhayu}\
  \bibnamefont {Chatterjee}}\ and\ \bibinfo {author} {\bibfnamefont {Subir}\
  \bibnamefont {Sachdev}},\ }\bibfield  {title} {\enquote {\bibinfo {title}
  {{Fractionalized Fermi Liquid with Bosonic Chargons as a Candidate for the
  Pseudogap Metal}},}\ }\href@noop {} {\bibfield  {journal} {\bibinfo
  {journal} {Phys. Rev. B}\ }\textbf {\bibinfo {volume} {94}},\ \bibinfo
  {pages} {205117} (\bibinfo {year} {2016})}\BibitemShut {NoStop}%
\bibitem [{\citenamefont {Efetov}\ \emph {et~al.}(2013)\citenamefont {Efetov},
  \citenamefont {Meier},\ and\ \citenamefont {P\'{e}pin}}]{Efetov:2013}%
  \BibitemOpen
  \bibfield  {author} {\bibinfo {author} {\bibfnamefont {K.~B.}\ \bibnamefont
  {Efetov}}, \bibinfo {author} {\bibfnamefont {H.}~\bibnamefont {Meier}}, \
  and\ \bibinfo {author} {\bibfnamefont {C.}~\bibnamefont {P\'{e}pin}},\
  }\bibfield  {title} {\enquote {\bibinfo {title} {{Pseudogap State Near a
  Quantum Critical Point}},}\ }\href {\doibase 10.1038/nphys2641} {\bibfield
  {journal} {\bibinfo  {journal} {Nature Phys.}\ }\textbf {\bibinfo {volume}
  {9}},\ \bibinfo {pages} {442--446} (\bibinfo {year} {2013})}\BibitemShut
  {NoStop}%
\bibitem [{\citenamefont {Hayward}\ \emph {et~al.}(2014)\citenamefont
  {Hayward}, \citenamefont {Hawthorn}, \citenamefont {Melko},\ and\
  \citenamefont {Sachdev}}]{Hayward:2014eo}%
  \BibitemOpen
  \bibfield  {author} {\bibinfo {author} {\bibfnamefont {L.~E.}\ \bibnamefont
  {Hayward}}, \bibinfo {author} {\bibfnamefont {D.~G.}\ \bibnamefont
  {Hawthorn}}, \bibinfo {author} {\bibfnamefont {R.~G.}\ \bibnamefont {Melko}},
  \ and\ \bibinfo {author} {\bibfnamefont {S.}~\bibnamefont {Sachdev}},\
  }\bibfield  {title} {\enquote {\bibinfo {title} {{Angular Fluctuations of a
  Multicomponent Order Describe the Pseudogap of YBa$_2$Cu$_3$O$_{6+x}$}},}\
  }\href@noop {} {\bibfield  {journal} {\bibinfo  {journal} {Science}\ }\textbf
  {\bibinfo {volume} {343}},\ \bibinfo {pages} {1336--1339} (\bibinfo {year}
  {2014})}\BibitemShut {NoStop}%
\bibitem [{\citenamefont {Tsvelik}\ and\ \citenamefont
  {Chubukov}(2014)}]{Tsvelik:2014ce}%
  \BibitemOpen
  \bibfield  {author} {\bibinfo {author} {\bibfnamefont {A.~M.}\ \bibnamefont
  {Tsvelik}}\ and\ \bibinfo {author} {\bibfnamefont {A.~V.}\ \bibnamefont
  {Chubukov}},\ }\bibfield  {title} {\enquote {\bibinfo {title} {{Composite
  Charge Order in the Pseudogap Region of the Cuprates}},}\ }\href@noop {}
  {\bibfield  {journal} {\bibinfo  {journal} {Phys. Rev. B}\ }\textbf {\bibinfo
  {volume} {89}},\ \bibinfo {pages} {184515} (\bibinfo {year}
  {2014})}\BibitemShut {NoStop}%
\bibitem [{\citenamefont {Wang}\ and\ \citenamefont
  {Chubukov}(2014)}]{Wang:2014fr}%
  \BibitemOpen
  \bibfield  {author} {\bibinfo {author} {\bibfnamefont {Yuxuan}\ \bibnamefont
  {Wang}}\ and\ \bibinfo {author} {\bibfnamefont {Andrey}\ \bibnamefont
  {Chubukov}},\ }\bibfield  {title} {\enquote {\bibinfo {title}
  {{Charge-Density-Wave Order With Momentum $(2Q,0)$ and $(0,2Q)$ Within The
  Spin-Fermion Model: Continuous and Discrete Symmetry Breaking, Preemptive
  Composite Order, and Relation to Pseudogap in Hole-Doped Cuprates}},}\
  }\href@noop {} {\bibfield  {journal} {\bibinfo  {journal} {Phys. Rev. B}\
  }\textbf {\bibinfo {volume} {90}},\ \bibinfo {pages} {035149} (\bibinfo
  {year} {2014})}\BibitemShut {NoStop}%
\bibitem [{\citenamefont {Wang}\ \emph {et~al.}(2015)\citenamefont {Wang},
  \citenamefont {Agterberg},\ and\ \citenamefont {Chubukov}}]{Wang:2015iq}%
  \BibitemOpen
  \bibfield  {author} {\bibinfo {author} {\bibfnamefont {Yuxuan}\ \bibnamefont
  {Wang}}, \bibinfo {author} {\bibfnamefont {Daniel~F.}\ \bibnamefont
  {Agterberg}}, \ and\ \bibinfo {author} {\bibfnamefont {Andrey}\ \bibnamefont
  {Chubukov}},\ }\bibfield  {title} {\enquote {\bibinfo {title} {{Coexistence
  of Charge-Density-Wave and Pair-Density-Wave Orders in Underdoped
  Cuprates}},}\ }\href@noop {} {\bibfield  {journal} {\bibinfo  {journal}
  {Phys. Rev. Lett.}\ }\textbf {\bibinfo {volume} {114}},\ \bibinfo {pages}
  {197001} (\bibinfo {year} {2015})}\BibitemShut {NoStop}%
\bibitem [{\citenamefont {Atkinson}\ \emph {et~al.}(2016)\citenamefont
  {Atkinson}, \citenamefont {Kampf},\ and\ \citenamefont
  {Bulut}}]{Atkinson:2016ba}%
  \BibitemOpen
  \bibfield  {author} {\bibinfo {author} {\bibfnamefont {W.~A.}\ \bibnamefont
  {Atkinson}}, \bibinfo {author} {\bibfnamefont {A.~P.}\ \bibnamefont {Kampf}},
  \ and\ \bibinfo {author} {\bibfnamefont {S.}~\bibnamefont {Bulut}},\
  }\bibfield  {title} {\enquote {\bibinfo {title} {{Emergence of Charge Order
  in a Staggered Loop-Current Phase of Cuprate High-Temperature
  Superconductors}},}\ }\href@noop {} {\bibfield  {journal} {\bibinfo
  {journal} {Phys. Rev. B}\ }\textbf {\bibinfo {volume} {93}},\ \bibinfo
  {pages} {134517} (\bibinfo {year} {2016})}\BibitemShut {NoStop}%
\bibitem [{\citenamefont {Hoffman}\ \emph {et~al.}(2002)\citenamefont
  {Hoffman}, \citenamefont {Hudson}, \citenamefont {Lang}, \citenamefont
  {Madhavan}, \citenamefont {Eisaki}, \citenamefont {Uchida},\ and\
  \citenamefont {Davis}}]{Hoffman:2002bk}%
  \BibitemOpen
  \bibfield  {author} {\bibinfo {author} {\bibfnamefont {J.~E.}\ \bibnamefont
  {Hoffman}}, \bibinfo {author} {\bibfnamefont {E.~W.}\ \bibnamefont {Hudson}},
  \bibinfo {author} {\bibfnamefont {K.~M.}\ \bibnamefont {Lang}}, \bibinfo
  {author} {\bibfnamefont {V.}~\bibnamefont {Madhavan}}, \bibinfo {author}
  {\bibfnamefont {H.}~\bibnamefont {Eisaki}}, \bibinfo {author} {\bibfnamefont
  {S.}~\bibnamefont {Uchida}}, \ and\ \bibinfo {author} {\bibfnamefont {J.~C.}\
  \bibnamefont {Davis}},\ }\bibfield  {title} {\enquote {\bibinfo {title} {{A
  Four Unit Cell Periodic Pattern of Quasi-Particle States Surrounding Vortex
  Cores in Bi$_2$Sr$_2$CaCu$_2$O$_{8+\delta}$}},}\ }\href@noop {} {\bibfield
  {journal} {\bibinfo  {journal} {Science}\ }\textbf {\bibinfo {volume}
  {295}},\ \bibinfo {pages} {466} (\bibinfo {year} {2002})}\BibitemShut
  {NoStop}%
\bibitem [{\citenamefont {Kohsaka}\ \emph {et~al.}(2007)\citenamefont
  {Kohsaka}, \citenamefont {Taylor}, \citenamefont {Fujita}, \citenamefont
  {Schmidt}, \citenamefont {Lupien}, \citenamefont {Hanaguri}, \citenamefont
  {Azuma}, \citenamefont {Takano}, \citenamefont {Eisaki}, \citenamefont
  {Takagi}, \citenamefont {Uchida},\ and\ \citenamefont
  {Davis}}]{Kohsaka:2007hx}%
  \BibitemOpen
  \bibfield  {author} {\bibinfo {author} {\bibfnamefont {Y.}~\bibnamefont
  {Kohsaka}}, \bibinfo {author} {\bibfnamefont {C.}~\bibnamefont {Taylor}},
  \bibinfo {author} {\bibfnamefont {K.}~\bibnamefont {Fujita}}, \bibinfo
  {author} {\bibfnamefont {A.}~\bibnamefont {Schmidt}}, \bibinfo {author}
  {\bibfnamefont {C.}~\bibnamefont {Lupien}}, \bibinfo {author} {\bibfnamefont
  {T.}~\bibnamefont {Hanaguri}}, \bibinfo {author} {\bibfnamefont
  {M.}~\bibnamefont {Azuma}}, \bibinfo {author} {\bibfnamefont
  {M.}~\bibnamefont {Takano}}, \bibinfo {author} {\bibfnamefont
  {H.}~\bibnamefont {Eisaki}}, \bibinfo {author} {\bibfnamefont
  {H.}~\bibnamefont {Takagi}}, \bibinfo {author} {\bibfnamefont
  {S.}~\bibnamefont {Uchida}}, \ and\ \bibinfo {author} {\bibfnamefont {J.~C.}\
  \bibnamefont {Davis}},\ }\bibfield  {title} {\enquote {\bibinfo {title} {{An
  Intrinsic Bond-Centered Electronic Glass with Unidirectional Domains in
  Underdoped Cuprates}},}\ }\href@noop {} {\bibfield  {journal} {\bibinfo
  {journal} {Science}\ }\textbf {\bibinfo {volume} {315}},\ \bibinfo {pages}
  {1380--1385} (\bibinfo {year} {2007})}\BibitemShut {NoStop}%
\bibitem [{\citenamefont {Wise}\ \emph {et~al.}(2008)\citenamefont {Wise},
  \citenamefont {Boyer}, \citenamefont {Chatterjee}, \citenamefont {Kondo},
  \citenamefont {Takeuchi}, \citenamefont {Ikuta}, \citenamefont {Wang},\ and\
  \citenamefont {Hudson}}]{Wise:2008cd}%
  \BibitemOpen
  \bibfield  {author} {\bibinfo {author} {\bibfnamefont {W.~D.}\ \bibnamefont
  {Wise}}, \bibinfo {author} {\bibfnamefont {M.~C.}\ \bibnamefont {Boyer}},
  \bibinfo {author} {\bibfnamefont {Kamalesh}\ \bibnamefont {Chatterjee}},
  \bibinfo {author} {\bibfnamefont {Takeshi}\ \bibnamefont {Kondo}}, \bibinfo
  {author} {\bibfnamefont {T.}~\bibnamefont {Takeuchi}}, \bibinfo {author}
  {\bibfnamefont {H.}~\bibnamefont {Ikuta}}, \bibinfo {author} {\bibfnamefont
  {Yayu}\ \bibnamefont {Wang}}, \ and\ \bibinfo {author} {\bibfnamefont
  {E.~W.}\ \bibnamefont {Hudson}},\ }\bibfield  {title} {\enquote {\bibinfo
  {title} {{Charge-Density-Wave Origin of Cuprate Checkerboard Visualized by
  Scanning Tunnelling Microscopy}},}\ }\href@noop {} {\bibfield  {journal}
  {\bibinfo  {journal} {Nature Phys.}\ }\textbf {\bibinfo {volume} {4}},\
  \bibinfo {pages} {696--699} (\bibinfo {year} {2008})}\BibitemShut {NoStop}%
\bibitem [{\citenamefont {Fujita}\ \emph {et~al.}(2014)\citenamefont {Fujita},
  \citenamefont {Hamidian}, \citenamefont {Edkins}, \citenamefont {Kim},
  \citenamefont {Kohsaka}, \citenamefont {Azuma}, \citenamefont {Takano},
  \citenamefont {Takagi}, \citenamefont {Eisaki}, \citenamefont {Uchida},
  \citenamefont {Allais}, \citenamefont {Lawler}, \citenamefont {Kim},
  \citenamefont {Sachdev},\ and\ \citenamefont {Davis}}]{Fujita:2014kg}%
  \BibitemOpen
  \bibfield  {author} {\bibinfo {author} {\bibfnamefont {K.}~\bibnamefont
  {Fujita}}, \bibinfo {author} {\bibfnamefont {M.~H.}\ \bibnamefont
  {Hamidian}}, \bibinfo {author} {\bibfnamefont {S.~D.}\ \bibnamefont
  {Edkins}}, \bibinfo {author} {\bibfnamefont {C.~K.}\ \bibnamefont {Kim}},
  \bibinfo {author} {\bibfnamefont {Y.}~\bibnamefont {Kohsaka}}, \bibinfo
  {author} {\bibfnamefont {M.}~\bibnamefont {Azuma}}, \bibinfo {author}
  {\bibfnamefont {M.}~\bibnamefont {Takano}}, \bibinfo {author} {\bibfnamefont
  {H.}~\bibnamefont {Takagi}}, \bibinfo {author} {\bibfnamefont
  {H.}~\bibnamefont {Eisaki}}, \bibinfo {author} {\bibfnamefont {S.-I}\
  \bibnamefont {Uchida}}, \bibinfo {author} {\bibfnamefont {A.}~\bibnamefont
  {Allais}}, \bibinfo {author} {\bibfnamefont {M.~J.}\ \bibnamefont {Lawler}},
  \bibinfo {author} {\bibfnamefont {E.~A.}\ \bibnamefont {Kim}}, \bibinfo
  {author} {\bibfnamefont {S.}~\bibnamefont {Sachdev}}, \ and\ \bibinfo
  {author} {\bibfnamefont {J.~C.~S.}\ \bibnamefont {Davis}},\ }\bibfield
  {title} {\enquote {\bibinfo {title} {{Direct Phase-Sensitive Identification
  of a $d$-Form Factor Density Wave in Underdoped Cuprates}},}\ }\href@noop {}
  {\bibfield  {journal} {\bibinfo  {journal} {Proc. Nature Acad. Sci.}\
  }\textbf {\bibinfo {volume} {111}},\ \bibinfo {pages} {E3026--E3032}
  (\bibinfo {year} {2014})}\BibitemShut {NoStop}%
\bibitem [{\citenamefont {Mesaros}\ \emph {et~al.}(2016)\citenamefont
  {Mesaros}, \citenamefont {Fujita}, \citenamefont {Edkins}, \citenamefont
  {Hamidian}, \citenamefont {Eisaki}, \citenamefont {Uchida}, \citenamefont
  {Davis}, \citenamefont {Lawler},\ and\ \citenamefont {Kim}}]{Mesaros:2016uz}%
  \BibitemOpen
  \bibfield  {author} {\bibinfo {author} {\bibfnamefont {A.}~\bibnamefont
  {Mesaros}}, \bibinfo {author} {\bibfnamefont {K.}~\bibnamefont {Fujita}},
  \bibinfo {author} {\bibfnamefont {S.~D.}\ \bibnamefont {Edkins}}, \bibinfo
  {author} {\bibfnamefont {M.~H.}\ \bibnamefont {Hamidian}}, \bibinfo {author}
  {\bibfnamefont {H.}~\bibnamefont {Eisaki}}, \bibinfo {author} {\bibfnamefont
  {S.}~\bibnamefont {Uchida}}, \bibinfo {author} {\bibfnamefont
  {J.~C.~S{\'e}amus}\ \bibnamefont {Davis}}, \bibinfo {author} {\bibfnamefont
  {M.~J.}\ \bibnamefont {Lawler}}, \ and\ \bibinfo {author} {\bibfnamefont
  {Eun-Ah}\ \bibnamefont {Kim}},\ }\bibfield  {title} {\enquote {\bibinfo
  {title} {{Commensurate $4a_0$ period Charge Density Modulations throughout
  the Bi$_2$Sr$_2$CaCu$_2$O$_{8+x}$ Pseudogap Regime}},}\ }\href@noop {}
  {\bibfield  {journal} {\bibinfo  {journal} {Proc. Nature Acad. Sci.}\
  }\textbf {\bibinfo {volume} {113}},\ \bibinfo {pages} {12661--12666}
  (\bibinfo {year} {2016})}\BibitemShut {NoStop}%
\bibitem [{\citenamefont {Comin}\ \emph
  {et~al.}(2015{\natexlab{a}})\citenamefont {Comin}, \citenamefont {Sutarto},
  \citenamefont {He}, \citenamefont {Neto}, \citenamefont {Chauviere},
  \citenamefont {Frano}, \citenamefont {Liang}, \citenamefont {Hardy},
  \citenamefont {Bonn}, \citenamefont {Yoshida}, \citenamefont {Eisaki},
  \citenamefont {Hoffman}, \citenamefont {Keimer}, \citenamefont {Sawatzky},\
  and\ \citenamefont {Damascelli}}]{Comin:2014vq}%
  \BibitemOpen
  \bibfield  {author} {\bibinfo {author} {\bibfnamefont {R.}~\bibnamefont
  {Comin}}, \bibinfo {author} {\bibfnamefont {R.}~\bibnamefont {Sutarto}},
  \bibinfo {author} {\bibfnamefont {F.}~\bibnamefont {He}}, \bibinfo {author}
  {\bibfnamefont {E.~da~Silva}\ \bibnamefont {Neto}}, \bibinfo {author}
  {\bibfnamefont {L.}~\bibnamefont {Chauviere}}, \bibinfo {author}
  {\bibfnamefont {A.}~\bibnamefont {Frano}}, \bibinfo {author} {\bibfnamefont
  {R.}~\bibnamefont {Liang}}, \bibinfo {author} {\bibfnamefont {W.~N.}\
  \bibnamefont {Hardy}}, \bibinfo {author} {\bibfnamefont {D.}~\bibnamefont
  {Bonn}}, \bibinfo {author} {\bibfnamefont {Y.}~\bibnamefont {Yoshida}},
  \bibinfo {author} {\bibfnamefont {H.}~\bibnamefont {Eisaki}}, \bibinfo
  {author} {\bibfnamefont {J.~E.}\ \bibnamefont {Hoffman}}, \bibinfo {author}
  {\bibfnamefont {B.}~\bibnamefont {Keimer}}, \bibinfo {author} {\bibfnamefont
  {G.~A.}\ \bibnamefont {Sawatzky}}, \ and\ \bibinfo {author} {\bibfnamefont
  {A.}~\bibnamefont {Damascelli}},\ }\bibfield  {title} {\enquote {\bibinfo
  {title} {{The Symmetry of Charge Order in Cuprates}},}\ }\href@noop {}
  {\bibfield  {journal} {\bibinfo  {journal} {Nat. Mater.}\ }\textbf {\bibinfo
  {volume} {14]}},\ \bibinfo {pages} {796--800} (\bibinfo {year}
  {2015}{\natexlab{a}})}\BibitemShut {NoStop}%
\bibitem [{\citenamefont {Comin}\ and\ \citenamefont
  {Damascelli}(2016)}]{Comin:2016}%
  \BibitemOpen
  \bibfield  {author} {\bibinfo {author} {\bibfnamefont {Riccardo}\
  \bibnamefont {Comin}}\ and\ \bibinfo {author} {\bibfnamefont {Andrea}\
  \bibnamefont {Damascelli}},\ }\bibfield  {title} {\enquote {\bibinfo {title}
  {Resonant x-ray scattering studies of charge order in cuprates},}\
  }\href@noop {} {\bibfield  {journal} {\bibinfo  {journal} {Ann. Rev. Condens.
  Mat.}\ }\textbf {\bibinfo {volume} {7}},\ \bibinfo {pages} {369--405}
  (\bibinfo {year} {2016})}\BibitemShut {NoStop}%
\bibitem [{\citenamefont {Atkinson}\ and\ \citenamefont
  {Kampf}(2015)}]{Atkinson:2015hd}%
  \BibitemOpen
  \bibfield  {author} {\bibinfo {author} {\bibfnamefont {W.~A.}\ \bibnamefont
  {Atkinson}}\ and\ \bibinfo {author} {\bibfnamefont {A.~P.}\ \bibnamefont
  {Kampf}},\ }\bibfield  {title} {\enquote {\bibinfo {title} {{Effect of
  Pointlike Impurities on $d_{x^2-y^2}$ Charge-Density Waves in Cuprate
  Superconductors}},}\ }\href {\doibase 10.1103/PhysRevB.91.104509} {\bibfield
  {journal} {\bibinfo  {journal} {Phys. Rev. B}\ }\textbf {\bibinfo {volume}
  {91}},\ \bibinfo {pages} {104509} (\bibinfo {year} {2015})}\BibitemShut
  {NoStop}%
\bibitem [{\citenamefont {Ghiringhelli}\ \emph {et~al.}(2012)\citenamefont
  {Ghiringhelli}, \citenamefont {Le~Tacon}, \citenamefont {Minola},
  \citenamefont {Blanco-Canosa}, \citenamefont {Mazzoli}, \citenamefont
  {Brookes}, \citenamefont {De~Luca}, \citenamefont {Frano}, \citenamefont
  {Hawthorn}, \citenamefont {He}, \citenamefont {Loew}, \citenamefont {Sala},
  \citenamefont {Peets}, \citenamefont {Salluzzo}, \citenamefont {Schierle},
  \citenamefont {Sutarto}, \citenamefont {Sawatzky}, \citenamefont {Weschke},
  \citenamefont {Keimer},\ and\ \citenamefont
  {Braicovich}}]{Ghiringhelli:2012bw}%
  \BibitemOpen
  \bibfield  {author} {\bibinfo {author} {\bibfnamefont {G.}~\bibnamefont
  {Ghiringhelli}}, \bibinfo {author} {\bibfnamefont {M.}~\bibnamefont
  {Le~Tacon}}, \bibinfo {author} {\bibfnamefont {M.}~\bibnamefont {Minola}},
  \bibinfo {author} {\bibfnamefont {S.}~\bibnamefont {Blanco-Canosa}}, \bibinfo
  {author} {\bibfnamefont {C.}~\bibnamefont {Mazzoli}}, \bibinfo {author}
  {\bibfnamefont {N.~B.}\ \bibnamefont {Brookes}}, \bibinfo {author}
  {\bibfnamefont {G.~M.}\ \bibnamefont {De~Luca}}, \bibinfo {author}
  {\bibfnamefont {A.}~\bibnamefont {Frano}}, \bibinfo {author} {\bibfnamefont
  {D.~G.}\ \bibnamefont {Hawthorn}}, \bibinfo {author} {\bibfnamefont
  {F.}~\bibnamefont {He}}, \bibinfo {author} {\bibfnamefont {T.}~\bibnamefont
  {Loew}}, \bibinfo {author} {\bibfnamefont {M.~M.}\ \bibnamefont {Sala}},
  \bibinfo {author} {\bibfnamefont {D.~C.}\ \bibnamefont {Peets}}, \bibinfo
  {author} {\bibfnamefont {M.}~\bibnamefont {Salluzzo}}, \bibinfo {author}
  {\bibfnamefont {E.}~\bibnamefont {Schierle}}, \bibinfo {author}
  {\bibfnamefont {R.}~\bibnamefont {Sutarto}}, \bibinfo {author} {\bibfnamefont
  {G.~A.}\ \bibnamefont {Sawatzky}}, \bibinfo {author} {\bibfnamefont
  {E.}~\bibnamefont {Weschke}}, \bibinfo {author} {\bibfnamefont
  {B.}~\bibnamefont {Keimer}}, \ and\ \bibinfo {author} {\bibfnamefont
  {L.}~\bibnamefont {Braicovich}},\ }\bibfield  {title} {\enquote {\bibinfo
  {title} {{Long-Range Incommensurate Charge Fluctuations in
  (Y,Nd)Ba$_2$Cu$_3$O$_{6+x}$}},}\ }\href@noop {} {\bibfield  {journal}
  {\bibinfo  {journal} {Science}\ }\textbf {\bibinfo {volume} {337}},\ \bibinfo
  {pages} {821--825} (\bibinfo {year} {2012})}\BibitemShut {NoStop}%
\bibitem [{\citenamefont {Chang}\ \emph {et~al.}(2012)\citenamefont {Chang},
  \citenamefont {Blackburn}, \citenamefont {Holmes}, \citenamefont
  {Christensen}, \citenamefont {Larsen}, \citenamefont {Mesot}, \citenamefont
  {Liang}, \citenamefont {Bonn}, \citenamefont {Hardy}, \citenamefont
  {Watenphul}, \citenamefont {Zimmermann}, \citenamefont {Forgan},\ and\
  \citenamefont {Hayden}}]{Chang:2012vf}%
  \BibitemOpen
  \bibfield  {author} {\bibinfo {author} {\bibfnamefont {J.}~\bibnamefont
  {Chang}}, \bibinfo {author} {\bibfnamefont {E.}~\bibnamefont {Blackburn}},
  \bibinfo {author} {\bibfnamefont {A.~T.}\ \bibnamefont {Holmes}}, \bibinfo
  {author} {\bibfnamefont {N.~B.}\ \bibnamefont {Christensen}}, \bibinfo
  {author} {\bibfnamefont {J.}~\bibnamefont {Larsen}}, \bibinfo {author}
  {\bibfnamefont {J.}~\bibnamefont {Mesot}}, \bibinfo {author} {\bibfnamefont
  {Ruixing}\ \bibnamefont {Liang}}, \bibinfo {author} {\bibfnamefont {D.~A.}\
  \bibnamefont {Bonn}}, \bibinfo {author} {\bibfnamefont {W.~N.}\ \bibnamefont
  {Hardy}}, \bibinfo {author} {\bibfnamefont {A.}~\bibnamefont {Watenphul}},
  \bibinfo {author} {\bibfnamefont {M.~V.}\ \bibnamefont {Zimmermann}},
  \bibinfo {author} {\bibfnamefont {E.~M.}\ \bibnamefont {Forgan}}, \ and\
  \bibinfo {author} {\bibfnamefont {S.~M.}\ \bibnamefont {Hayden}},\ }\bibfield
   {title} {\enquote {\bibinfo {title} {{Direct Observation of Competition
  Between Superconductivity and Charge Density Wave Order in
  YBa$_2$Cu$_3$O$_y$}},}\ }\href {\doibase 10.1038/nphys2456} {\bibfield
  {journal} {\bibinfo  {journal} {Nat. Phys.}\ }\textbf {\bibinfo {volume}
  {8}},\ \bibinfo {pages} {871--876} (\bibinfo {year} {2012})}\BibitemShut
  {NoStop}%
\bibitem [{\citenamefont {Blackburn}\ \emph {et~al.}(2013)\citenamefont
  {Blackburn}, \citenamefont {Chang}, \citenamefont {H\"ucker}, \citenamefont
  {Holmes}, \citenamefont {Christensen}, \citenamefont {Liang}, \citenamefont
  {Bonn}, \citenamefont {Hardy}, \citenamefont {R\"utt}, \citenamefont
  {Gutowski}, \citenamefont {Zimmermann}, \citenamefont {Forgan},\ and\
  \citenamefont {Hayden}}]{Blackburn:2013bs}%
  \BibitemOpen
  \bibfield  {author} {\bibinfo {author} {\bibfnamefont {E.}~\bibnamefont
  {Blackburn}}, \bibinfo {author} {\bibfnamefont {J.}~\bibnamefont {Chang}},
  \bibinfo {author} {\bibfnamefont {M.}~\bibnamefont {H\"ucker}}, \bibinfo
  {author} {\bibfnamefont {A.~T.}\ \bibnamefont {Holmes}}, \bibinfo {author}
  {\bibfnamefont {N.~B.}\ \bibnamefont {Christensen}}, \bibinfo {author}
  {\bibfnamefont {Ruixing}\ \bibnamefont {Liang}}, \bibinfo {author}
  {\bibfnamefont {D.~A.}\ \bibnamefont {Bonn}}, \bibinfo {author}
  {\bibfnamefont {W.~N.}\ \bibnamefont {Hardy}}, \bibinfo {author}
  {\bibfnamefont {U.}~\bibnamefont {R\"utt}}, \bibinfo {author} {\bibfnamefont
  {O.}~\bibnamefont {Gutowski}}, \bibinfo {author} {\bibfnamefont {M.~v.}\
  \bibnamefont {Zimmermann}}, \bibinfo {author} {\bibfnamefont {E.~M.}\
  \bibnamefont {Forgan}}, \ and\ \bibinfo {author} {\bibfnamefont {S.~M.}\
  \bibnamefont {Hayden}},\ }\bibfield  {title} {\enquote {\bibinfo {title}
  {{X-Ray Diffraction Observations of a Charge-Density-Wave Order in
  Superconducting Ortho-II
  ${\mathrm{YBa}}_{2}{\mathrm{Cu}}_{3}\mathrm{O}_{6.54}$ Single Crystals in
  Zero Magnetic Field}},}\ }\href {\doibase 10.1103/PhysRevLett.110.137004}
  {\bibfield  {journal} {\bibinfo  {journal} {Phys. Rev. Lett.}\ }\textbf
  {\bibinfo {volume} {110}},\ \bibinfo {pages} {137004} (\bibinfo {year}
  {2013})}\BibitemShut {NoStop}%
\bibitem [{\citenamefont {Blanco-Canosa}\ \emph {et~al.}(2014)\citenamefont
  {Blanco-Canosa}, \citenamefont {Frano}, \citenamefont {Schierle},
  \citenamefont {Porras}, \citenamefont {Loew}, \citenamefont {Minola},
  \citenamefont {Bluschke}, \citenamefont {Weschke}, \citenamefont {Keimer},\
  and\ \citenamefont {Le~Tacon}}]{BlancoCanosa:2014ul}%
  \BibitemOpen
  \bibfield  {author} {\bibinfo {author} {\bibfnamefont {S.}~\bibnamefont
  {Blanco-Canosa}}, \bibinfo {author} {\bibfnamefont {A.}~\bibnamefont
  {Frano}}, \bibinfo {author} {\bibfnamefont {E.}~\bibnamefont {Schierle}},
  \bibinfo {author} {\bibfnamefont {J.}~\bibnamefont {Porras}}, \bibinfo
  {author} {\bibfnamefont {T.}~\bibnamefont {Loew}}, \bibinfo {author}
  {\bibfnamefont {M.}~\bibnamefont {Minola}}, \bibinfo {author} {\bibfnamefont
  {M.}~\bibnamefont {Bluschke}}, \bibinfo {author} {\bibfnamefont
  {E.}~\bibnamefont {Weschke}}, \bibinfo {author} {\bibfnamefont
  {B.}~\bibnamefont {Keimer}}, \ and\ \bibinfo {author} {\bibfnamefont
  {M.}~\bibnamefont {Le~Tacon}},\ }\bibfield  {title} {\enquote {\bibinfo
  {title} {{Resonant X-Ray Scattering Study of Charge-Density Wave Correlations
  in $\mathrm{YBa}_{2}\mathrm{Cu}_{3}\mathrm{O}_{6+x}$}},}\ }\href {\doibase
  10.1103/PhysRevB.90.054513} {\bibfield  {journal} {\bibinfo  {journal} {Phys.
  Rev. B}\ }\textbf {\bibinfo {volume} {90}},\ \bibinfo {pages} {054513}
  (\bibinfo {year} {2014})}\BibitemShut {NoStop}%
\bibitem [{\citenamefont {Chang}\ \emph {et~al.}(2016)\citenamefont {Chang},
  \citenamefont {Blackburn}, \citenamefont {Ivashko}, \citenamefont {Holmes},
  \citenamefont {Christensen}, \citenamefont {H{\"u}cker}, \citenamefont
  {Liang}, \citenamefont {Bonn}, \citenamefont {Hardy}, \citenamefont
  {R{\"u}tt}, \citenamefont {Zimmermann}, \citenamefont {Forgan},\ and\
  \citenamefont {Hayden}}]{Chang:2016gz}%
  \BibitemOpen
  \bibfield  {author} {\bibinfo {author} {\bibfnamefont {J.}~\bibnamefont
  {Chang}}, \bibinfo {author} {\bibfnamefont {E.}~\bibnamefont {Blackburn}},
  \bibinfo {author} {\bibfnamefont {O.}~\bibnamefont {Ivashko}}, \bibinfo
  {author} {\bibfnamefont {A.~T.}\ \bibnamefont {Holmes}}, \bibinfo {author}
  {\bibfnamefont {N.~B.}\ \bibnamefont {Christensen}}, \bibinfo {author}
  {\bibfnamefont {M.}~\bibnamefont {H{\"u}cker}}, \bibinfo {author}
  {\bibfnamefont {Ruixing}\ \bibnamefont {Liang}}, \bibinfo {author}
  {\bibfnamefont {D.~A.}\ \bibnamefont {Bonn}}, \bibinfo {author}
  {\bibfnamefont {W.~N.}\ \bibnamefont {Hardy}}, \bibinfo {author}
  {\bibfnamefont {U.}~\bibnamefont {R{\"u}tt}}, \bibinfo {author}
  {\bibfnamefont {M.~V.}\ \bibnamefont {Zimmermann}}, \bibinfo {author}
  {\bibfnamefont {E.~M.}\ \bibnamefont {Forgan}}, \ and\ \bibinfo {author}
  {\bibfnamefont {S.~M.}\ \bibnamefont {Hayden}},\ }\bibfield  {title}
  {\enquote {\bibinfo {title} {{Magnetic Field Controlled Charge Density Wave
  Coupling in Underdoped $\mathrm{YBa}_{2}\mathrm{Cu}_{3}\mathrm{O}_{6+\delta}$
  }},}\ }\href@noop {} {\bibfield  {journal} {\bibinfo  {journal} {Nat.
  Commun.}\ }\textbf {\bibinfo {volume} {7}},\ \bibinfo {pages} {11494}
  (\bibinfo {year} {2016})}\BibitemShut {NoStop}%
\bibitem [{\citenamefont {Tabis}\ \emph {et~al.}(2014)\citenamefont {Tabis},
  \citenamefont {Li}, \citenamefont {Le~Tacon}, \citenamefont {Braicovich},
  \citenamefont {Kreyssig}, \citenamefont {Minola}, \citenamefont {Dellea},
  \citenamefont {Weschke}, \citenamefont {Veit}, \citenamefont {Ramazanoglu},
  \citenamefont {Goldman}, \citenamefont {Schmitt}, \citenamefont
  {Ghiringhelli}, \citenamefont {Bari{\u s}i\'c}, \citenamefont {Chan},
  \citenamefont {Dorow}, \citenamefont {Yu}, \citenamefont {Zhao},
  \citenamefont {Keimer},\ and\ \citenamefont {Greven}}]{Tabis:2014kb}%
  \BibitemOpen
  \bibfield  {author} {\bibinfo {author} {\bibfnamefont {W.}~\bibnamefont
  {Tabis}}, \bibinfo {author} {\bibfnamefont {Y.}~\bibnamefont {Li}}, \bibinfo
  {author} {\bibfnamefont {M.}~\bibnamefont {Le~Tacon}}, \bibinfo {author}
  {\bibfnamefont {L.}~\bibnamefont {Braicovich}}, \bibinfo {author}
  {\bibfnamefont {A.}~\bibnamefont {Kreyssig}}, \bibinfo {author}
  {\bibfnamefont {M.}~\bibnamefont {Minola}}, \bibinfo {author} {\bibfnamefont
  {G.}~\bibnamefont {Dellea}}, \bibinfo {author} {\bibfnamefont
  {E.}~\bibnamefont {Weschke}}, \bibinfo {author} {\bibfnamefont {M.~J.}\
  \bibnamefont {Veit}}, \bibinfo {author} {\bibfnamefont {M.}~\bibnamefont
  {Ramazanoglu}}, \bibinfo {author} {\bibfnamefont {A.~I.}\ \bibnamefont
  {Goldman}}, \bibinfo {author} {\bibfnamefont {T.}~\bibnamefont {Schmitt}},
  \bibinfo {author} {\bibfnamefont {G.}~\bibnamefont {Ghiringhelli}}, \bibinfo
  {author} {\bibfnamefont {N.}~\bibnamefont {Bari{\u s}i\'c}}, \bibinfo
  {author} {\bibfnamefont {M.~K.}\ \bibnamefont {Chan}}, \bibinfo {author}
  {\bibfnamefont {C.~J.}\ \bibnamefont {Dorow}}, \bibinfo {author}
  {\bibfnamefont {G.}~\bibnamefont {Yu}}, \bibinfo {author} {\bibfnamefont
  {X.}~\bibnamefont {Zhao}}, \bibinfo {author} {\bibfnamefont {B.}~\bibnamefont
  {Keimer}}, \ and\ \bibinfo {author} {\bibfnamefont {M.}~\bibnamefont
  {Greven}},\ }\bibfield  {title} {\enquote {\bibinfo {title} {{Charge Order
  and Its Connection with Fermi-Liquid Charge Transport in a Pristine
  High-$T_c$ Cuprate}},}\ }\href@noop {} {\bibfield  {journal} {\bibinfo
  {journal} {Nat. Commun.}\ }\textbf {\bibinfo {volume} {5}},\ \bibinfo {pages}
  {5875} (\bibinfo {year} {2014})}\BibitemShut {NoStop}%
\bibitem [{\citenamefont {Tabis}\ \emph {et~al.}(2017)\citenamefont {Tabis},
  \citenamefont {Yu}, \citenamefont {Bialo}, \citenamefont {Bluschke},
  \citenamefont {Kolodziej}, \citenamefont {Kozlowski}, \citenamefont
  {Blackburn}, \citenamefont {Sen}, \citenamefont {Forgan}, \citenamefont
  {Zimmermann}, \citenamefont {Tang}, \citenamefont {Weschke}, \citenamefont
  {Vignolle}, \citenamefont {Hepting}, \citenamefont {Gretarsson},
  \citenamefont {Sutarto}, \citenamefont {He}, \citenamefont {Le~Tacon},
  \citenamefont {Bari{\u s}i\'c},
  \citenamefont {Yu},\ and\ \citenamefont {Greven}}]{Tabis:2017vp}%
  \BibitemOpen
  \bibfield  {author} {\bibinfo {author} {\bibfnamefont {W}~\bibnamefont
  {Tabis}}, \bibinfo {author} {\bibfnamefont {B}~\bibnamefont {Yu}}, \bibinfo
  {author} {\bibfnamefont {I}~\bibnamefont {Bialo}}, \bibinfo {author}
  {\bibfnamefont {M}~\bibnamefont {Bluschke}}, \bibinfo {author} {\bibfnamefont
  {T}~\bibnamefont {Kolodziej}}, \bibinfo {author} {\bibfnamefont
  {A}~\bibnamefont {Kozlowski}}, \bibinfo {author} {\bibfnamefont
  {E}~\bibnamefont {Blackburn}}, \bibinfo {author} {\bibfnamefont
  {K}~\bibnamefont {Sen}}, \bibinfo {author} {\bibfnamefont {E~M}\ \bibnamefont
  {Forgan}}, \bibinfo {author} {\bibfnamefont {M~v}\ \bibnamefont
  {Zimmermann}}, \bibinfo {author} {\bibfnamefont {Y}~\bibnamefont {Tang}},
  \bibinfo {author} {\bibfnamefont {E}~\bibnamefont {Weschke}}, \bibinfo
  {author} {\bibfnamefont {B.}~\bibnamefont {Vignolle}}, \bibinfo {author}
  {\bibfnamefont {M}~\bibnamefont {Hepting}}, \bibinfo {author} {\bibfnamefont
  {H}~\bibnamefont {Gretarsson}}, \bibinfo {author} {\bibfnamefont
  {R}~\bibnamefont {Sutarto}}, \bibinfo {author} {\bibfnamefont
  {F}~\bibnamefont {He}}, \bibinfo {author} {\bibfnamefont {M}~\bibnamefont
  {Le~Tacon}}, \bibinfo {author} {\bibfnamefont {N}~\bibnamefont {Bari{\u s}i\'c}}, \bibinfo {author}
  {\bibfnamefont {G}~\bibnamefont {Yu}}, \ and\ \bibinfo {author}
  {\bibfnamefont {M.}~\bibnamefont {Greven}},\ }\bibfield  {title} {\enquote
  {\bibinfo {title} {{Synchrotron x-ray scattering study of charge-density-wave
  order in $\mathrm{HgBa_{2}}\mathrm{CuO_{4+\ensuremath\delta}}$}},}\
  }\href@noop {} {\bibfield  {journal} {\bibinfo  {journal} {Physical Review
  B}\ }\textbf {\bibinfo {volume} {96}},\ \bibinfo {pages} {134510} (\bibinfo
  {year} {2017})}\BibitemShut {NoStop}%
\bibitem [{\citenamefont {Comin}\ \emph {et~al.}(2014)\citenamefont {Comin},
  \citenamefont {Frano}, \citenamefont {Yee}, \citenamefont {Yoshida},
  \citenamefont {Eisaki}, \citenamefont {Schierle}, \citenamefont {Weschke},
  \citenamefont {Sutarto}, \citenamefont {He}, \citenamefont {Soumyanarayanan},
  \citenamefont {He}, \citenamefont {Le~Tacon}, \citenamefont {Elfimov},
  \citenamefont {Hoffman}, \citenamefont {Sawatzky}, \citenamefont {Keimer},\
  and\ \citenamefont {Damascelli}}]{Comin:2013ck}%
  \BibitemOpen
  \bibfield  {author} {\bibinfo {author} {\bibfnamefont {R.}~\bibnamefont
  {Comin}}, \bibinfo {author} {\bibfnamefont {A.}~\bibnamefont {Frano}},
  \bibinfo {author} {\bibfnamefont {M.~M.}\ \bibnamefont {Yee}}, \bibinfo
  {author} {\bibfnamefont {Y.}~\bibnamefont {Yoshida}}, \bibinfo {author}
  {\bibfnamefont {H.}~\bibnamefont {Eisaki}}, \bibinfo {author} {\bibfnamefont
  {E.}~\bibnamefont {Schierle}}, \bibinfo {author} {\bibfnamefont
  {E.}~\bibnamefont {Weschke}}, \bibinfo {author} {\bibfnamefont
  {R.}~\bibnamefont {Sutarto}}, \bibinfo {author} {\bibfnamefont
  {F.}~\bibnamefont {He}}, \bibinfo {author} {\bibfnamefont {A.}~\bibnamefont
  {Soumyanarayanan}}, \bibinfo {author} {\bibfnamefont {Yang}\ \bibnamefont
  {He}}, \bibinfo {author} {\bibfnamefont {M.}~\bibnamefont {Le~Tacon}},
  \bibinfo {author} {\bibfnamefont {I.~S.}\ \bibnamefont {Elfimov}}, \bibinfo
  {author} {\bibfnamefont {Jennifer~E.}\ \bibnamefont {Hoffman}}, \bibinfo
  {author} {\bibfnamefont {G.~A.}\ \bibnamefont {Sawatzky}}, \bibinfo {author}
  {\bibfnamefont {B.}~\bibnamefont {Keimer}}, \ and\ \bibinfo {author}
  {\bibfnamefont {A.}~\bibnamefont {Damascelli}},\ }\bibfield  {title}
  {\enquote {\bibinfo {title} {Charge order driven by {Fermi}-arc instability
  in {Bi$_2$Sr$_{2-x}$La$_x$CuO$_{6+\delta}$}},}\ }\href {\doibase
  10.1126/science.1242996} {\bibfield  {journal} {\bibinfo  {journal}
  {Science}\ }\textbf {\bibinfo {volume} {343}},\ \bibinfo {pages} {390--392}
  (\bibinfo {year} {2014})}\BibitemShut {NoStop}%
\bibitem [{\citenamefont {Peng}\ \emph {et~al.}(2016)\citenamefont {Peng},
  \citenamefont {Salluzzo}, \citenamefont {Sun}, \citenamefont {Ponti},
  \citenamefont {Betto}, \citenamefont {Ferretti}, \citenamefont {Fumagalli},
  \citenamefont {Kummer}, \citenamefont {Le~Tacon}, \citenamefont {Zhou},
  \citenamefont {Brookes}, \citenamefont {Braicovich},\ and\ \citenamefont
  {Ghiringhelli}}]{Peng:2016jr}%
  \BibitemOpen
  \bibfield  {author} {\bibinfo {author} {\bibfnamefont {Y.~Y.}\ \bibnamefont
  {Peng}}, \bibinfo {author} {\bibfnamefont {M.}~\bibnamefont {Salluzzo}},
  \bibinfo {author} {\bibfnamefont {X.}~\bibnamefont {Sun}}, \bibinfo {author}
  {\bibfnamefont {A.}~\bibnamefont {Ponti}}, \bibinfo {author} {\bibfnamefont
  {D.}~\bibnamefont {Betto}}, \bibinfo {author} {\bibfnamefont {A.~M.}\
  \bibnamefont {Ferretti}}, \bibinfo {author} {\bibfnamefont {F.}~\bibnamefont
  {Fumagalli}}, \bibinfo {author} {\bibfnamefont {K.}~\bibnamefont {Kummer}},
  \bibinfo {author} {\bibfnamefont {M.}~\bibnamefont {Le~Tacon}}, \bibinfo
  {author} {\bibfnamefont {X.~J.}\ \bibnamefont {Zhou}}, \bibinfo {author}
  {\bibfnamefont {N.~B.}\ \bibnamefont {Brookes}}, \bibinfo {author}
  {\bibfnamefont {L.}~\bibnamefont {Braicovich}}, \ and\ \bibinfo {author}
  {\bibfnamefont {G.}~\bibnamefont {Ghiringhelli}},\ }\bibfield  {title}
  {\enquote {\bibinfo {title} {{Direct Observation of Charge Order in
  Underdoped and Optimally Doped {Bi$_2$(Sr,La)$_2$Cuo$_{6+\Delta}$} by
  Resonant Inelastic X-Ray Scattering}},}\ }\href@noop {} {\bibfield  {journal}
  {\bibinfo  {journal} {Phys. Rev. B}\ }\textbf {\bibinfo {volume} {94}},\
  \bibinfo {pages} {184511} (\bibinfo {year} {2016})}\BibitemShut {NoStop}%
\bibitem [{\citenamefont {Forgan}\ \emph {et~al.}(2015)\citenamefont {Forgan},
  \citenamefont {Blackburn}, \citenamefont {Holmes}, \citenamefont {Briffa},
  \citenamefont {Chang}, \citenamefont {Bouchenoire}, \citenamefont {Brown},
  \citenamefont {Liang}, \citenamefont {Bonn}, \citenamefont {Hardy},
  \citenamefont {Christensen}, \citenamefont {Zimmermann}, \citenamefont
  {H{\"u}cker},\ and\ \citenamefont {Hayden}}]{Forgan:2015ux}%
  \BibitemOpen
  \bibfield  {author} {\bibinfo {author} {\bibfnamefont {E.~M.}\ \bibnamefont
  {Forgan}}, \bibinfo {author} {\bibfnamefont {E.}~\bibnamefont {Blackburn}},
  \bibinfo {author} {\bibfnamefont {A.~T.}\ \bibnamefont {Holmes}}, \bibinfo
  {author} {\bibfnamefont {A.~K.~R.}\ \bibnamefont {Briffa}}, \bibinfo {author}
  {\bibfnamefont {J.}~\bibnamefont {Chang}}, \bibinfo {author} {\bibfnamefont
  {L.}~\bibnamefont {Bouchenoire}}, \bibinfo {author} {\bibfnamefont {S.~D.}\
  \bibnamefont {Brown}}, \bibinfo {author} {\bibfnamefont {Ruixing}\
  \bibnamefont {Liang}}, \bibinfo {author} {\bibfnamefont {D.}~\bibnamefont
  {Bonn}}, \bibinfo {author} {\bibfnamefont {W.~N.}\ \bibnamefont {Hardy}},
  \bibinfo {author} {\bibfnamefont {N.~B.}\ \bibnamefont {Christensen}},
  \bibinfo {author} {\bibfnamefont {M.~V.}\ \bibnamefont {Zimmermann}},
  \bibinfo {author} {\bibfnamefont {M.}~\bibnamefont {H{\"u}cker}}, \ and\
  \bibinfo {author} {\bibfnamefont {S.~M.}\ \bibnamefont {Hayden}},\ }\bibfield
   {title} {\enquote {\bibinfo {title} {{The Microscopic Structure of Charge
  Density Waves in Underdoped YBa$_2$Cu$_3$O$_{6.54}$ Revealed by X-Ray
  Diffraction}},}\ }\href@noop {} {\bibfield  {journal} {\bibinfo  {journal}
  {Nat. Commun.}\ }\textbf {\bibinfo {volume} {6}},\ \bibinfo {pages} {10064}
  (\bibinfo {year} {2015})}\BibitemShut {NoStop}%
\bibitem [{\citenamefont {Miao}\ \emph {et~al.}(2017)\citenamefont {Miao},
  \citenamefont {Lorenzana}, \citenamefont {Seibold}, \citenamefont {Peng},
  \citenamefont {Amorese}, \citenamefont {Yakhou-Harris}, \citenamefont
  {Kummer}, \citenamefont {Brookes}, \citenamefont {Konik}, \citenamefont
  {Thampy}, \citenamefont {Gu}, \citenamefont {Ghiringhelli}, \citenamefont
  {Braicovich},\ and\ \citenamefont {Dean}}]{Miao:2017bv}%
  \BibitemOpen
  \bibfield  {author} {\bibinfo {author} {\bibfnamefont {H.}~\bibnamefont
  {Miao}}, \bibinfo {author} {\bibfnamefont {J.}~\bibnamefont {Lorenzana}},
  \bibinfo {author} {\bibfnamefont {G.}~\bibnamefont {Seibold}}, \bibinfo
  {author} {\bibfnamefont {Y.~Y.}\ \bibnamefont {Peng}}, \bibinfo {author}
  {\bibfnamefont {A.}~\bibnamefont {Amorese}}, \bibinfo {author} {\bibfnamefont
  {F.}~\bibnamefont {Yakhou-Harris}}, \bibinfo {author} {\bibfnamefont
  {K.}~\bibnamefont {Kummer}}, \bibinfo {author} {\bibfnamefont {N.~B.}\
  \bibnamefont {Brookes}}, \bibinfo {author} {\bibfnamefont {R.~M.}\
  \bibnamefont {Konik}}, \bibinfo {author} {\bibfnamefont {V.}~\bibnamefont
  {Thampy}}, \bibinfo {author} {\bibfnamefont {G.~D.}\ \bibnamefont {Gu}},
  \bibinfo {author} {\bibfnamefont {G.}~\bibnamefont {Ghiringhelli}}, \bibinfo
  {author} {\bibfnamefont {L.}~\bibnamefont {Braicovich}}, \ and\ \bibinfo
  {author} {\bibfnamefont {M.~P.~M.}\ \bibnamefont {Dean}},\ }\bibfield
  {title} {\enquote {\bibinfo {title} {{High-temperature charge density wave
  correlations in La$_{1.875}$Ba$_{0.125}$CuO$_4$ without spin--charge
  locking}},}\ }\href@noop {} {\bibfield  {journal} {\bibinfo  {journal} {Proc.
  Nat. Acad. Sci.}\ }\textbf {\bibinfo {volume} {114}},\ \bibinfo {pages}
  {12430--12435} (\bibinfo {year} {2017})}\BibitemShut {NoStop}%
\bibitem [{\citenamefont {Vojta}(2009)}]{Vojta:2009}%
  \BibitemOpen
  \bibfield  {author} {\bibinfo {author} {\bibfnamefont {Matthias}\
  \bibnamefont {Vojta}},\ }\bibfield  {title} {\enquote {\bibinfo {title}
  {Lattice symmetry breaking in cuprate superconductors: stripes, nematics, and
  superconductivity},}\ }\href {\doibase 10.1080/00018730903122242} {\bibfield
  {journal} {\bibinfo  {journal} {Advances in Physics}\ }\textbf {\bibinfo
  {volume} {58}},\ \bibinfo {pages} {699--820} (\bibinfo {year}
  {2009})}\BibitemShut {NoStop}%
\bibitem [{\citenamefont {Wu}\ \emph {et~al.}(2011)\citenamefont {Wu},
  \citenamefont {Mayaffre}, \citenamefont {Kr{\"a}mer}, \citenamefont
  {Horvati{\'c}}, \citenamefont {Berthier}, \citenamefont {Hardy},
  \citenamefont {Liang}, \citenamefont {Bonn},\ and\ \citenamefont
  {Julien}}]{Wu:2011ke}%
  \BibitemOpen
  \bibfield  {author} {\bibinfo {author} {\bibfnamefont {Tao}\ \bibnamefont
  {Wu}}, \bibinfo {author} {\bibfnamefont {Hadrien}\ \bibnamefont {Mayaffre}},
  \bibinfo {author} {\bibfnamefont {Steffen}\ \bibnamefont {Kr{\"a}mer}},
  \bibinfo {author} {\bibfnamefont {Mladen}\ \bibnamefont {Horvati{\'c}}},
  \bibinfo {author} {\bibfnamefont {Claude}\ \bibnamefont {Berthier}}, \bibinfo
  {author} {\bibfnamefont {W~N}\ \bibnamefont {Hardy}}, \bibinfo {author}
  {\bibfnamefont {Ruixing}\ \bibnamefont {Liang}}, \bibinfo {author}
  {\bibfnamefont {D~A}\ \bibnamefont {Bonn}}, \ and\ \bibinfo {author}
  {\bibfnamefont {Marc-Henri}\ \bibnamefont {Julien}},\ }\bibfield  {title}
  {\enquote {\bibinfo {title} {{Magnetic-Field-Induced Charge-Stripe Order in
  the High-Temperature Superconductor YBa$_2$Cu$_3$O$_y$}},}\ }\href@noop {}
  {\bibfield  {journal} {\bibinfo  {journal} {Nature}\ }\textbf {\bibinfo
  {volume} {477}},\ \bibinfo {pages} {191--194} (\bibinfo {year}
  {2011})}\BibitemShut {NoStop}%
\bibitem [{\citenamefont {Wu}\ \emph {et~al.}(2013)\citenamefont {Wu},
  \citenamefont {Mayaffre}, \citenamefont {Kr\"{a}mer}, \citenamefont
  {Horvati\'{c}}, \citenamefont {Berthier}, \citenamefont {Kuhns},
  \citenamefont {Reyes}, \citenamefont {Liang}, \citenamefont {Hardy},
  \citenamefont {Bonn},\ and\ \citenamefont {Julien}}]{Wu:2013}%
  \BibitemOpen
  \bibfield  {author} {\bibinfo {author} {\bibfnamefont {Tao}\ \bibnamefont
  {Wu}}, \bibinfo {author} {\bibfnamefont {Hadrien}\ \bibnamefont {Mayaffre}},
  \bibinfo {author} {\bibfnamefont {Steffen}\ \bibnamefont {Kr\"{a}mer}},
  \bibinfo {author} {\bibfnamefont {Mladen}\ \bibnamefont {Horvati\'{c}}},
  \bibinfo {author} {\bibfnamefont {Claude}\ \bibnamefont {Berthier}}, \bibinfo
  {author} {\bibfnamefont {Philip~L.}\ \bibnamefont {Kuhns}}, \bibinfo {author}
  {\bibfnamefont {Arneil~P.}\ \bibnamefont {Reyes}}, \bibinfo {author}
  {\bibfnamefont {Ruixing}\ \bibnamefont {Liang}}, \bibinfo {author}
  {\bibfnamefont {W.~N.}\ \bibnamefont {Hardy}}, \bibinfo {author}
  {\bibfnamefont {D.~A.}\ \bibnamefont {Bonn}}, \ and\ \bibinfo {author}
  {\bibfnamefont {Marc-Henri}\ \bibnamefont {Julien}},\ }\bibfield  {title}
  {\enquote {\bibinfo {title} {{Emergence of Charge Order from the Vortex State
  of a High-Temperature Superconductor}},}\ }\href {\doibase
  10.1038/ncomms3113} {\bibfield  {journal} {\bibinfo  {journal} {Nat.
  Commun.}\ }\textbf {\bibinfo {volume} {4}},\ \bibinfo {pages} {2113}
  (\bibinfo {year} {2013})}\BibitemShut {NoStop}%
\bibitem [{\citenamefont {Wu}\ \emph {et~al.}(2015)\citenamefont {Wu},
  \citenamefont {Mayaffre}, \citenamefont {Kr{\"a}mer}, \citenamefont
  {Horvati{\'c}}, \citenamefont {Berthier}, \citenamefont {Hardy},
  \citenamefont {Liang}, \citenamefont {Bonn},\ and\ \citenamefont
  {Julien}}]{Wu:2015bt}%
  \BibitemOpen
  \bibfield  {author} {\bibinfo {author} {\bibfnamefont {Tao}\ \bibnamefont
  {Wu}}, \bibinfo {author} {\bibfnamefont {Hadrien}\ \bibnamefont {Mayaffre}},
  \bibinfo {author} {\bibfnamefont {Steffen}\ \bibnamefont {Kr{\"a}mer}},
  \bibinfo {author} {\bibfnamefont {Mladen}\ \bibnamefont {Horvati{\'c}}},
  \bibinfo {author} {\bibfnamefont {Claude}\ \bibnamefont {Berthier}}, \bibinfo
  {author} {\bibfnamefont {W~N}\ \bibnamefont {Hardy}}, \bibinfo {author}
  {\bibfnamefont {Ruixing}\ \bibnamefont {Liang}}, \bibinfo {author}
  {\bibfnamefont {D~A}\ \bibnamefont {Bonn}}, \ and\ \bibinfo {author}
  {\bibfnamefont {Marc-Henri}\ \bibnamefont {Julien}},\ }\bibfield  {title}
  {\enquote {\bibinfo {title} {{Incipient Charge Order Observed by NMR in the
  Normal State of YBa$_2$Cu$_3$O$_{y}$}},}\ }\href@noop {} {\bibfield
  {journal} {\bibinfo  {journal} {Nat. Commun.}\ }\textbf {\bibinfo {volume}
  {6}},\ \bibinfo {pages} {6438} (\bibinfo {year} {2015})}\BibitemShut
  {NoStop}%
\bibitem [{\citenamefont {Zhou}\ \emph {et~al.}(2017)\citenamefont {Zhou},
  \citenamefont {Hirata}, \citenamefont {Wu}, \citenamefont {Vinograd},
  \citenamefont {Mayaffre}, \citenamefont {Kr\"amer}, \citenamefont
  {Horvati\'c}, \citenamefont {Berthier}, \citenamefont {Reyes}, \citenamefont
  {Kuhns}, \citenamefont {Liang}, \citenamefont {Hardy}, \citenamefont {Bonn},\
  and\ \citenamefont {Julien}}]{Zhou:2016}%
  \BibitemOpen
  \bibfield  {author} {\bibinfo {author} {\bibfnamefont {R.}~\bibnamefont
  {Zhou}}, \bibinfo {author} {\bibfnamefont {M.}~\bibnamefont {Hirata}},
  \bibinfo {author} {\bibfnamefont {T.}~\bibnamefont {Wu}}, \bibinfo {author}
  {\bibfnamefont {I.}~\bibnamefont {Vinograd}}, \bibinfo {author}
  {\bibfnamefont {H.}~\bibnamefont {Mayaffre}}, \bibinfo {author}
  {\bibfnamefont {S.}~\bibnamefont {Kr\"amer}}, \bibinfo {author}
  {\bibfnamefont {M.}~\bibnamefont {Horvati\'c}}, \bibinfo {author}
  {\bibfnamefont {C.}~\bibnamefont {Berthier}}, \bibinfo {author}
  {\bibfnamefont {A.~P.}\ \bibnamefont {Reyes}}, \bibinfo {author}
  {\bibfnamefont {P.~L.}\ \bibnamefont {Kuhns}}, \bibinfo {author}
  {\bibfnamefont {R.}~\bibnamefont {Liang}}, \bibinfo {author} {\bibfnamefont
  {W.~N.}\ \bibnamefont {Hardy}}, \bibinfo {author} {\bibfnamefont {D.~A.}\
  \bibnamefont {Bonn}}, \ and\ \bibinfo {author} {\bibfnamefont {M.-H.}\
  \bibnamefont {Julien}},\ }\bibfield  {title} {\enquote {\bibinfo {title}
  {{Observation of Electronic Bound States in Charge-Ordered
  {YBa$_2$Cu$_3$O$_y$}}},}\ }\href@noop {} {\bibfield  {journal} {\bibinfo
  {journal} {Phys. Rev. Lett.}\ }\textbf {\bibinfo {volume} {118}},\ \bibinfo
  {pages} {017001} (\bibinfo {year} {2017})}\BibitemShut {NoStop}%
\bibitem [{\citenamefont {Kawasaki}\ \emph {et~al.}(2017)\citenamefont
  {Kawasaki}, \citenamefont {Li}, \citenamefont {Kitahashi}, \citenamefont
  {Lin}, \citenamefont {Kuhns}, \citenamefont {Reyes},\ and\ \citenamefont
  {Zheng}}]{Kawasaki:2017ue}%
  \BibitemOpen
  \bibfield  {author} {\bibinfo {author} {\bibfnamefont {S.}~\bibnamefont
  {Kawasaki}}, \bibinfo {author} {\bibfnamefont {Z.}~\bibnamefont {Li}},
  \bibinfo {author} {\bibfnamefont {M.}~\bibnamefont {Kitahashi}}, \bibinfo
  {author} {\bibfnamefont {C.~T.}\ \bibnamefont {Lin}}, \bibinfo {author}
  {\bibfnamefont {P.~L.}\ \bibnamefont {Kuhns}}, \bibinfo {author}
  {\bibfnamefont {A.~P.}\ \bibnamefont {Reyes}}, \ and\ \bibinfo {author}
  {\bibfnamefont {Guo-Qing}\ \bibnamefont {Zheng}},\ }\href@noop {} {\enquote
  {\bibinfo {title} {{Charge-Density-Wave Order Takes Over Antiferromagnetism
  in Bi$_2$Sr$_{2-x}$La$_x$CuO$_{6}$ Superconductors}},}\ } (\bibinfo {year}
  {2017}),\ \Eprint {http://arxiv.org/abs/https://arxiv.org/abs/1704.06169}
  {https://arxiv.org/abs/1704.06169} \BibitemShut {NoStop}%
\bibitem [{\citenamefont {Butaud}\ \emph {et~al.}(1985)\citenamefont {Butaud},
  \citenamefont {S{\'e}gransan}, \citenamefont {Berthier}, \citenamefont
  {Dumas},\ and\ \citenamefont {Schlenker}}]{Butaud:1985eh}%
  \BibitemOpen
  \bibfield  {author} {\bibinfo {author} {\bibfnamefont {P.}~\bibnamefont
  {Butaud}}, \bibinfo {author} {\bibfnamefont {P.}~\bibnamefont
  {S{\'e}gransan}}, \bibinfo {author} {\bibfnamefont {C.}~\bibnamefont
  {Berthier}}, \bibinfo {author} {\bibfnamefont {J.}~\bibnamefont {Dumas}}, \
  and\ \bibinfo {author} {\bibfnamefont {C.}~\bibnamefont {Schlenker}},\
  }\bibfield  {title} {\enquote {\bibinfo {title} {{NMR Study of the
  Charge-Density Wave in Rb$_{0.30}$MoO$_3$ Single Crystal}},}\ }\href@noop {}
  {\bibfield  {journal} {\bibinfo  {journal} {Phys. Rev. Lett.}\ }\textbf
  {\bibinfo {volume} {55}},\ \bibinfo {pages} {253--256} (\bibinfo {year}
  {1985})}\BibitemShut {NoStop}%
\bibitem [{\citenamefont {Ross}\ \emph {et~al.}(1990)\citenamefont {Ross},
  \citenamefont {Wang},\ and\ \citenamefont {Slichter}}]{Ross:1990in}%
  \BibitemOpen
  \bibfield  {author} {\bibinfo {author} {\bibfnamefont {Joseph~H.}\
  \bibnamefont {Ross}}, \bibinfo {author} {\bibfnamefont {Zhiyue}\ \bibnamefont
  {Wang}}, \ and\ \bibinfo {author} {\bibfnamefont {Charles~P.}\ \bibnamefont
  {Slichter}},\ }\bibfield  {title} {\enquote {\bibinfo {title} {{NMR Studies
  of NbSe$_3$: Electronic Structures, Static Charge-Density-Wave Measurements,
  and Observations of the Moving Charge-Density Wave}},}\ }\href@noop {}
  {\bibfield  {journal} {\bibinfo  {journal} {Phys. Rev. B}\ }\textbf {\bibinfo
  {volume} {41}},\ \bibinfo {pages} {2722--2734} (\bibinfo {year}
  {1990})}\BibitemShut {NoStop}%
\bibitem [{\citenamefont {Skripov}\ \emph {et~al.}(1995)\citenamefont
  {Skripov}, \citenamefont {Sibirtsev}, \citenamefont {Cherepanov},\ and\
  \citenamefont {Aleksashin}}]{Skripov:1995ts}%
  \BibitemOpen
  \bibfield  {author} {\bibinfo {author} {\bibfnamefont {A.~V.}\ \bibnamefont
  {Skripov}}, \bibinfo {author} {\bibfnamefont {D.~S.}\ \bibnamefont
  {Sibirtsev}}, \bibinfo {author} {\bibfnamefont {Yu~G.}\ \bibnamefont
  {Cherepanov}}, \ and\ \bibinfo {author} {\bibfnamefont {B.~A.}\ \bibnamefont
  {Aleksashin}},\ }\bibfield  {title} {\enquote {\bibinfo {title} {{$^{77}$Se
  NMR Study of the Charge Density Wave State in 2H-NbSe$_{2 }$and 1T-VSe$_{2
  }$}},}\ }\href@noop {} {\bibfield  {journal} {\bibinfo  {journal} {J. Phys.:
  Cond. Mat.}\ }\textbf {\bibinfo {volume} {7}},\ \bibinfo {pages} {4479--4487}
  (\bibinfo {year} {1995})}\BibitemShut {NoStop}%
\bibitem [{\citenamefont {Berthier}\ \emph {et~al.}(2001)\citenamefont
  {Berthier}, \citenamefont {Jerome},\ and\ \citenamefont
  {Molinie}}]{Berthier:2001ga}%
  \BibitemOpen
  \bibfield  {author} {\bibinfo {author} {\bibfnamefont {C.}~\bibnamefont
  {Berthier}}, \bibinfo {author} {\bibfnamefont {D.}~\bibnamefont {Jerome}}, \
  and\ \bibinfo {author} {\bibfnamefont {P.}~\bibnamefont {Molinie}},\
  }\bibfield  {title} {\enquote {\bibinfo {title} {{NMR Study on a 2H-NbSe$_2$
  Single Crystal: A Microscopic Investigation of the Charge Density Waves
  State}},}\ }\href@noop {} {\bibfield  {journal} {\bibinfo  {journal} {J.
  Phys. C: Sol. State Phys.}\ }\textbf {\bibinfo {volume} {11}},\ \bibinfo
  {pages} {797--814} (\bibinfo {year} {2001})}\BibitemShut {NoStop}%
\bibitem [{\citenamefont {Ghoshray}\ \emph {et~al.}(2009)\citenamefont
  {Ghoshray}, \citenamefont {Pahari}, \citenamefont {Ghoshray}, \citenamefont
  {Eremenko}, \citenamefont {Sirenko},\ and\ \citenamefont
  {Suits}}]{Ghoshray:2009dq}%
  \BibitemOpen
  \bibfield  {author} {\bibinfo {author} {\bibfnamefont {K.}~\bibnamefont
  {Ghoshray}}, \bibinfo {author} {\bibfnamefont {B.}~\bibnamefont {Pahari}},
  \bibinfo {author} {\bibfnamefont {A.}~\bibnamefont {Ghoshray}}, \bibinfo
  {author} {\bibfnamefont {V.~V.}\ \bibnamefont {Eremenko}}, \bibinfo {author}
  {\bibfnamefont {V.~A.}\ \bibnamefont {Sirenko}}, \ and\ \bibinfo {author}
  {\bibfnamefont {B.~H.}\ \bibnamefont {Suits}},\ }\bibfield  {title} {\enquote
  {\bibinfo {title} {{$^{93}$Nb NMR Study of the Charge Density Wave State in
  NbSe$_2$ }},}\ }\href@noop {} {\bibfield  {journal} {\bibinfo  {journal} {J.
  Phys: Cond. Mat.}\ }\textbf {\bibinfo {volume} {21}},\ \bibinfo {pages}
  {155701} (\bibinfo {year} {2009})}\BibitemShut {NoStop}%
\bibitem [{\citenamefont {Kharkov}\ and\ \citenamefont
  {Sushkov}(2016)}]{Kharkov:2016tf}%
  \BibitemOpen
  \bibfield  {author} {\bibinfo {author} {\bibfnamefont {Yaroslav~A.}\
  \bibnamefont {Kharkov}}\ and\ \bibinfo {author} {\bibfnamefont {Oleg~P.}\
  \bibnamefont {Sushkov}},\ }\bibfield  {title} {\enquote {\bibinfo {title}
  {{The Amplitudes and the Structure of the Charge Density Wave in YBCO}},}\
  }\href@noop {} {\bibfield  {journal} {\bibinfo  {journal} {Scientific
  Reports}\ }\textbf {\bibinfo {volume} {6}},\ \bibinfo {pages} {34551}
  (\bibinfo {year} {2016})}\BibitemShut {NoStop}%
\bibitem [{\citenamefont {Haase}\ \emph {et~al.}(2000)\citenamefont {Haase},
  \citenamefont {Slichter}, \citenamefont {Stern}, \citenamefont {Milling},\
  and\ \citenamefont {Hinks}}]{Haase:2000vb}%
  \BibitemOpen
  \bibfield  {author} {\bibinfo {author} {\bibfnamefont {J.}~\bibnamefont
  {Haase}}, \bibinfo {author} {\bibfnamefont {C.~P.}\ \bibnamefont {Slichter}},
  \bibinfo {author} {\bibfnamefont {R.}~\bibnamefont {Stern}}, \bibinfo
  {author} {\bibfnamefont {C.~T.}\ \bibnamefont {Milling}}, \ and\ \bibinfo
  {author} {\bibfnamefont {D.~G.}\ \bibnamefont {Hinks}},\ }\bibfield  {title}
  {\enquote {\bibinfo {title} {{NMR Evidence for Spatial Modulations in the
  Cuprates}},}\ }\href@noop {} {\bibfield  {journal} {\bibinfo  {journal} {J.
  Supercond.}\ }\textbf {\bibinfo {volume} {13}},\ \bibinfo {pages} {723}
  (\bibinfo {year} {2000})}\BibitemShut {NoStop}%
\bibitem [{\citenamefont {Campi}\ \emph {et~al.}(2015)\citenamefont {Campi},
  \citenamefont {Bianconi}, \citenamefont {Poccia}, \citenamefont {Bianconi},
  \citenamefont {Barba}, \citenamefont {Arrighetti}, \citenamefont {Innocenti},
  \citenamefont {Karpinski}, \citenamefont {Zhigadlo}, \citenamefont {Kazakov},
  \citenamefont {Burghammer}, \citenamefont {Zimmermann}, \citenamefont
  {Sprung},\ and\ \citenamefont {Ricci}}]{Campi:2015cva}%
  \BibitemOpen
  \bibfield  {author} {\bibinfo {author} {\bibfnamefont {G.}~\bibnamefont
  {Campi}}, \bibinfo {author} {\bibfnamefont {A.}~\bibnamefont {Bianconi}},
  \bibinfo {author} {\bibfnamefont {N.}~\bibnamefont {Poccia}}, \bibinfo
  {author} {\bibfnamefont {G.}~\bibnamefont {Bianconi}}, \bibinfo {author}
  {\bibfnamefont {L.}~\bibnamefont {Barba}}, \bibinfo {author} {\bibfnamefont
  {G.}~\bibnamefont {Arrighetti}}, \bibinfo {author} {\bibfnamefont
  {D.}~\bibnamefont {Innocenti}}, \bibinfo {author} {\bibfnamefont
  {J.}~\bibnamefont {Karpinski}}, \bibinfo {author} {\bibfnamefont {N.~D.}\
  \bibnamefont {Zhigadlo}}, \bibinfo {author} {\bibfnamefont {S.~M.}\
  \bibnamefont {Kazakov}}, \bibinfo {author} {\bibfnamefont {M.}~\bibnamefont
  {Burghammer}}, \bibinfo {author} {\bibfnamefont {M.~V.}\ \bibnamefont
  {Zimmermann}}, \bibinfo {author} {\bibfnamefont {M.}~\bibnamefont {Sprung}},
  \ and\ \bibinfo {author} {\bibfnamefont {A.}~\bibnamefont {Ricci}},\
  }\bibfield  {title} {\enquote {\bibinfo {title} {{Inhomogeneity of
  Charge-Density-Wave Order and Quenched Disorder in a High-$T_c$
  Superconductor}},}\ }\href@noop {} {\bibfield  {journal} {\bibinfo  {journal}
  {Nature}\ }\textbf {\bibinfo {volume} {525}},\ \bibinfo {pages} {359--362}
  (\bibinfo {year} {2015})}\BibitemShut {NoStop}%
\bibitem [{\citenamefont {Caplan}\ \emph {et~al.}(2015)\citenamefont {Caplan},
  \citenamefont {Wachtel},\ and\ \citenamefont {Orgad}}]{Caplan:2015jm}%
  \BibitemOpen
  \bibfield  {author} {\bibinfo {author} {\bibfnamefont {Yosef}\ \bibnamefont
  {Caplan}}, \bibinfo {author} {\bibfnamefont {Gideon}\ \bibnamefont
  {Wachtel}}, \ and\ \bibinfo {author} {\bibfnamefont {Dror}\ \bibnamefont
  {Orgad}},\ }\bibfield  {title} {\enquote {\bibinfo {title} {{Long-Range Order
  and Pinning of Charge-Density Waves in Competition with
  Superconductivity}},}\ }\href@noop {} {\bibfield  {journal} {\bibinfo
  {journal} {Phys. Rev. B}\ }\textbf {\bibinfo {volume} {92}},\ \bibinfo
  {pages} {224504} (\bibinfo {year} {2015})}\BibitemShut {NoStop}%
\bibitem [{\citenamefont {Metlitski}\ and\ \citenamefont
  {Sachdev}(2010)}]{Metlitski:2010vf}%
  \BibitemOpen
  \bibfield  {author} {\bibinfo {author} {\bibfnamefont {Max}\ \bibnamefont
  {Metlitski}}\ and\ \bibinfo {author} {\bibfnamefont {Subir}\ \bibnamefont
  {Sachdev}},\ }\bibfield  {title} {\enquote {\bibinfo {title} {{Instabilities
  Near the Onset of Spin Density Wave Order in Metals}},}\ }\href@noop {}
  {\bibfield  {journal} {\bibinfo  {journal} {New J. Phys.}\ }\textbf {\bibinfo
  {volume} {12}},\ \bibinfo {pages} {105007} (\bibinfo {year}
  {2010})}\BibitemShut {NoStop}%
\bibitem [{\citenamefont {Holder}\ and\ \citenamefont
  {Metzner}(2012)}]{Holder:2012ks}%
  \BibitemOpen
  \bibfield  {author} {\bibinfo {author} {\bibfnamefont {Tobias}\ \bibnamefont
  {Holder}}\ and\ \bibinfo {author} {\bibfnamefont {Walter}\ \bibnamefont
  {Metzner}},\ }\bibfield  {title} {\enquote {\bibinfo {title} {{Incommensurate
  Nematic Fluctuations in Two-Dimensional Metals}},}\ }\href@noop {} {\bibfield
   {journal} {\bibinfo  {journal} {Phys. Rev. B}\ }\textbf {\bibinfo {volume}
  {85}},\ \bibinfo {pages} {165130} (\bibinfo {year} {2012})}\BibitemShut
  {NoStop}%
\bibitem [{\citenamefont {Sachdev}\ and\ \citenamefont
  {La~Placa}(2013)}]{Sachdev:2013bo}%
  \BibitemOpen
  \bibfield  {author} {\bibinfo {author} {\bibfnamefont {Subir}\ \bibnamefont
  {Sachdev}}\ and\ \bibinfo {author} {\bibfnamefont {Rolando}\ \bibnamefont
  {La~Placa}},\ }\bibfield  {title} {\enquote {\bibinfo {title} {{Bond Order in
  Two-Dimensional Metals with Antiferromagnetic Exchange Interactions}},}\
  }\href@noop {} {\bibfield  {journal} {\bibinfo  {journal} {Phys. Rev. Lett.}\
  }\textbf {\bibinfo {volume} {111}},\ \bibinfo {pages} {027202} (\bibinfo
  {year} {2013})}\BibitemShut {NoStop}%
\bibitem [{\citenamefont {Sau}\ and\ \citenamefont
  {Sachdev}(2014)}]{Sau:2013vw}%
  \BibitemOpen
  \bibfield  {author} {\bibinfo {author} {\bibfnamefont {Jay~D.}\ \bibnamefont
  {Sau}}\ and\ \bibinfo {author} {\bibfnamefont {Subir}\ \bibnamefont
  {Sachdev}},\ }\bibfield  {title} {\enquote {\bibinfo {title} {Mean-field
  theory of competing orders in metals with antiferromagnetic exchange
  interactions},}\ }\href {\doibase 10.1103/PhysRevB.89.075129} {\bibfield
  {journal} {\bibinfo  {journal} {Phys. Rev. B}\ }\textbf {\bibinfo {volume}
  {89}},\ \bibinfo {pages} {075129} (\bibinfo {year} {2014})}\BibitemShut
  {NoStop}%
\bibitem [{\citenamefont {Bulut}\ \emph {et~al.}(2013)\citenamefont {Bulut},
  \citenamefont {Atkinson},\ and\ \citenamefont {Kampf}}]{Bulut:2013bz}%
  \BibitemOpen
  \bibfield  {author} {\bibinfo {author} {\bibfnamefont {S.}~\bibnamefont
  {Bulut}}, \bibinfo {author} {\bibfnamefont {W.~A.}\ \bibnamefont {Atkinson}},
  \ and\ \bibinfo {author} {\bibfnamefont {A.~P.}\ \bibnamefont {Kampf}},\
  }\bibfield  {title} {\enquote {\bibinfo {title} {{Spatially Modulated
  Electronic Nematicity in the Three-Band Model of Cuprate Superconductors}},}\
  }\href {\doibase 10.1103/PhysRevB.88.155132} {\bibfield  {journal} {\bibinfo
  {journal} {Phys. Rev. B}\ }\textbf {\bibinfo {volume} {88}},\ \bibinfo
  {pages} {155132} (\bibinfo {year} {2013})}\BibitemShut {NoStop}%
\bibitem [{\citenamefont {P\'epin}\ \emph {et~al.}(2014)\citenamefont
  {P\'epin}, \citenamefont {de~Carvalho}, \citenamefont {Kloss},\ and\
  \citenamefont {Montiel}}]{Pepin:2014tb}%
  \BibitemOpen
  \bibfield  {author} {\bibinfo {author} {\bibfnamefont {C.}~\bibnamefont
  {P\'epin}}, \bibinfo {author} {\bibfnamefont {V.~S.}\ \bibnamefont
  {de~Carvalho}}, \bibinfo {author} {\bibfnamefont {T.}~\bibnamefont {Kloss}},
  \ and\ \bibinfo {author} {\bibfnamefont {X.}~\bibnamefont {Montiel}},\
  }\bibfield  {title} {\enquote {\bibinfo {title} {Pseudogap, charge order, and
  pairing density wave at the hot spots in cuprate superconductors},}\ }\href
  {\doibase 10.1103/PhysRevB.90.195207} {\bibfield  {journal} {\bibinfo
  {journal} {Phys. Rev. B}\ }\textbf {\bibinfo {volume} {90}},\ \bibinfo
  {pages} {195207} (\bibinfo {year} {2014})}\BibitemShut {NoStop}%
\bibitem [{\citenamefont {Yamakawa}\ and\ \citenamefont
  {Kontani}(2015)}]{Yamakawa:2015hb}%
  \BibitemOpen
  \bibfield  {author} {\bibinfo {author} {\bibfnamefont {Youichi}\ \bibnamefont
  {Yamakawa}}\ and\ \bibinfo {author} {\bibfnamefont {Hiroshi}\ \bibnamefont
  {Kontani}},\ }\bibfield  {title} {\enquote {\bibinfo {title}
  {{Spin-Fluctuation-Driven Nematic Charge-Density Wave in Cuprate
  Superconductors: Impact of Aslamazov-Larkin Vertex Corrections}},}\
  }\href@noop {} {\bibfield  {journal} {\bibinfo  {journal} {Phys. Rev. Lett.}\
  }\textbf {\bibinfo {volume} {114}},\ \bibinfo {pages} {257001} (\bibinfo
  {year} {2015})}\BibitemShut {NoStop}%
\bibitem [{\citenamefont {Tsuchiizu}\ \emph {et~al.}(2016)\citenamefont
  {Tsuchiizu}, \citenamefont {Yamakawa},\ and\ \citenamefont
  {Kontani}}]{Tsuchiizu:2015vz}%
  \BibitemOpen
  \bibfield  {author} {\bibinfo {author} {\bibfnamefont {Masahisa}\
  \bibnamefont {Tsuchiizu}}, \bibinfo {author} {\bibfnamefont {Youichi}\
  \bibnamefont {Yamakawa}}, \ and\ \bibinfo {author} {\bibfnamefont {Hiroshi}\
  \bibnamefont {Kontani}},\ }\bibfield  {title} {\enquote {\bibinfo {title}
  {{$p$-Orbital Density Wave with $d$ Symmetry in High-$T_c$ Cuprate
  Superconductors}},}\ }\href@noop {} {\bibfield  {journal} {\bibinfo
  {journal} {Phys. Rev. B}\ }\textbf {\bibinfo {volume} {93}},\ \bibinfo
  {pages} {155148} (\bibinfo {year} {2016})}\BibitemShut {NoStop}%
\bibitem [{\citenamefont {Wang}\ \emph {et~al.}(2017)\citenamefont {Wang},
  \citenamefont {Wang}, \citenamefont {Schattner}, \citenamefont {Berg},\ and\
  \citenamefont {Fernandes}}]{Wang:2017wk}%
  \BibitemOpen
  \bibfield  {author} {\bibinfo {author} {\bibfnamefont {Xiaoyu}\ \bibnamefont
  {Wang}}, \bibinfo {author} {\bibfnamefont {Yuxuan}\ \bibnamefont {Wang}},
  \bibinfo {author} {\bibfnamefont {Yoni}\ \bibnamefont {Schattner}}, \bibinfo
  {author} {\bibfnamefont {Erez}\ \bibnamefont {Berg}}, \ and\ \bibinfo
  {author} {\bibfnamefont {Rafael~M}\ \bibnamefont {Fernandes}},\ }\href@noop
  {} {\enquote {\bibinfo {title} {{Is Charge Order Induced Near an
  Antiferromagnetic Quantum Critical Point?}}}\ } (\bibinfo {year} {2017}),\
  \Eprint {http://arxiv.org/abs/https://arxiv.org/abs/1710.02158}
  {https://arxiv.org/abs/1710.02158} \BibitemShut {NoStop}%
\bibitem [{\citenamefont {Allais}\ \emph {et~al.}(2014)\citenamefont {Allais},
  \citenamefont {Bauer},\ and\ \citenamefont {Sachdev}}]{Allais:2014kg}%
  \BibitemOpen
  \bibfield  {author} {\bibinfo {author} {\bibfnamefont {Andrea}\ \bibnamefont
  {Allais}}, \bibinfo {author} {\bibfnamefont {Johannes}\ \bibnamefont
  {Bauer}}, \ and\ \bibinfo {author} {\bibfnamefont {Subir}\ \bibnamefont
  {Sachdev}},\ }\bibfield  {title} {\enquote {\bibinfo {title} {{Density wave
  instabilities in a correlated two-dimensional metal}},}\ }\href@noop {}
  {\bibfield  {journal} {\bibinfo  {journal} {Phys. Rev. B}\ }\textbf {\bibinfo
  {volume} {90}},\ \bibinfo {pages} {155114} (\bibinfo {year}
  {2014})}\BibitemShut {NoStop}%
\bibitem [{\citenamefont {Chowdhury}\ and\ \citenamefont
  {Sachdev}(2014)}]{Chowdhury:2014}%
  \BibitemOpen
  \bibfield  {author} {\bibinfo {author} {\bibfnamefont {Debanjan}\
  \bibnamefont {Chowdhury}}\ and\ \bibinfo {author} {\bibfnamefont {Subir}\
  \bibnamefont {Sachdev}},\ }\bibfield  {title} {\enquote {\bibinfo {title}
  {Feedback of superconducting fluctuations on charge order in the underdoped
  cuprates},}\ }\href@noop {} {\bibfield  {journal} {\bibinfo  {journal} {Phys.
  Rev. B}\ }\textbf {\bibinfo {volume} {90}},\ \bibinfo {pages} {134516}
  (\bibinfo {year} {2014})}\BibitemShut {NoStop}%
\bibitem [{\citenamefont {Atkinson}\ \emph {et~al.}(2015)\citenamefont
  {Atkinson}, \citenamefont {Kampf},\ and\ \citenamefont
  {Bulut}}]{Atkinson:2014}%
  \BibitemOpen
  \bibfield  {author} {\bibinfo {author} {\bibfnamefont {W.~A.}\ \bibnamefont
  {Atkinson}}, \bibinfo {author} {\bibfnamefont {A.~P.}\ \bibnamefont {Kampf}},
  \ and\ \bibinfo {author} {\bibfnamefont {S.}~\bibnamefont {Bulut}},\
  }\bibfield  {title} {\enquote {\bibinfo {title} {Charge order in the
  pseudogap phase of cuprate superconductors},}\ }\href@noop {} {\bibfield
  {journal} {\bibinfo  {journal} {New Journal of Physics}\ }\textbf {\bibinfo
  {volume} {17}},\ \bibinfo {pages} {013025} (\bibinfo {year}
  {2015})}\BibitemShut {NoStop}%
\bibitem [{cav()}]{caveat2}%
  \BibitemOpen
  \href@noop {} {}\bibinfo {note} {It was pointed out in
  Ref.~\onlinecite{Verret:2017th} that the CDW does not by itself open a gap at
  the Fermi energy for general modulation wavevectors $\bq$; however, the
  formation of a CDW is only energetically favorable if at least a partial gap
  opens, and the value of $\bq$ that optimizes the CDW in self-consistent
  calculations is therefore also that which produces the largest gap in the
  density of states.}\BibitemShut {Stop}%
\bibitem [{\citenamefont {Wu}\ \emph {et~al.}()\citenamefont {Wu} \emph
  {et~al.}}]{Wu:2017}%
  \BibitemOpen
  \bibfield  {author} {\bibinfo {author} {\bibfnamefont {T.}~\bibnamefont {Wu}}
  \emph {et~al.},\ }\href@noop {} {}\bibinfo {note} {Unpublished}\BibitemShut
  {NoStop}%
\bibitem [{\citenamefont {Haase}\ \emph {et~al.}(2004)\citenamefont {Haase},
  \citenamefont {Sushkov}, \citenamefont {Horsch},\ and\ \citenamefont
  {Williams}}]{Haase:2004jk}%
  \BibitemOpen
  \bibfield  {author} {\bibinfo {author} {\bibfnamefont {J.}~\bibnamefont
  {Haase}}, \bibinfo {author} {\bibfnamefont {O.~P.}\ \bibnamefont {Sushkov}},
  \bibinfo {author} {\bibfnamefont {P.}~\bibnamefont {Horsch}}, \ and\ \bibinfo
  {author} {\bibfnamefont {G.~V.~M.}\ \bibnamefont {Williams}},\ }\bibfield
  {title} {\enquote {\bibinfo {title} {{Planar Cu and O Hole Densities in
  High-$T_c$ Cuprates Determined with NMR}},}\ }\href@noop {} {\bibfield
  {journal} {\bibinfo  {journal} {Phys. Rev. B}\ }\textbf {\bibinfo {volume}
  {69}},\ \bibinfo {pages} {094504} (\bibinfo {year} {2004})}\BibitemShut
  {NoStop}%
\bibitem [{\citenamefont {Grissonnanche}\ \emph {et~al.}(2014)\citenamefont
  {Grissonnanche}, \citenamefont {Cyr-Choini{\`e}re}, \citenamefont
  {Lalibert{\'e}}, \citenamefont {Ren{\'e}~de Cotret}, \citenamefont
  {Juneau-Fecteau}, \citenamefont {Dufour-Beaus{\'e}jour}, \citenamefont
  {Delage}, \citenamefont {LeBoeuf}, \citenamefont {Chang}, \citenamefont
  {Ramshaw}, \citenamefont {Bonn}, \citenamefont {Hardy}, \citenamefont
  {Liang}, \citenamefont {Adachi}, \citenamefont {Hussey}, \citenamefont
  {Vignolle}, \citenamefont {Proust}, \citenamefont {Sutherland}, \citenamefont
  {Kr{\"a}mer}, \citenamefont {Park}, \citenamefont {Graf}, \citenamefont
  {Doiron-Leyraud},\ and\ \citenamefont {Taillefer}}]{Grissonnanche:us}%
  \BibitemOpen
  \bibfield  {author} {\bibinfo {author} {\bibfnamefont {G.}~\bibnamefont
  {Grissonnanche}}, \bibinfo {author} {\bibfnamefont {O.}~\bibnamefont
  {Cyr-Choini{\`e}re}}, \bibinfo {author} {\bibfnamefont {F.}~\bibnamefont
  {Lalibert{\'e}}}, \bibinfo {author} {\bibfnamefont {S.}~\bibnamefont
  {Ren{\'e}~de Cotret}}, \bibinfo {author} {\bibfnamefont {A.}~\bibnamefont
  {Juneau-Fecteau}}, \bibinfo {author} {\bibfnamefont {S.}~\bibnamefont
  {Dufour-Beaus{\'e}jour}}, \bibinfo {author} {\bibfnamefont {M.~{\`E}.}\
  \bibnamefont {Delage}}, \bibinfo {author} {\bibfnamefont {D.}~\bibnamefont
  {LeBoeuf}}, \bibinfo {author} {\bibfnamefont {J.}~\bibnamefont {Chang}},
  \bibinfo {author} {\bibfnamefont {B.~J.}\ \bibnamefont {Ramshaw}}, \bibinfo
  {author} {\bibfnamefont {D.~A.}\ \bibnamefont {Bonn}}, \bibinfo {author}
  {\bibfnamefont {W.~N.}\ \bibnamefont {Hardy}}, \bibinfo {author}
  {\bibfnamefont {R.}~\bibnamefont {Liang}}, \bibinfo {author} {\bibfnamefont
  {S.}~\bibnamefont {Adachi}}, \bibinfo {author} {\bibfnamefont {N.~E.}\
  \bibnamefont {Hussey}}, \bibinfo {author} {\bibfnamefont {B.}~\bibnamefont
  {Vignolle}}, \bibinfo {author} {\bibfnamefont {C.}~\bibnamefont {Proust}},
  \bibinfo {author} {\bibfnamefont {M.}~\bibnamefont {Sutherland}}, \bibinfo
  {author} {\bibfnamefont {S.}~\bibnamefont {Kr{\"a}mer}}, \bibinfo {author}
  {\bibfnamefont {J.~H.}\ \bibnamefont {Park}}, \bibinfo {author}
  {\bibfnamefont {D.}~\bibnamefont {Graf}}, \bibinfo {author} {\bibfnamefont
  {N.}~\bibnamefont {Doiron-Leyraud}}, \ and\ \bibinfo {author} {\bibfnamefont
  {Louis}\ \bibnamefont {Taillefer}},\ }\bibfield  {title} {\enquote {\bibinfo
  {title} {{Direct measurement of the upper critical field in cuprate
  superconductors}},}\ }\href@noop {} {\bibfield  {journal} {\bibinfo
  {journal} {Nat. Commun.}\ }\textbf {\bibinfo {volume} {5}},\ \bibinfo {pages}
  {3280} (\bibinfo {year} {2014})}\BibitemShut {NoStop}%
\bibitem [{\citenamefont {Zhou}\ \emph {et~al.}()\citenamefont {Zhou},
  \citenamefont {Hirata}, \citenamefont {Wu}, \citenamefont {Vinograd},
  \citenamefont {Mayaffre}, \citenamefont {Kr\"amer}, \citenamefont {Reyes},
  \citenamefont {Kuhns}, \citenamefont {Liang}, \citenamefont {Hardy},
  \citenamefont {Bonn},\ and\ \citenamefont {Julien}}]{Zhou:2017}%
  \BibitemOpen
  \bibfield  {author} {\bibinfo {author} {\bibfnamefont {Rui}\ \bibnamefont
  {Zhou}}, \bibinfo {author} {\bibfnamefont {M.}~\bibnamefont {Hirata}},
  \bibinfo {author} {\bibfnamefont {T.}~\bibnamefont {Wu}}, \bibinfo {author}
  {\bibfnamefont {I.}~\bibnamefont {Vinograd}}, \bibinfo {author}
  {\bibfnamefont {H.}~\bibnamefont {Mayaffre}}, \bibinfo {author}
  {\bibfnamefont {S.}~\bibnamefont {Kr\"amer}}, \bibinfo {author}
  {\bibfnamefont {A.P.}\ \bibnamefont {Reyes}}, \bibinfo {author}
  {\bibfnamefont {P.L.}\ \bibnamefont {Kuhns}}, \bibinfo {author}
  {\bibfnamefont {R.}~\bibnamefont {Liang}}, \bibinfo {author} {\bibfnamefont
  {W.N.}\ \bibnamefont {Hardy}}, \bibinfo {author} {\bibfnamefont {D.~A.}\
  \bibnamefont {Bonn}}, \ and\ \bibinfo {author} {\bibfnamefont {M.-H.}\
  \bibnamefont {Julien}},\ }\href@noop {} {\enquote {\bibinfo {title} {{Spin
  susceptibility of charge ordered YBa$_2$Cu$_3$O$_y$ across the upper critical
  field}},}\ }\bibinfo {note} {Unpublished},\ \Eprint
  {http://arxiv.org/abs/https://arxiv.org/abs/1711.00109}
  {https://arxiv.org/abs/1711.00109} \BibitemShut {NoStop}%
\bibitem [{\citenamefont {Hunt}\ \emph {et~al.}(2001)\citenamefont {Hunt},
  \citenamefont {Singer}, \citenamefont {Cederstr{\"o}m},\ and\ \citenamefont
  {Imai}}]{Hunt:2001jp}%
  \BibitemOpen
  \bibfield  {author} {\bibinfo {author} {\bibfnamefont {A~W}\ \bibnamefont
  {Hunt}}, \bibinfo {author} {\bibfnamefont {P~M}\ \bibnamefont {Singer}},
  \bibinfo {author} {\bibfnamefont {A~F}\ \bibnamefont {Cederstr{\"o}m}}, \
  and\ \bibinfo {author} {\bibfnamefont {T}~\bibnamefont {Imai}},\ }\bibfield
  {title} {\enquote {\bibinfo {title} {{Glassy slowing of stripe modulation in
  (La,Eu,Nd)$_{2-x}$(Sr,Ba)$_x$CuO4:â ${}^{63}$Cuand ${}^{139}$La NQR study
  down to 350 mK}},}\ }\href@noop {} {\bibfield  {journal} {\bibinfo  {journal}
  {Phys. Rev. B}\ }\textbf {\bibinfo {volume} {64}},\ \bibinfo {pages} {134525}
  (\bibinfo {year} {2001})}\BibitemShut {NoStop}%
\bibitem [{\citenamefont {Ouazi}\ \emph {et~al.}(2006)\citenamefont {Ouazi},
  \citenamefont {Bobroff}, \citenamefont {Alloul}, \citenamefont {Le~Tacon},
  \citenamefont {Blanchard}, \citenamefont {Collin}, \citenamefont {Julien},
  \citenamefont {Horvati{\'c}},\ and\ \citenamefont {Berthier}}]{Ouazi:2006ep}%
  \BibitemOpen
  \bibfield  {author} {\bibinfo {author} {\bibfnamefont {S.}~\bibnamefont
  {Ouazi}}, \bibinfo {author} {\bibfnamefont {J.}~\bibnamefont {Bobroff}},
  \bibinfo {author} {\bibfnamefont {H.}~\bibnamefont {Alloul}}, \bibinfo
  {author} {\bibfnamefont {M.}~\bibnamefont {Le~Tacon}}, \bibinfo {author}
  {\bibfnamefont {N.}~\bibnamefont {Blanchard}}, \bibinfo {author}
  {\bibfnamefont {G.}~\bibnamefont {Collin}}, \bibinfo {author} {\bibfnamefont
  {M.~H.}\ \bibnamefont {Julien}}, \bibinfo {author} {\bibfnamefont
  {M.}~\bibnamefont {Horvati{\'c}}}, \ and\ \bibinfo {author} {\bibfnamefont
  {C.}~\bibnamefont {Berthier}},\ }\bibfield  {title} {\enquote {\bibinfo
  {title} {{Impurity-Induced Local Magnetism and Density of States in the
  Superconducting State of YBa$_2$Cu$_3$O$_7$}},}\ }\href@noop {} {\bibfield
  {journal} {\bibinfo  {journal} {Phys. Rev. Lett.}\ }\textbf {\bibinfo
  {volume} {96}},\ \bibinfo {pages} {127005} (\bibinfo {year}
  {2006})}\BibitemShut {NoStop}%
\bibitem [{\citenamefont {Harter}\ \emph {et~al.}(2007)\citenamefont {Harter},
  \citenamefont {Andersen}, \citenamefont {Bobroff}, \citenamefont {Gabay},\
  and\ \citenamefont {Hirschfeld}}]{Harter:2007da}%
  \BibitemOpen
  \bibfield  {author} {\bibinfo {author} {\bibfnamefont {J~W}\ \bibnamefont
  {Harter}}, \bibinfo {author} {\bibfnamefont {B.~M.}\ \bibnamefont
  {Andersen}}, \bibinfo {author} {\bibfnamefont {J}~\bibnamefont {Bobroff}},
  \bibinfo {author} {\bibfnamefont {M}~\bibnamefont {Gabay}}, \ and\ \bibinfo
  {author} {\bibfnamefont {P.~J.}\ \bibnamefont {Hirschfeld}},\ }\bibfield
  {title} {\enquote {\bibinfo {title} {{Antiferromagnetic Correlations and
  Impurity Broadening of NMR Linewidths in Cuprate Superconductors}},}\
  }\href@noop {} {\bibfield  {journal} {\bibinfo  {journal} {Phys. Rev. B}\
  }\textbf {\bibinfo {volume} {75}},\ \bibinfo {pages} {054520} (\bibinfo
  {year} {2007})}\BibitemShut {NoStop}%
\bibitem [{\citenamefont {Andersen}\ \emph {et~al.}(1995)\citenamefont
  {Andersen}, \citenamefont {Liechtenstein}, \citenamefont {Jepsen},\ and\
  \citenamefont {Paulsen}}]{Andersen:1995}%
  \BibitemOpen
  \bibfield  {author} {\bibinfo {author} {\bibfnamefont {O.~K.}\ \bibnamefont
  {Andersen}}, \bibinfo {author} {\bibfnamefont {A.~I.}\ \bibnamefont
  {Liechtenstein}}, \bibinfo {author} {\bibfnamefont {O.}~\bibnamefont
  {Jepsen}}, \ and\ \bibinfo {author} {\bibfnamefont {F.}~\bibnamefont
  {Paulsen}},\ }\bibfield  {title} {\enquote {\bibinfo {title} {{LDA Energy
  Bands, Low-Energy Hamiltonians, $t'$, $t''$, $t_\perp({\bf k})$, and
  $J_\perp$}},}\ }\href {\doibase
  http://dx.doi.org/10.1016/0022-3697(95)00269-3} {\bibfield  {journal}
  {\bibinfo  {journal} {J. Phys. Chem. Solids}\ }\textbf {\bibinfo {volume}
  {56}},\ \bibinfo {pages} {1573 -- 1591} (\bibinfo {year} {1995})},\ \bibinfo
  {note} {proceedings of the Conference on Spectroscopies in Novel
  Superconductors}\BibitemShut {NoStop}%
\bibitem [{\citenamefont {McMillan}(1975)}]{McMillan:1975}%
  \BibitemOpen
  \bibfield  {author} {\bibinfo {author} {\bibfnamefont {W.~L.}\ \bibnamefont
  {McMillan}},\ }\bibfield  {title} {\enquote {\bibinfo {title} {Landau theory
  of charge-density waves in transition-metal dichalcogenides},}\ }\href
  {\doibase 10.1103/PhysRevB.12.1187} {\bibfield  {journal} {\bibinfo
  {journal} {Phys. Rev. B}\ }\textbf {\bibinfo {volume} {12}},\ \bibinfo
  {pages} {1187--1196} (\bibinfo {year} {1975})}\BibitemShut {NoStop}%
\bibitem [{\citenamefont {Arguello}\ \emph {et~al.}(2014)\citenamefont
  {Arguello}, \citenamefont {Chockalingam}, \citenamefont {Rosenthal},
  \citenamefont {Zhao}, \citenamefont {Guti\'errez}, \citenamefont {Kang},
  \citenamefont {Chung}, \citenamefont {Fernandes}, \citenamefont {Jia},
  \citenamefont {Millis}, \citenamefont {Cava},\ and\ \citenamefont
  {Pasupathy}}]{Arguello:2014}%
  \BibitemOpen
  \bibfield  {author} {\bibinfo {author} {\bibfnamefont {C.~J.}\ \bibnamefont
  {Arguello}}, \bibinfo {author} {\bibfnamefont {S.~P.}\ \bibnamefont
  {Chockalingam}}, \bibinfo {author} {\bibfnamefont {E.~P.}\ \bibnamefont
  {Rosenthal}}, \bibinfo {author} {\bibfnamefont {L.}~\bibnamefont {Zhao}},
  \bibinfo {author} {\bibfnamefont {C.}~\bibnamefont {Guti\'errez}}, \bibinfo
  {author} {\bibfnamefont {J.~H.}\ \bibnamefont {Kang}}, \bibinfo {author}
  {\bibfnamefont {W.~C.}\ \bibnamefont {Chung}}, \bibinfo {author}
  {\bibfnamefont {R.~M.}\ \bibnamefont {Fernandes}}, \bibinfo {author}
  {\bibfnamefont {S.}~\bibnamefont {Jia}}, \bibinfo {author} {\bibfnamefont
  {A.~J.}\ \bibnamefont {Millis}}, \bibinfo {author} {\bibfnamefont {R.~J.}\
  \bibnamefont {Cava}}, \ and\ \bibinfo {author} {\bibfnamefont {A.~N.}\
  \bibnamefont {Pasupathy}},\ }\bibfield  {title} {\enquote {\bibinfo {title}
  {Visualizing the charge density wave transition in $2h$-${\text{nbse}}_{2}$
  in real space},}\ }\href {\doibase 10.1103/PhysRevB.89.235115} {\bibfield
  {journal} {\bibinfo  {journal} {Phys. Rev. B}\ }\textbf {\bibinfo {volume}
  {89}},\ \bibinfo {pages} {235115} (\bibinfo {year} {2014})}\BibitemShut
  {NoStop}%
\bibitem [{ske()}]{skewness}%
  \BibitemOpen
  \href@noop {} {}\bibinfo {note} {In finite-size systems, the skewness is
  sensitive to how it is calculated. Here, the local density of states is
  calculated using a lorentzian broadening $\eta$ that must be larger than the
  level spacing $\Delta \sim (N_0 L^2 N_k^2)^{-1}$ and smaller than relevant
  features, such as the CDW gap. The LDOS distributions are calculated with
  $\eta = 0.002$ and $N_k = 16$, corresponding to $\Delta = 2\times
  10^{-5}$.}\BibitemShut {Stop}%
\bibitem [{\citenamefont {Comin}\ \emph
  {et~al.}(2015{\natexlab{b}})\citenamefont {Comin}, \citenamefont {Sutarto},
  \citenamefont {da~Silva~Neto}, \citenamefont {Chauviere}, \citenamefont
  {Liang}, \citenamefont {Hardy}, \citenamefont {Bonn}, \citenamefont {He},
  \citenamefont {Sawatzky},\ and\ \citenamefont {Damascelli}}]{Comin:2015fo}%
  \BibitemOpen
  \bibfield  {author} {\bibinfo {author} {\bibfnamefont {R.}~\bibnamefont
  {Comin}}, \bibinfo {author} {\bibfnamefont {R.}~\bibnamefont {Sutarto}},
  \bibinfo {author} {\bibfnamefont {E.~H.}\ \bibnamefont {da~Silva~Neto}},
  \bibinfo {author} {\bibfnamefont {L.}~\bibnamefont {Chauviere}}, \bibinfo
  {author} {\bibfnamefont {R.}~\bibnamefont {Liang}}, \bibinfo {author}
  {\bibfnamefont {W.~N.}\ \bibnamefont {Hardy}}, \bibinfo {author}
  {\bibfnamefont {D.~A.}\ \bibnamefont {Bonn}}, \bibinfo {author}
  {\bibfnamefont {F.}~\bibnamefont {He}}, \bibinfo {author} {\bibfnamefont
  {G.~A.}\ \bibnamefont {Sawatzky}}, \ and\ \bibinfo {author} {\bibfnamefont
  {A.}~\bibnamefont {Damascelli}},\ }\bibfield  {title} {\enquote {\bibinfo
  {title} {{Broken Translational and Rotational Symmetry Via Charge Stripe
  Order in Underdoped YBa$_{2}$Cu$_{3}$O$_{6+y}$}},}\ }\href@noop {} {\bibfield
   {journal} {\bibinfo  {journal} {Science}\ }\textbf {\bibinfo {volume}
  {347}},\ \bibinfo {pages} {1335--1339} (\bibinfo {year}
  {2015}{\natexlab{b}})}\BibitemShut {NoStop}%
\bibitem [{\citenamefont {Sebastian}\ \emph {et~al.}(2012)\citenamefont
  {Sebastian}, \citenamefont {Harrison},\ and\ \citenamefont
  {Lonzarich}}]{Sebastian:2012wh}%
  \BibitemOpen
  \bibfield  {author} {\bibinfo {author} {\bibfnamefont {Suchitra~E.}\
  \bibnamefont {Sebastian}}, \bibinfo {author} {\bibfnamefont {N.}~\bibnamefont
  {Harrison}}, \ and\ \bibinfo {author} {\bibfnamefont {Gilbert}\ \bibnamefont
  {Lonzarich}},\ }\bibfield  {title} {\enquote {\bibinfo {title} {{Towards
  Resolution of the Fermi Surface in Underdoped High-Tc Superconductors }},}\
  }\href@noop {} {\bibfield  {journal} {\bibinfo  {journal} {Rep. Prog. Phys.}\
  }\textbf {\bibinfo {volume} {75}},\ \bibinfo {pages} {102501} (\bibinfo
  {year} {2012})}\BibitemShut {NoStop}%
\end{thebibliography}%

\end{document}